\documentclass[12pt]{report}
\pdfoutput=1
\usepackage[margin=3cm]{geometry}
\usepackage{setspace}
\usepackage{appendix}
\usepackage[font=small]{caption}
\usepackage{graphicx}
\usepackage{amsmath,amsfonts,amssymb}
\usepackage{braket,cancel,slashed}
\usepackage{mathrsfs}
\usepackage[numbers,merge,sort&compress]{natbib}
\def\lsim{\mathrel{\vcenter{\hbox{$<$}\nointerlineskip\hbox{$\sim$}}}}
\def\gsim{\mathrel{\vcenter{\hbox{$>$}\nointerlineskip\hbox{$\sim$}}}}

\begin{document}

\title{\LARGE Neutrino Mass and Proton Lifetime in a Realistic 
SUSY $SO(10)$ Model}
\author{\Large {\bf Matthew Severson}\\
Dissertation for Doctor of Philosophy\\
Maryland Center for Fundamental Physics\\ Department of Physics\\
University of Maryland, College Park, MD 20742, USA\\}
\date{Dissertation Committee:\\ Rabindra Mohapatra, Chair\\
Jeffrey Adams, Dean's Representative\\ Kaustubh Agashe\\ Zackaria
Chacko\\ Gregory Sullivan}

\maketitle

\begin{abstract}
This work presents a complete analysis of fermion fitting and proton
decay in a supersymmetric $SO(10)$ model previously suggested by
Dutta, Mimura, and Mohapatra.

A key question in any grand unified theory is whether it satisfies the
stringent experimental lower limits on the partial lifetimes of the
proton. In more generic models, substantial fine-tuning is required
among GUT-scale parameters to satisfy the limits. In the proposed
model, the {\bf 10}, $\overline{\bf{126}}$, and {\bf 120} Yukawa
couplings contributing to fermion masses have restricted textures
intended to give favorable results for proton lifetime, while still
giving rise to a realistic fermion sector, without the need for
fine-tuning, even for large $\tan\beta$, and for either type-I or
type-II dominance in the neutrino mass matrix.

In this thesis, I investigate the above hypothesis at a strict
numerical level of scrutiny; I obtain a valid fit for the entire
fermion sector for both types of seesaw dominance, including
$\theta_{13}$ in good agreement with the most recent data. For the
case with type-II seesaw, I find that, using the Yukawa couplings
fixed by the successful fermion sector fit, proton partial lifetime
limits are readily satisfied for all but one of the pertinent decay
modes for nearly arbitrary values of the triplet-Higgs mixing
parameters, with the $K^+ \bar\nu$ mode requiring a minor ${\cal
O}(10^{-1})$ cancellation in order to satisfy its limit.  I also find
a maximum partial lifetime for that mode of $\tau(K^+ \bar\nu) \sim
10^{36}$\,years. For the type-I seesaw case, I find that $K^+ \bar\nu$
decay mode is satisfied for any values of the triplet mixing
parameters giving no major enhancement, and all other modes are easily
satisfied for arbitrary mixing values; I also find a maximum partial
lifetime for $K^+ \bar\nu$ of nearly $10^{38}$\,years, which is
largely sub-dominant to gauge boson decay channels.
\end{abstract}

\pagestyle{empty}
\pagenumbering{none}

\-\vspace{4em}
\begin{center} 
  \emph{For Erin, for my family, and for all the buds. \\[4mm] Booj.} 
\end{center} 


\pagestyle{plain}
\onehalfspacing
\pagenumbering{roman} \setcounter{page}{3}
\begin{center} \large Acknowledgments \end{center} 

\vspace{1cm}

I owe my nearly all of my competence in physics as well as the
completion of this thesis to a great number of individuals within the
University of Maryland community, to whom I am so indebted.

I would first like to deeply thank my advisor, Rabindra ``Rabi''
Mohapatra, for his many years of commitment to my endeavors at
Maryland, and for sharing with me his pioneering intuitions for the
wonderous structures of Grand Unified theory. His instruction and
guidance were freely offered as often as needed (and repeated\ldots
and repeated), and his kindness and light-hearted patience throughout
are primary reasons I was not defeated by frustration or
discouragement. 

I would like to extend thanks to several other faculty members of UMD
physics and its nearest neighbors as well. Kaustubh Agashe and
Zackaria Chacko provided guidance in primary coursework, service on my
dissertation committee, and many important insights in the years
between. As members of the committee from outside MCFP, Gregory
Sullivan and Jeffrey Adams offered new perspectives surprisingly close
to the scope of this work and yet previously so far from my sight.
Thanks go as well to Raman Sundrum, Jonathan Rosenberg, and William
Linch for sharing their tremdendous insights in various lectures and
courses.

I further owe gratitude to a great number of fellow and former
students whose assistance was truly crucial to my efforts to
understand quantum field theory, the Standard Model and beyond, and
modern physics in general. The umdphys08 group provided not only a
great deal of instruction and guidance in early coursework, but also
solidarity in the dark days of first year and, ultimately, a great
deal of guidance in re-training my mind to be more reasonable. William
Donnelly and Evan Berkowitz were especially patient in answering a
myriad of questions about mathematics. Bhupal Dev filled in countless
blanks as I began research and my quest to study beyond the Stardard
Model physics. Sim\'on Riquelme has been invaluable in furthering my
knowledge, scope, and intuition for field theory and gravity. Numerous
additional colleagues were always willing to discuss or distract in
times of confusion.

Additional thanks go to the entire Department of Physics staff, who
were always happy to keep my professional life from falling to pieces,
and to the National Science Foundation and the University of Maryland
for financial support throughout my enrollment in this program.

I would also like to acknowledge my undergraduate professor, Justin
Sanders, who is perhaps the most personally influential physics
instructor I had in all my years of study; it is safe to say I would
likely still be quite uncomfortable with quantum mechanics if not for
his guidance.

Finally, special thanks go to Michael Richman for his tireless efforts
in teaching me Linux and Python as well as how to code beyond the for
loop, how to say what I mean, and how to train myself to be better at
training myself.

Outside the scope of physics, I will always be thankful for the
support of so many others as I endured the long struggle to complete
this work; without them, I simply would not have made it.

My parents, Mike and Robin Severson, believed in my abilities so
strongly that they hardly flinched when I announced I would be
starting college over to study physics. I can never thank them enough
for their patience and confidence in me, or for passing on to me the
ability to love so intensely. Thanks also go to Becky Severson and Pat
Mulroney for their support and encouragement, and for their commitment
to my parents' happiness. 

The deepest of thanks to my brother, Nick, for giving me the ability
to believe in myself, and for teaching me how to limitlessly expand
the world through imagination.

Thanks go as well to my grandparents, to the entire Hohn and Severson
families, and to all of my closest friends back home and around the
country for their encouragement and enthusiasm over the years, as well
as for enduring my lengthy absence as I committed to completion of
this thesis. Special thanks to my grandmother, Vivian Hohn, for her
passion and confidence in me, to my uncle, Brad Ellison of LSU
physics, for his interest and support, and to Zach Wilkerson for his
commitment to growing along with me throughout this endeavor.

I am grateful, too, for the many new friendships that developed during
my time at UMD. In addition to Mike Richman and Sim\'on Riquelme, Bill
McConville, Paul Schmidt, Crystal Wheaden, and Mike Azatov have helped
make life in College Park entirely bearable, especially through music.

Finally, to my wife, Erin Moody, words cannot express how deeply 
grateful I am for her unending love, compassion, encouragement, and 
commitment during even the darkest of times throughout this experience.

\singlespacing
\tableofcontents

\newpage
\onehalfspacing
\pagenumbering{arabic} \setcounter{page}{1}

%
%

\chapter{Introduction}\label{1-intro}

The Standard Model of particle physics \cite{glashow-SM,*salam-ward,
*salam,*weinberg} is among the most fascinating of modern marvels,
though it is an inconspicuous one. Its mathematical structure is
capable of describing, with unparalleled precision, virtually every
aspect of the statistical behavior of the elementary particles
composing normal matter. With its last key aspects discovered by the
early 1970's, the completed model emerged as the culmination of some
forty years of effort to solve the many mysteries generated by the
discoveries of quantum mechanics and relativity in the early 20th
century.

Yet even as the final pieces were being put in place, physicists were
already certain the model and its implications gave an incomplete
version of the story of our universe: for as many questions as it
answered with the utmost of elegance, the Standard Model (SM) left
many mysteries unsolved and also gave rise to a few new ones. The
model gives no indication as to why, in light of electroweak
unification, there were still three separate forces in nature; in
fact, it quite conspicuously gives no description of gravity, and
further gives no explanation for dark matter or matter-antimatter
asymmetry. Additionally it suggests that electric charge is quantized
but provides no explanation for why it should be, nor does it
relatedly give any reason for the values of hypercharge.

Furthermore, empirical evidence for other failures of the model were
coming to light even before its completion. One important example of
such evidence indicated a discrepancy in solar neutrino flux, which
would ultimately come to be understood as a consequence of the
oscillation of propagating neutrinos from one flavor to another
\cite{kajita,*superk-atm,*sno}. It was already known at the time that
such oscillations occur only among particles having mass, whereas the
SM predicted neutrinos to be massless.

Thus, theorists began working to find an extension of the model that
would solve its problems without disrupting the beautiful predictions
of its existing framework. One of the first notions to lead to some
success was \emph{Grand Unification} \cite{su5,pati-salam}, which
nests the symmetry group of the SM in a higher dimensional group by
expanding the potential (or superpotential) to include terms allowed
by the higher dimensional symmetry; the new potential typically
introduces heavy Higgs-like bosons and may include new multiplets of
existing particles. Such a mathematical extension of the model is
phenomenologically justified through the assumption that the
``larger'' symmetry of the Grand Unified theory would have been
present at higher energies typical in the early universe, and that the
SM symmetry would \emph{emerge} at low energies through a
\emph{spontaneous breaking} of the larger symmetry. Grand Unified
theory (GUT) provided understanding for some of the mysteries of the
SM, and, when combined with the seesaw mechanism (see below) a few
years later, it led to a nicely self-consistent and potentially
testable explanation for neutrino masses and their apparent smallness.
GUT framework again created some new questions of its own, and it also
gave some curious predictions, such as the existence of proton decay
\cite{su5}.

Over the past few decades, and through the inclusion of Supersymmetry
(SUSY) \cite{golfand,*wess-zumino1,*wess-zumino2, volkov,*gervais}, a
few classes of GUT models, especially those based on the $SO(10)$
symmetry group \cite{minkowski,*georgi}, have come to be realized as
significantly more complete descriptions of our universe than the
Standard Model. One of the more basic yet intriguing features of these
models is the ability to naturally accommodate a right-handed
neutrino, consequently allowing for a well-motivated implementation of
the seesaw mechanism for neutrino mass \cite{mink-seesaw,*yanagida,
*gell-slansky,*glashow-seesaw,rabi-seesaw}, a long-uncontested ansatz
that dynamically explains the smallness of left-handed neutrino
masses. The seesaw was originally implemented in the framework of SUSY
$SO(10)$ with {\bf 10}- and {\bf 126}-dimensional Higgs multiplets
coupling to fermions \cite{rabi-aulakh,kuo}; the vacuum expectation
value (vev) of the {\bf 126} field plays the role of both breaking
$B\!-\!L$ and triggering the seesaw mechanism, thereby creating a deep
mathematical connection between the smallness of neutrino masses and
the other fermion masses. This seemingly limited yet elegant approach
yielded a realistic neutrino sector, including an accurate prediction
of the value of $\theta_{13}$ \cite{goh-ng,babu-mace}, long before
experiments were measuring its value. In the SUSY context, it further
provides a clear candidate for dark matter. This so-called ``minimal''
$SO(10)$ model has been explored much more thoroughly over the years
by many authors with the arrival of precision measurements
\cite{rabi-babu,babu-mace,lazarides, *schechter,*rabi-senj,bajc,
goh-ng,bertolini1,*bertolini2, *bertolini3,*joshipura,fukuyama1,
*fukuyama2, *fukuyama3,*fukuyama4, altarelli}, and it remains a viable
predictor of the neutrino sector parameters.

Many of the remaining concerns associated with GUT models are on the
verge of being addressed experimentally. Theorists and
phenomenologists have made extensive effort to carefully explore and
catalogue in the vast number of feasible options available when
constructing such a model, because each choice leads to a distinct set
of favorable and unfavorable phenomenological features. It seems that
within the next 10-20 years, this formidable tree of models will
finally be pruned substantially as experiments close in on precise
values for the phenomenological outputs whose predictions may
distinguish one model from the next, including the remaining
parameters of neutrino oscillation \cite{dune} and the lifetime of
the proton \cite{hyperk}.

Proton decay is arguably the most problematic feature common to nearly
all GUT models. In all $SU(5)$ and $SO(10)$ models, heavy gauge boson
exchanges give rise to effective higher-dimensional operators that
allow for quark-lepton mixing and, consequently, nonzero probabilities
for proton decay widths. Furthermore, in SUSY GUT models, although one
sees an decrease in the decay widths following from gauge boson
exchange, several additional decay modes are available, as each of the
GUT-scale Higgs superfields contains colored Higgs triplets that
allows for proton decay through exchange of Higgsino superpartners.

No one yet knows whether protons do in fact decay at all; if the
answer turns out to be no, that will of course be the end of the line
for GUT models without some new mechanism. So far, the lower limit on
proton lifetime is known to be at least $\sim\!\!10^{33}$ years, and
partial lifetimes for the various decay modes have been continually
rising through the findings of experiments \cite{superk-pdecay}. Thus,
if any $SO(10)$ model is to be trusted, its prediction for the proton
lifetime must be at least so high a number. Most minimal $SU(5)$
models have already been virtually ruled out by such limits. 

There are ways in which the proton lifetime goal can be achieved
within the framework of a given model, but doing so typically requires
substantial fine-tuning, which occurs via rather extreme cancellations
($\gsim \mathcal{O}(10^{-4})$) among the mixing parameters of the
color-triplet Higgsinos exchanged in the decay. The values of those
mixings cannot be reasonably recognized as more than arbitrary free
parameters, so to expect multiple instances of very sensitive
relationships among them requires putting much faith in either unknown
dynamics or extremely good luck. Restricting the SUSY vev ratio
$v_u/v_d$, conventionally parametrized as $\tan\beta$, to small values
can provide some relief without cancellation for Higgsino-mediated
decay channels, but such an assumption is still \emph{ad hoc} and may
ultimately be inconsistent with experimental findings; hence it is
strongly preferable to construct a model which is tractable for any
feasible $\tan\beta$.

If however the GUT Yukawas, which are $3\!\times\!3$ matrices in
generation space, have some key elements naturally small or zero, then
extreme cancellations can be largely avoided by eliminating most of
the dominant contributions to proton decay width. A paper by
Dutta, Mimura, and Mohapatra \cite{dmm1302} proposed such a Yukawa
texture for the $SO(10)$ model that includes a \textbf{120}
coupling in addition to the \textbf{10} and $\overline{\bf{126}}$
Higgs contributions to fermion masses. The authors suggested that
proton decay limits may be satisfied, especially for model with
type-II seesaw dominance and sketched the relationships between key
fermion fit parameters and proton partial lifetimes; however, the work
gave mainly heuristic arguments and leading-order estimates to only
tentatively support the hypothesis.

The work I present in this thesis revisits the above hypothesis and
exposes it to robust testing by providing a careful and complete
analysis of the characteristics of proton decay in the model. I
grounded the analysis in conservative assumptions, including large
$\tan\beta$, and performed a comprehensive numerical calculation
relying on as few approximations as necessary. Furthermore, I extended
the cursory work from ref. \cite{dmm1302} for type-I seesaw to fully
consider both the type-I and II seesaw dominance cases. The modes of
proton decay that I checked for sufficiency are those known to be most
problematic: $p \rightarrow K^+ \bar{\nu}$, $K^0 \ell^+$, $\pi^+
\bar{\nu}$, and $\pi^0 \ell^+$, where $\ell = e,\mu$.

The calculation consisted of two components: first I found a stable
numerical fit to all fermion mass and mixing parameters, including the
neutrino sector (where values are predictions of the model); then,
using the Yukawa couplings fixed by the fermion fit as input, I
searched the parameter space of heavy color triplet mixing
parameters for areas that lead to adequately large partial lifetimes
for the dominant modes of proton decay. 

The results not only give satisfactory predictions for the neutrino
sector based on corresponding charged sector fits, but also adequately
predict sufficiently long-lived protons without relying on the usual
large degree of tuning. I find that the ansatz is \emph{completely
successful} in satisfying the proton lifetime limits without any need
for cancellation for the type-I seesaw scenario; a modest
$\mathcal{O}(10^{-1})$ cancellation is needed in the type-II case to
satisfy the partial lifetime limit of the often-problematic $p
\rightarrow K^+ \bar{\nu}$ mode. These results for type-I versus
type-II are contrary to the tentative expectations of the authors in
\cite{dmm1302}; the discrepancy is due mainly to the unexpected
significance of the effect of rotation to mass basis on the results of
the decay width calculations, combined with the numerical details of
the rotation matrices arising from the charged sector mass and CKM
fit. 

The thesis is organized as follows. In chapter \ref{2-sm}, I give an
introduction to the Standard Model of particle physics and discuss its
strengths and weaknesses. In chapter \ref{3-susy}, I give an
introduction to supersymmetry and the Minimally Supersymmetric
Standard Model (MSSM) and again discuss its strengths and weaknesses.
In chapter \ref{4-so10}, I give an overview of Grand Unified theories
and their strengths and weaknesses and an introduction to $SO(10)$
models; I also introduce the details of the model on which this work
focuses, including the superpotential and the fermion mass matrices
following from it, and the details of the Yukawa texture ansatz. In
chapter \ref{5-pdecay}, I expand further on the model specifics and
examine general GUT proton-decay logistics in order to derive the
needed partial decay widths. In chapter \ref{6-results}, I present the
fermion sector results of the numerical fitting to the measured masses
and mixings, and I present the results of the calculation of the
important partial lifetimes of the proton. In chapter
\ref{7-conclusion}, I discuss the implications of the results and give
my conclusions.

\chapter{The Standard Model}\label{2-sm}

\section{The Structure of the Standard Model}
Strictly speaking, the Standard Model (SM) is a
\emph{spontaneously-broken non-Abelian gauge theory of quantum
fields}. This extremely content-laden tagline can be parsed as
follows.

A \emph{quantum field} is a function over some space or spacetime that
assigns an algebraic operator, rather than a numerical value, to each
point in the space. Such an operator typically acts on elements of a
separate internal vector space; that action creates (or destroys)
discrete excited states of the underlying field called \emph{quanta}.
The actions of multiple operators are not generally commutative.
    
In relativistic quantum field theory, \emph{elementary particles} are
realized as excitations in \emph{Fock space}, which is a
generalization of the (non-relativistic) quantum-mechanical Hilbert
space that allows for the accommodation of multi-particle states in
which the number of particles is not fixed. The ``value'' of a typical
(scalar) quantum field $\phi$ at a spacetime point $x$ goes like
$\mathrm{e}^{i p \cdot x}\, \hat{a}^\dagger \ket{0}$ or
$\mathrm{e}^{-i p \cdot x}\, \hat{a} \ket{0}$, where $\hat{a}^\dagger$
is the raising operator (like that of a harmonic oscillator) whose
action on the Fock space \emph{ground state} $\ket{0}$ (``the
vacuum'') creates a single quantum of the field. The new state
$\hat{a}^\dagger \ket{0}$, explicitly notated as ``$\ket{1}$'' or,
more commonly, ``$\ket{p}$'', is identified with a plane wave carrying
momentum $p$, ``pinned'' to spacetime at the point $x$, and it can be
further associated with a \emph{representation} of the \emph{Lorentz
group}, $SO(1,3)$, which I will describe in detail shortly. The
lowering operator $\hat{a}$ acting on $\ket{p}$ destroys a single
field quantum, while $\hat{a}^\dagger \hat{a}^\dagger \ket{0}$ creates
two quanta, corresponding to a two-particle state $\ket{p_1 p_2}$, and
so on. Note though that states of more than one identical particles
are forbidden for fermionic fields due to the Pauli exclusion
principle. As with any lowering operator, $\hat{a} \ket{0} = 0$. 

Both ``non-Abelian'' and ``gauge'' theories of quantum fields are
types of \emph{group theories}. A \emph{group} is a set of elements,
together with an associative operation, that
\begin{itemize}
  \item is \emph{closed} under the action of the operation on any two 
    elements
  \item contains a unique \emph{identity} element
  \item contains a unique \emph{inverse} for every element.
\end{itemize}
The set of elements of a group can be finite and discrete,
countably infinite, or a continuous spectrum. A simple example of a
group is the integers with the addition operation $\{\mathbb{Z}, +\}$,
where zero is the identity element and negative integers are the
inverse elements of positive integers (and vice
versa).

If the elements of a continuous group of also form a topological
\emph{manifold} (\emph{i.e.}, if the space is ``smooth'', or
continuous and differentiable throughout), then the group is known as
a \emph{Lie group}.

A \emph{non-Abelian} group is a group (finite or continuous) for
which the group operation is non-commutative on two elements; i.e.,
for elements $a,b$ of a group $\{G,\cdot\}$, $a \cdot b \neq b \cdot
a$.

Before I can give proper discussions of the remaining terms in this
``mathematical name'' for the Standard Model, I will need to introduce
quite a bit of additional terminology.

A group \emph{representation} is a map from a group $G$ to a set of
linear transformations on a vector space $V$. More explicitly, the
map $\pi$ is a homomorphism
\begin{equation*}
  \pi: G \longrightarrow GL(V)
\end{equation*}
with the property
\begin{equation*}
  \pi(g \cdot h) = \pi(g) \circ \pi(h) \quad {\rm for}~g,h \in G;
\end{equation*}
$GL(V)$ is the general linear group (a group in its own right)
consisting of all $N \times N$ matrices acting on an $N$-dimensional
vector space $V$; thus the representation of a group $\pi(G)$ is
always some subgroup of $GL(V)$. If the homomorphism $\pi$ is
one-to-one, (\emph{injective}), then the map is an isomorphism: $G
\cong \pi(G)$, and the representation is said to be \emph{faithful}.

A representation is conventionally named simply with a bold numeral
indicating its dimension, as in, for example, the ``\textbf{2}'' or
the ``\textbf{3}'' representation of $SU(2)$. In a mild abuse of
terminology, physicists are quite prone to referring to a vector $v
\in V$, on which the elements of a group representation act, as a
``representation'' of the group as well; in fact, I will often do so
in this work.

When a mathematical system is left unchanged by the simultaneous
action of a group on each of the components of the system, the group
is called a \emph{symmetry} of the system, and the system is said to
be \emph{invariant} under the group action.

To qualify the above concepts in the pertinent context, let me point
out that the Lagrangian of the Standard Model is invariant under the
action of the continuous group
\begin{equation*}
    SU(3)_C \times SU(2)_L \times U(1)_Y \times \left(\mathbb{R}^{1,3}
\rtimes SO(1,3)\right),
\end{equation*}
where
\begin{itemize}
  \item $SO(N)$ is the non-Abelian group of \emph{orthogonal} (i.e., 
    length-preserving) rotations in $N$-dimensions, with elements $O$
    such that $O^T O = \mathbb{I}~ \forall O \in SO(N)$; it is
    naturally equipped with the \emph{fundamental} or
    \emph{standard}\footnote {``Standard representation'' is the
    conventional term among mathematicians.} representation of
    $N\times N$ matrices satisfying the above property and with
    determinant $1$, which act on vectors in the space $\mathbb{R}^N$.

  \item $SU(N)$ is the analogous group of complex \emph{unitary} 
    rotations with elements $U$ such that $U^\dagger U = \mathbb{I}~
    \forall U \in SU(N)$, and with fundamental representation acting
    on elements of the \emph{complex} space $\mathbb{C}^N$.

  \item $U(1)$ is the Abelian group of rotations by a complex phase 
    ${\rm e}^{i\theta}$ for some real number $\theta$, which acts on
    single elements of $\mathbb{C}$, \emph{i.e.}, complex numbers.

  \item the direct products ``$\times$'' indicate that, although 
    the individual groups are generally non-Abelian, the actions of
    the groups commute with one another.

  \item $\mathbb{R}^{1,3} \rtimes SO(1,3)$ is the \emph{Poincar\'e} 
    group, the ``spacetime part'' of the SM symmetry. Poincar\'e
    invariance is what makes the SM consistent with the principles of
    special relativity.  $\mathbb{R}^{1,3}$ gives the translational
    symmetry of any SM process (i.e., the physics is the same whether
    some interaction happens at point $x$ or point $y$), and
    $SO(1,3)$, the \emph{Lorentz} group, contains ordinary rotations
    in 3D space plus boosts (time-space mixing rotations). The
    presence of the \emph{ semi-direct product}, ``$\rtimes$'', is due
    to the fact that the product of an $SO(1,3)$ transformation and an
    $\mathbb{R}^{1,3}$ translation is another translation in a
    different reference frame; hence, for a general spacetime
    translation $U \sim \mathrm{e}^{i p \cdot x}$ and a general
    spacetime rotation $\Lambda \in SO(1,3)$, the commutator $U \cdot
    \Lambda - \Lambda \cdot U \sim U^\prime$ is nonzero (i.e., they do
    not commute).  The \emph{signature} ``1,3'' carries the
    distinction between timelike and spacelike directions; the two
    have opposite-sign contributions to the metric $\eta_{\mu\nu}$
    used to calculate inner products between elements of the
    Poincar\'e group, which creates the potential for \emph{null}, or
    ``light-like'' propagation, for which the invariant spacetime
    \emph{interval} $ds^2 \equiv \eta_{\mu\nu} x^\mu x^\nu = dt^2 -
    dx^2 = 0$.\footnote{I will use the ``mostly minus'' signature, with
    spacelike elements of the metric negative, {\it i.e.} $\eta \equiv
    {\rm diag}\, (1, -1, -1, -1)$.}
\end{itemize}
Note that $SO(N)$, $SU(N)$, and $U(1)$ are all Lie groups.

A \emph{Lie algebra} $\mathfrak{g}$ is related to the Lie group $G$ by
the following rule: for all $N \times N$ matrices $X \in \mathfrak{g}$
and $\theta \in \mathbb{R}$, $U = \mathrm{e}^{i \theta X} \in G$. Note
that the factor of $i$ is a practical convention used by physicists.
The real parameter $\theta$ sets the magnitude for the group
transformation (extraction of this factor from $X$ is not necessary,
but it is convenient and will be easier to generalize later); in the
cases of orthogonal or unitary transformations, it can be interpreted
as a rotation angle. If $\theta \ll 1$, then $U$ can be simplified
using the infinitesimal form of the exponential $U \approx 1 + i
\theta X$.

The \emph{generators} of a Lie algebra $t^a$ are the basis elements
through which all $X \in \mathfrak{g}$ can be constructed; i.e., $X =
\sum \alpha^a t^a~\forall X \in \mathfrak{g}$, with $\alpha^a \in
\mathbb{R}$. By the relationship given in the previous paragraph, any
element of the group can be written as $U = \mathrm{e}^{i \alpha^a
t^a}$, where the rotation angle has been absorbed into the constants
$\alpha$. This is a general form for the elements of $SU(N)$ in
the SM; their action on fermion fields is $\psi \rightarrow U \psi$.

The $N(N-1)/2$ generators of the Lie algebra $\mathfrak{so}(N)$ are
antisymmetric, and the $N^2-1$ generators of $\mathfrak{su}(N)$ are
Hermitian. The closure of $G$ is guaranteed if the generators of
$\mathfrak{g}$ satisfy the commutator relationship
\begin{equation} 
  \left[\, t^a, t^b\, \right] \equiv t^a t^b - t^b t^a = i
    {f^{ab}}_c t^c, \nonumber 
\end{equation}
where ${f^{ab}}_c$ are called the \emph{structure constants} of the
algebra. The structure constants are simply numbers that
determine the exactly how one generator is constructed from the
others. It is naturally the case that many of the structure constants
for a particular Lie algebra are zero.

Here I can finally return to the defining the terms appearing in the
opening sentence. A \emph{gauge symmetry} is an invariance under
\emph{local} group transformations, as opposed to \emph{global}
transformations.  In a global transformation, the rotation parameters
$\alpha^a$ are constant real numbers, as described above. In a local
transformation, the parameters are instead functions of spacetime,
$\alpha^a = \alpha^a \!(x)$, which is actually a stronger condition
({\it i.e.}, local symmetry implies global symmetry).

This promotion of transformations has surprising effects on the nature
of a theory. Before trying to understand gauge symmetry in a quantum
field theory, I will consider a simple example from classical
electromagnetism. One may recall that an electromagnetic wave has only
two degrees of freedom, namely the polarizations of $\mathbf{E}$ and
$\mathbf{B}$; yet, the four-vector potential $A_\mu$, whose spacetime
derivatives give rise to those fields, seemingly comes equipped with
four degrees of freedom. Thus it seems the potential has some
intrinsic redundancy; in fact, that redundancy follows directly from
the ambiguity in its definition:
\begin{equation}
  A_\mu \rightarrow A_\mu + \partial_\mu \alpha,
  \label{eq:Agauge}
\end{equation}
where $\alpha(x)$ is some scalar function (the degeneracy of this
notation with that of the gauge transformation parameters is
intentional). Furthermore, the Lagrangian for $A_\mu$, from which
Maxwell's equations follow, $\mathcal{L} = -\frac{1}{4} F_{\mu\nu}
F^{\mu\nu}$, is invariant under the redefinition (\ref{eq:Agauge}).
This is a simple example of a gauge symmetry.

As it turns out, $A_\mu$ is a representation of the Lorentz group, and
precisely that which one would promote to an operator if looking to
quantize electromagnetism. If one naively attempts to do so by, for
instance, following procedure analogous to that for a scalar field,
serious difficulties arise presently. Given the equation of motion for
the classical photon-to-be,
\begin{equation*}
  \partial_\mu F^{\mu\nu} = \partial_\mu (\partial_\mu A_\nu -
  \partial_\nu A_\mu) = J^\nu,
\end{equation*}
or, after Fourier transform,
\begin{equation*}
  (- p^2 g_{\mu\nu} + p_\mu p_\nu) A_\mu = J^\nu,
\end{equation*}
one finds that the naive choice for the corresponding propagator is
ill-defined. However, one can utilize the ambiguity in
(\ref{eq:Agauge}) to resolve the issue by adding a term that depends
on the ``choice of gauge'', \emph{i.e.} the form of $\alpha (x)$ (or,
traditionally, an analogous function). In the end one sees that the
lack of an ``ordinary'' propagator is a consequence of neglecting the
redundancy of the extraneous degrees of freedom. Therefore, any
quantized theory of electromagnetism will necessarily also be a gauge
theory.

One important consequence of this generalization is that terms in the
Lagrangian containing derivatives of matter fields are no longer
invariant under group transformations. For the Abelian group
$U(1)_{\mathrm{em}}$ of proper quantum electrodynamics (QED), a term
involving matter fields such as $\bar\psi \psi$ (more on this form
later\ldots) is unchanged by the transformation $\psi \rightarrow
\mathrm{e}^{i \alpha} \psi$ even after the ``gauging'' of the
symmetry, $\alpha \rightarrow \alpha(x)$, because the transformation
factors enter as conjugates and simply cancel; however, the derivative
transformation picks up an extra term:
\begin{equation*}
  \partial_\mu \psi \rightarrow \mathrm{e}^{i \alpha(x)} \partial_\mu
  \psi + i \partial_\mu \alpha \, \mathrm{e}^{i \alpha(x)} \psi.
\end{equation*}
In order to restore invariance to derivative terms in the
Lagrangian, one must introduce the \emph{gauge covariant derivative}
$D_\mu \equiv \partial_\mu + iA_\mu$. Using this form in place of the
normal derivative, as well as the transformations for both $A_\mu$ and
$\psi$, one finds that $D_\mu \psi \rightarrow \mathrm{e}^{i
\alpha(x)} D_\mu \psi$, as desired.  The details of the Lagrangian in
light of this formulation will be discussed in more detail later. The
generalization of this process to non-Abelian groups is relatively
straightforward.

As the final topic from my opening remark, a \emph{spontaneously broken}
symmetry is a symmetry of the Lagrangian that is not respected by the
ground state of the theory. In the case of the SM, the $SU(2)_L \times
U(1)_Y$ electroweak symmetry is not a symmetry of the vacuum.  The
symmetry is ``broken'' (really more like obscured) specifically by the
Higgs field via the \emph{Higgs mechanism} at the electroweak scale
$\sim$100 GeV.  I will discuss the Higgs mechanism and the
implications of this symmetry breaking in more detail shortly.

At this point, all of the terminology I used at the start of the
chapter to name the mathematical structure of the SM has been
introduced. Before discussing the Lagrangian and the interactions at
the heart of the model, I will discuss the details of representations
of the SM fields.

\subsection{The Representations of Standard Model Fields}

The SM includes the following quantum fields:
\begin{itemize}
  \item three copies of four fermionic fields: $3\, \times\,2$
    \emph{quark} fields, $\{u,c,t\}$, and $\{d,s,b\}$, and $3\,
    \times\,2$ \emph{lepton} fields, $\{\nu_e, \nu_\mu, \nu_\tau\}$,
    and $\{e,\mu,\tau\}$; the ``copies'', known as \emph{generations},
    differ only in mass and have the same quantum numbers otherwise;
  \item four force-carrying bosonic fields: the photon, $A_\mu$
    (often notated as ``$\gamma$"), the gluons, $G_\mu^a$ (often
    notated as ``$g$''), and the $W_\mu^\pm$ and $Z_\mu$ weak bosons;
  \item one Higgs boson field, $\phi$.
\end{itemize}
The force-carrying bosons named here are the physical
particles, of definite mass, which differ from the massless fields
found in the model prior to spontaneous symmetry breaking. Those
fields will be discussed shortly, and their relationships to the above
particles will be made clear when I discuss symmetry breaking in more
detail.

Each field above is associated to a particular representation of the
SM gauge group (gauge bosons) or the vector spaces on which it acts
(matter fermions and Higgs). Differences in representation are what
give the fields unique properties, which lead to our observation of
several unique types of elementary particles.  Below I will discuss
the representations for each field.

\subsubsection{Spacetime Representations} The different classes of
fields listed above experience spacetime transformations as different
representations of the Poincar\'e group, which, in a sense, gives rise
to the simplest definition of \emph{elementary particle}: a state
whose degrees of freedom mix only with each other, as elements of a
single representation, under the action of the Poincar\'e group,
\cite{schwartz}. Furthermore, the nature of translation is generic to
all of the fields, so it is specifically the Lorentz representation of
a particle that determines the nature of the interactions it may have,
and even the nature of its free propagation through empty space.

\paragraph*{Lorentz Scalars.} The most basic and uninteresting
Lorentz representation is the \emph{trivial} representation; fields in
this representation are invariant under group transformations and are
consequently scalars in the formalism of the group.\footnote{Note the
concept of a trivial representation is general to all groups and is
not a special feature of the Lorentz group. \smallskip} The Higgs
boson is the only Lorentz scalar field in the SM.

\paragraph*{Lorentz Vectors.} The force-carrier gauge bosons of the
SM are Lorentz \emph{four-vectors}, i.e., 3+1-dimensional elements of
the fundamental representation; for the Lorentz group, this implies
transformation via the same $4 \times 4$ boost or rotation matrices as
$x^\mu$, $p^\mu$, etc. one sees in basic index-notated special
relativity: $A^\prime_\nu = {\Lambda_\nu}^\mu A_\mu$.

\paragraph*{Spinors.} The matter fermions of the SM are Lorentz or
Dirac \emph{spinors}. A spinor representation is also realized as
matrices acting on multiplets in a vector space, but it is a different
vector space, of generally different dimension, from that of the
fundamental representation. The relationship between the two spaces is
an interesting one. The group $Spin(N)$, whose elements act on the
spinors, is a \emph{double cover} of the orthogonal group $SO(N)$,
meaning there are two ``copies'' of the $SO(N)$ manifold in that of
$Spin(N)$, and there is a 2-to-1 map from the latter onto the former.
As a result, for any rotation of a vector in the space of the $SO(N)$
fundamental, there are two topologically \emph{distinct} continuous
paths, from the same initial state to the same final state, through
which the spinor can be rotated.  Another important result of this
relationship is that an ordinary spatial rotation of a spinor through
$2\pi$ results in the \emph{negative} of the original state; a second
$2\pi$ rotation is required to return the spinor to its original
orientation.

For the Lorentz group, the double covering group is $Spin(1,3) \cong
SL(2,\mathbb{C})$, which is the \emph{special linear} group over
complex numbers, whose elements are $2 \times 2$ matrices with complex
entries and determinant 1. The action of $SL(2,\mathbb{C})$ is on
two-component \emph{Weyl} or \emph{chiral} spinors $\psi_{L,R}$; the
Dirac spinor more commonly associated with the Lorentz group is
actually a \emph{bispinor}, spinor $\oplus$ spinor; this reducibility
is manifest in the \emph{Weyl basis} for the gamma matrices, where the
bispinor corresponding to a SM fermion is the direct sum $\psi =
\psi_L \oplus \psi_R$; many interactions of bispinors, including those
in QED, decouple into left and right parts in that basis.
Four-component Dirac ``spinors'' are related to Weyl bispinors by a
change of basis.

Interaction of spinors with a Lorentz vector is realized through the
\emph{Dirac algebra}, which consists of $4\!\times\!4$ matrices
$\gamma^\mu$ that form an anti-commuting \emph{Clifford algebra},
meaning they satisfy
\begin{equation}
  \left\{ \gamma^\mu, \gamma^\nu \right\} = 2\, \eta^{\mu\nu}\,
  \mathbb{I}_4, \nonumber
\end{equation}
where $\mathbb{I}_4$ is the identity in the spin space. Note that
each matrix carries a Lorentz spacetime index, which can have values
$\mu = 0,1,2,3$ as one would expect; yet, the $\gamma$-matrices are
better thought of as a basis for representing four-vectors as group
elements in the spin space (\emph{i.e.}, matrix operators that act on
spinors), rather than as forming a spacetime four-vector themselves,
especially as they transform differently (and passively) under the
Lorentz group.

In analogy with non-relativistic angular momentum, the six objects
\begin{equation}
  S^{\mu\nu} \equiv \frac{1}{2} \gamma^{\mu\nu} \equiv \frac{i}{4}
  \left[\, \gamma^\mu, \gamma^\nu\, \right], \nonumber
\end{equation}
are the generators of angular momentum and boosts in the spin space;
accordingly, $S^{\mu\nu}$, rather than the $\gamma$-matrices
themselves, satisfy the Lie algebra $\mathfrak{so}(1,3)$, and hence
represent the group $Spin(1,3)$. The Lorentz transformation of a Dirac
spinor is given in terms of these generators:
\begin{equation}
  \psi \rightarrow \Lambda_\frac{1}{2} \psi = \exp\left(- \frac{i}{2}
  \omega_{\mu\nu} S^{\mu\nu} \right) \psi, \nonumber
\end{equation}
where $\omega_{\mu\nu}$ is an anti-symmetric tensor of constant
infinitesimal rotation parameters. This Lorentz transformation for
spinors is related to the vector transformation ${\Lambda^\mu}_\nu$
through the gamma matrices:
\begin{equation}
  \Lambda_\frac{1}{2}^{-1} \gamma^\mu \Lambda_\frac{1}{2} =
  {\Lambda^\mu}_\nu \gamma^\nu. \nonumber
\end{equation}

Before I move on, note that the Lorentz invariant contraction of
spinors is
\begin{equation*}
\bar\psi \psi \equiv \psi^\dagger \gamma^0 \psi = \psi^\dagger_R
\psi_L + \psi^\dagger_L \psi_R, 
\end{equation*}
rather than the naive choice of $\psi^\dagger \psi$. It will generally
be the case that Lorentz tensors constructed from spinors will involve
some product of gamma matrices sandwiched between $\bar\psi$ and
$\psi$: the vector $\bar\psi \gamma^\mu \psi$, which couples to
ordinary Lorentz vectors, the pseudo-vector $\bar\psi \gamma^\mu
\gamma^5 \psi$, the two-tensor $\bar\psi \gamma^{\mu\nu} \psi$, etc.

\subsubsection{Representations of the Internal Gauge Group} All
three components of the \emph{internal} symmetry group of the SM are
gauged groups. Fermionic matter fields transform under the action of
the fundamental representations of those groups; \emph{i.e.}, the
fields are components of an $N$-dimensional multiplet on which a group
$SU(N)$ acts in the form of an $N \times N$ matrix.

In particular, fermions with left-handed chirality are known to pair
off into doublets,
\begin{equation*}
  q \equiv \left(\!\begin{array}{c}u_L\\d_L\end{array}\!\right)
    \qquad
  \ell \equiv \left(\!\begin{array}{c}\nu_L\\e_L\end{array}\!\right),
\end{equation*}
which can be rotated by $SU(2)$ group elements; gauge covariance of
the group leads to interactions between the left-handed fermion
multiplets above and the $W$ bosons, giving rise to the weak force,
although the details are complicated a bit by electroweak symmetry
breaking (EWSB). The transformations are associated with left-handed
fermions having non-trivial \emph{weak isospin} charge,
$\boldsymbol{T}$.  Right-handed fermions, $u_R$, $d_R$, and $e_R$,
have $\boldsymbol{T} = 0$, and so each exists only in the trivial
representation of $SU(2)$. In analogy with ordinary spin, the
components of each doublet have eigenvalues $T^3 = \pm 1/2$.

Similarly, quarks possess an additional degree of freedom known as
\emph{color} and consequently form triplets,
\begin{equation*}
  u = \left(\!\begin{array}{c}u_r\\u_g\\u_b\end{array}\!\right)
    \qquad
  d = \left(\!\begin{array}{c}d_r\\d_g\\d_b\end{array}\!\right),
\end{equation*}
which can be rotated by $SU(3)$ group elements; gauge covariance of
the group gives rise to the strong force through interactions between
the quark multiplets above and the gluons. Leptons do not carry color
charge and so are found in the trivial representation of this group.
Interestingly enough, every known physical state involving quarks
which has been empirically verified is \emph{color neutral}, or
``white''; individual quarks do not freely propagate at low energies.
This property of quarks, known as \emph{confinement}, is perhaps not
yet fully understood, but is due in part to the fact that the strength
of the coupling constant $g_s$ for color interactions increases as
energy decreases.

Finally, all fermionic SM fields individually have nonzero \emph{weak
hypercharge}, $Y_w$, which is associated with rotations by group
elements of the $U(1)_Y$ symmetry; gauge covariance of the group
ultimately gives rise to the electromagnetic force through
interactions between fermions and photons, although, again, the
details are complicated by EWSB. The transformations act on individual
fields rather than multiplets, meaning the group elements are simply
complex numbers of unit magnitude. 

The corresponding antiparticle fields of the SM fermions, which are
the charge conjugates of the particle fields, are found in analogous
\emph{conjugate} representations, named ``$\boldsymbol{\overline
{2}}$'', ``$\boldsymbol{\overline{3}}$'', etc.; the antiparticle
partners themselves are named by one of a few conventions. One often
sees the notation $\psi^{\cal C} \equiv C \bar{\psi}^T = C \gamma^0
\psi^*$ to indicate antiparticle fields, where the $C$ is a unitary
matrix with $C^T = -C$; by this construction, the antiparticle
$\psi^{\cal C}$ has the same chirality as its partner $\psi$. Once I
move on from discussing the SM, I will normally use this notation.
Note though that if I want to give the antiparticle partners of the
$SU(2)_L$ doublets above, I would write something like
\begin{equation*}
  q^\dagger_R \equiv \left(u_R^\dagger, d_R^\dagger \right),
    \qquad
  \ell^\dagger_R \equiv \left(\nu_R^\dagger, e_R^\dagger \right),
\end{equation*}
to make manifest that only antiparticles with right-handed chirality
will form $SU(2)_L$ doublets that interact via the weak force. 

Force-carrier gauge bosons experience (and, in a way, exhibit) the
action of the internal symmetry groups of the SM as elements of the
\emph{adjoint} representations of the groups; the adjoint
representation is that which is exhibited by the generators of the Lie
algebra themselves; the group action on the generators is $t^a
\rightarrow g\, t^a g^{-1}$ for some $g \in$ group $G$; more
specifically for our purposes, $t^a \rightarrow U\, t^a\, U^\dagger$
for $U \in SU(N)$. The boson fields $A^a_\mu (x)$ associated with a
particular symmetry group will be in one-to-one correspondence with
the generators of the symmetry. For a gauge symmetry, the
transformation of the bosons mimics that of the generators, but with
an important extension: $A^a \rightarrow U\, A^a\, U^\dagger + dU\,
U^\dagger$; taking $U = \mathrm{e}^{-i \alpha^a t^a}$ as before, and
for infinitesimal transformations $\alpha (x) \ll 1$, this corresponds
to $A^a_\mu \rightarrow A^a_\mu + \partial_\mu \alpha^a - {f^a}_{bc}
\alpha^b A^c_\mu$, which is the generalization of
eq.\,(\ref{eq:Agauge}) for the abelian gauge field $A_\mu$ discussed
earlier. The generalized gauge covariant derivative for a non-Abelian
group utilizes the above properties to give the mapping of the boson
field into the vector space of the group: $D_\mu = \partial_\mu - ig
t^a A^a_\mu$, where $g$ is the coupling constant of the interaction
with other fields; interactions with matter fields arise through this
\emph{minimal coupling} of the gauge field to the derivative.

The vector bosons associated with the unbroken symmetry of the SM are the
single field $B_\mu$ for the Abelian group $U(1)_Y$, the three fields
$W^a_\mu$ for $SU(2)_L$, and the eight \emph{gluons} $G^{a'}_\mu$ for
$SU(3)_c$.

The scalar Higgs field $\phi$ is an $SU(2)_L$ doublet
\begin{equation}
  \phi = \left(\! \begin{array}{c} \phi^+\\ \phi^0 \end{array}
    \!\right),
\label{eq:Hdoub}
\end{equation}
with hypercharge $Y_w = 1/2$. Each component field is complex, so
$\phi$ generally has 4 degrees of freedom. The non-trivial $SU(2)_L$
representation enables electroweak symmetry breaking when the field
acquires a vacuum expectation value, which I will discuss in more
detail shortly. Additionally, the field belongs to the trivial
representation of $SU(3)_C$.

A summary of the charges of all the SM fields under each
symmetry group is given in Table \ref{table:SMcharges}.
 
\begin{table}[t]
\begin{center}
\begin{tabular}{||c|c|c|c||} \hline\hline
	& $SU(3)$ rep & $SU(2)$ rep & $Y_w$ \\ \hline
        $q^i_L$ & \textbf{3} & \textbf{2} & 1/6\\
	$u^i_R$ & \textbf{3} & \textbf{1} & 2/3\\
	$d^i_R$ & \textbf{3} & \textbf{1} & -1/3\\
        $\ell^i_L$ & \textbf{1} & \textbf{2} & -1/2\\
        $e^i_R$ & \textbf{1} & \textbf{1} & -1\\ \hline
        $B_\mu$ & \textbf{1} & \textbf{1} & 0\\
        $W_\mu^a$ & \textbf{1} & \textbf{3} (adj) & 0\\
        $G_\mu^{a'}$ & \textbf{8} (adj) & \textbf{1} & 0\\ \hline
        $\phi$ & \textbf{1} & \textbf{2} & 1/2\\
	\hline\hline
\end{tabular}
\caption[Representations and charges of SM fields under the gauge
symmetries of the model.]{Representations and charges of SM fields
under the internal gauge symmetries of the model.}
\label{table:SMcharges}
\end{center}
\end{table}

\subsection{Standard Model Interactions and Lagrangian}

In accordance with classical Lagrangian theory, the SM Lagrangian
should incorporate all of the allowed dynamics of its particles in
terms of only the fields and their spacetime derivatives. A properly
formed Lagrangian density $\mathcal{L}$ should be such that the action
$\mathcal{S} \equiv \int d^4x \,\mathcal{L}$ is invariant under a
general transformation of either the Poincar\'e group or the internal
SM gauge group (at least up to some total derivative), which implies
that each term in $\mathcal{L}$ should be written in such a way that
all of its components are contracted to result in a scalar under
general transformations. Also, it follows from $\mathcal{S}$ (and
$\hbar = 1$) that $\mathcal{L}$ must have dimensions of energy$^4$.

In classical field theory, kinetic terms are $\sim (d \Phi)^2$. For a
scalar quantum field $\phi$ (of dimension $[\phi] = 1$), the analogy
is exact: $\mathcal{L}_\mathrm{kin} = (\partial_\mu \phi)^2$, where
there is an implied sum over $\mu$ (note $[\partial_\mu] = [p^\mu] =
1$ also, so that $[\mathcal{L}_\mathrm{kin}] = 4$ as desired). The
generalization for a complex field (like the Higgs) is $\partial^\mu
\phi^* \partial_\mu \phi$. I mentioned the kinetic Lagrangian for the
Abelian $A_\mu$ field in the earlier discussion on gauge symmetry; the
generalization to non-Abelian bosons follows from $F^a_{\mu\nu} =
\partial_\mu A^a_\nu - \partial_\nu A^a_\mu + g {f^a}_{bc} A^b_\mu
A^c_\nu$. Note, one can see from this expression that non-Abelian
bosons interact among themselves, \emph{i.e.}, they carry charge under
the force they mediate, which is not the case for electrically-neutral
photons. The resulting kinetic terms for the SM Lagrangian are
\begin{align}
  \mathcal{L}_{SM} ~\ni ~-\frac{1}{4} G_{a'}^{\mu\nu} G^{a'}_{\mu\nu}
  ~-~\frac{1}{4} W_a^{\mu\nu} W^a_{\mu\nu} ~-~\frac{1}{4} B^{\mu\nu}
  B_{\mu\nu},
\label{eq:Lgauge}
\end{align}
where $B_{\mu\nu}$ is analogous to the Abelian electromagnetic field
strength tensor $F_{\mu\nu}$.

The kinetic term for fermion fields is a bit more tricky. For one,
Dirac spinors have dimension $[\psi] = 3/2$, so the operator in
question will need to contain only a single derivative; furthermore,
that derivative will still need to be contracted with another
vector-like object. The solution, courtesy of Dirac, turns out to be
$i \bar\psi \gamma^\mu \partial_\mu \psi$. Note one often sees
\emph{Feynman slash notation} $\slashed p = \gamma^\mu p_\mu$ for
contraction of four-vectors with the gamma matrices.

The interaction terms for scalars or spinors with the gauge bosons
follow straightforwardly from replacing the derivatives above with the
corresponding gauge covariant derivatives. The components of the
Lagrangian consistent with the representations described in the
previous section are
\begin{align}
  \mathcal{L}_{SM} ~\ni ~\bar q_L^i\, \gamma^\mu \left(i
  \partial_\mu + g_s \lambda^{a'} G^{a'}_\mu + g T^a W^a_\mu +
  \frac{1}{6}\, g' B_\mu \right) q_L^i \nonumber \\ + ~\bar u_R^i\,
  \gamma^\mu \left(i \partial_\mu + g_s \lambda^{a'} G^{a'}_\mu +
  \frac{2}{3}\, g' B_\mu \right) u_R^i \nonumber \\ + ~\bar d_R^i\,
  \gamma^\mu \left(i \partial_\mu + g_s \lambda^{a'} G^{a'}_\mu -
  \frac{1}{3}\, g' B_\mu \right) d_R^i \nonumber \\ + ~\bar\ell_L^i\,
  \gamma^\mu \left(i \partial_\mu + g T^a W^a_\mu - \frac{1}{2}\,
  g' B_\mu \right) \ell_L^i \nonumber \\ + ~\bar e_R^i\, \gamma^\mu
  \left(i \partial_\mu - g' B_\mu \right) e_R^i \nonumber \\ +
  ~\phi^\dagger \left(\partial^\mu + i g T^a W_a^\mu + \frac{i}{2}\,
  g' B^\mu \right) \left(\partial_\mu - i g T^b W^b_\mu -
  \frac{i}{2}\, g' B_\mu \right) \phi,
\label{eq:Lint}
\end{align}
where the generators $T^a \equiv \sigma^a/2$, with $a' = 1,2,3$, are
half the Pauli matrices; $\lambda^{a'}$, with $a = 1,\dots,8$, are the
analogous generators of $SU(3)$; and $i = 1,2,3$ are the generation
indices, for which all of the above interactions are diagonal (in the
unbroken, massless case). In this context the spinor fields $f_{L,R}$
with $f = u,d,e,\nu$ are four-component Dirac spinors, rather than
two-component Weyl spinors, but with with the left- or right-handed
components set to zero, which can be done using the \emph{chiral
projection operators} $P_{L,R} \equiv \frac{1}{2}(1 \mp \gamma_5\,)$
such that $f_{L,R} = P_{L,R} f$. Note the quark-lepton asymmetry due
to the absence of the right-handed neutrino field. The implicit
transpose in $\phi^\dagger$ is with respect to its $SU(2)$ components,
and the adjacent derivative acts on it to the left. Also note that the
indices for the internal spaces of $SU(2)$ and  $SU(3)$ have been
suppressed for clarity; for example, the fully notated version of the
quark doublet term above would be
\begin{equation*}
  \mathcal{L} \ni \bar q_L^{i \alpha \rho} \gamma^\mu \left(
  \delta_{\alpha\beta} \delta_{\rho\sigma} (i \partial_\mu +
  \frac{1}{6}\, g' B_\mu) + g_s \delta_{\alpha\beta}
  \lambda^{a'}_{\rho\sigma} G^{a'}_\mu + g \delta_{\rho\sigma}
  T^a_{\alpha\beta} W^a_\mu \right) q_L^{i \beta \sigma},
\end{equation*}
where $\alpha = 1,2$ are the internal $SU(2)$ indices, and $\rho =
1,2,3$ are those of $SU(3)$.

The Higgs field $\phi$ also interacts with the matter fields through
the \emph{Yukawa} terms, and has self-interactions allowed by the
freedom of the Lorentz scalar representation as well:
\begin{align}
  \mathcal{L}_{SM} ~\ni ~-&y_u^{ij} \epsilon_{\alpha\beta}\,
  \bar q_{Li}^\alpha \phi^{* \beta} u_{R j} ~- ~y_d^{ij}\,
  \bar q_{Li}^\alpha \phi^\alpha\, d_{R j} ~- ~y_e^{ij}\,
  \bar \ell_{Li}^\alpha \phi^\alpha\, e_{R j} \nonumber \\[2mm] 
  &~+ ~\mathrm{Hermitian\;conjugates} \nonumber \\ &~+
  ~\mu^2 \phi^\dagger \phi ~- ~\lambda \left(\phi^\dagger \phi
  \right)^2,
\label{eq:LHiggs}
\end{align}
where I've included the $SU(2)$ indices in the Yukawa terms due to
their non-triviality. Note that $\epsilon_{\alpha\beta} \phi^{*
\beta}$ (with $\epsilon_{12} = 1$) transforms identically to $\phi$
under $SU(2)$ but has the opposite hypercharge as well as the
necessary component structure needed to couple $\phi^+$ and $\phi^0$
to $u$ in the same way as $d$ and $e$.

The scalar self-coupling parameters $\mu$ and $\lambda$ are
unconstrained in principle. One would expect $\mu$ to function as a
mass for the field, but note that the term has opposite the expected
sign (assuming $\mu^2 > 0$); this subtlety has profound implications
for the potential of $\phi$, as I will discuss in the next section.

\subsection{Electroweak Symmetry Breaking and the \\ Broken
Lagrangian} 

\begin{figure}[t]
\begin{center}
    \includegraphics[width=7.8cm]{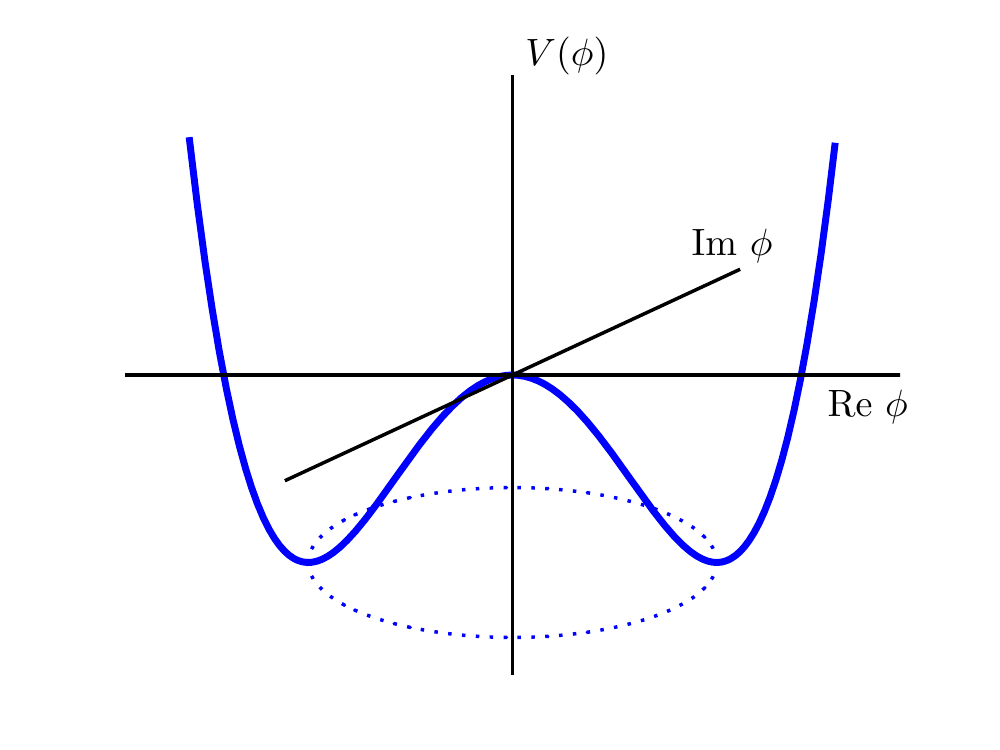} \caption[The classical
    potential for the Higgs field as a function of $\phi$.]{The
    classical potential for the Higgs field as a function of $\phi$.}
    \label{fig:HiggsV}
\end{center}
\end{figure}

Experimentally, matter fermions and weak gauge bosons are known to
have mass, yet I gave no explicit mass terms in the Lagrangian, as
stated in eqs.\,(\ref{eq:Lgauge})-(\ref{eq:LHiggs}). In fact, it is
not hard to convince oneself that (\emph{a}) a mass term like $M^2
A^\mu A_\mu$ for a gauge boson breaks its gauge symmetry, and
(\emph{b}) a Dirac mass term like $m (\bar\psi_L \psi_R$ + h.c.)\;for
a fermion is intractable in light of the inequivalent electroweak
quantum numbers ($T^3$ and $Y_w$) for left- and right-handed fields.
It \emph{is} completely tractable however to generate \emph{effective}
mass terms for both gauge bosons and fermions using a dynamic scalar
field with the appropriate characteristics.  This is the role of the
Higgs field in the SM; the details of the emergence of these masses
through the Higgs mechanism are as follows.

From a classical perspective, one can view the final two terms in
eq.\,(\ref{eq:LHiggs}), which describe the self-interaction of the Higgs
field, as a scalar potential\footnote{an additional symmetry $\phi
\rightarrow -\phi$ is imposed on the Higgs Lagrangian to guarantee the
presence of a stable minimum.}
\begin{equation}
  V(\phi) = -\mu^2 \phi^\dagger \phi ~+ ~\lambda \left(\phi^\dagger
  \phi \right)^2.
\end{equation}
In the alternate case where the $\mu^2$ term is instead positive,
this potential has a single minimum at $\phi_0 = 0$; however, for a
negative $\mu^2$ term and appropriate related values for $\mu$ and
$\lambda$, $V$ has the shape seen in Figure \ref{fig:HiggsV}. This
potential is seen to have a continuously degenerate minimum, with a
constant magnitude $\phi_0 = \mu /\sqrt{2 \lambda} \equiv v$ but
arbitrary phase.

From the perspective of quantum field theory, this nonvanishing
minimum corresponds to a \emph{vacuum expectation value} (vev)
$\langle \phi \rangle$ for the scalar field $\phi$; however, a field
with such a vev cannot be quantized in the usual manner using
creation/annihilation operators, which demands $\hat a \ket{0} = 0$;
yet, there is a simple way to bypass the issue: one can reparametrize
the Higgs doublet given in eq.\,(\ref{eq:Hdoub}) as \vspace{2mm}
\begin{equation}
  \phi (x) = \left(\! \begin{array}{c} 0\\ v + h^0 (x)
  \end{array} \!\right),
\label{eq:Hreparam}
\end{equation}
where the dynamical real scalar field $h^0 (x)$ can be quantized as
usual and treated as fluctuations about the nonvanishing but constant
vacuum $v$; an excitation of the field $h^0$ is the Higgs boson. The
alignment of $v$ with the $\phi^0$-direction can be accomplished
without loss of generality through a global $SU(2)_L$ transformation;
the complex scalar field $\phi^+$ and the imaginary part of $\phi^0$
have been set to zero using $SU(2)_L \times U(1)_Y$ gauge
transformations, and thus can be taken as unphysical. The above
construction explicitly breaks the $SU(2)_L \times U(1)_Y$ symmetry of
the theory. Substituting this parametrization for $\phi$ into
eq.\,(\ref{eq:LHiggs}), one finds masses proportional to $v$ have
emerged for the fermions as a result of the breaking: 
\begin{align}
  \mathcal{L}_{\cancel{SM}} ~\ni ~-&y_u^{ij} v\, \bar u_L^i  u_R^j
  ~- ~y_d^{ij} v\, \bar d_L^i d_R^j ~- ~y_e^{ij} v\, \bar e_L^i e_R^j
  ~+ ~\mathrm{h.c.}.
\label{eq:Lmf}
\end{align}
The same substitution in the final line of eq.\,(\ref{eq:Lint})
yields analogous terms for the gauge bosons, albeit with the presence
of non-trivial mixing among the massless fields:
\begin{align}
  \mathcal{L}_{\cancel{SM}} ~\ni ~\frac{v^2}{4} \left[\,g^2
  \left(W_1^\mu + i W_2^\mu \right) \left(W^1_\mu - i W^2_\mu
  \right) ~+ ~\left(-g W^3_\mu + g' B_\mu \right)^2\, \right].
\label{eq:LMg}
\end{align}
The combinations $W^1_\mu \mp i W^2_\mu \equiv \sqrt{2}\, W^\pm_\mu$
used here were chosen by our forefathers because the coupling of
$W^{1,2}_\mu$ to matter consistently appears in these pairings, as one
can see through the expansion of the $q$, $\ell$, and $\phi$ terms in
eq.\,(\ref{eq:Lint}); since $W_+^\mu W^-_\mu = (W^1_\mu)^2 +
(W^2_\mu)^2$, the mass eigenstates are equivalent. In contrast to
that, the combination $-g W^3_\mu\, +\, g' B_\mu$ appears as a result
of the diagonality of both the $T^3$ and $Y$ generators and cannot be
avoided. Rather than ponder the curious cross terms, one can view the
combination as a change of basis needed to describe the mass
eigenstates manifestly. In fact, these mixed states correspond to the
\emph{physical} particles observed in experiment; yet, there were four
bosons in the system prior to the breaking, so where has the fourth
state gone? Let me define the (properly normalized) mixed $W^3 + B$
state discussed above as
\begin{equation*}
  Z_\mu \equiv \frac{1}{\sqrt{g^2 + g'^2}} \left(g W^3_\mu - g' B_\mu
  \right),
\end{equation*}
and also introduce the angle $\theta_W$ such that $\tan \theta_W =
g'/g$, so that $Z_\mu = \cos \theta_W W^3_\mu - \sin \theta_W B_\mu$.
Then there should exist a state
\begin{equation*}
  A_\mu \equiv \frac{1}{\sqrt{g^2 + g'^2}} \left(g' W^3_\mu + g B_\mu
  \right) = \sin \theta_W W^3_\mu + \cos \theta_W B_\mu,
\end{equation*}
orthogonal to $Z_\mu$, which is also a result of the rotation by
$\theta_W$, and which apparently corresponds to the generator $T^3 +
Y$; if I write this generator as an $SU(2)$ element acting on the
Higgs doublet (recall $Y_{\phi} = +1/2$), one can see that it
annihilates the vacuum in spite of the vev:
\begin{equation*}
  \bra{0} \left(T^3 + Y \right) \phi \ket{0} = \frac{1}{2} \bra{0}
  \left(\sigma^3 + \mathbb{I}\, \right) \phi \ket{0} = \left(\!
  \begin{array}{cc} 1 & 0 \\ 0 & 0
  \end{array} \!\right) \left(\! \begin{array}{c} 0 \\ v \end{array}
    \!\right) = 0;
\end{equation*}
hence, $T^3 + Y$ generates an unbroken symmetry, whose corresponding
boson $A_\mu$ remains massless. As the generator is diagonal, the
unbroken symmetry is a $U(1)$, albeit a different one from that of
weak hypercharge. One can easily be convinced that this symmetry
corresponds to electromagnetism, with $A_\mu$ as the photon and the
electric charge as $Q \equiv T^3 + Y$.

In addition to the terms in eqs.\,(\ref{eq:Lmf}) and (\ref{eq:LMg}),
there is an otherwise identical set of terms with $v \rightarrow h^0$
that give the interactions of the massive fermions (excluding the
neutrino) and the gauge bosons with the neutral Higgs boson.

The covariant derivative in terms of the boson mass eigenstates is
\begin{equation*}
  D_\mu = \partial_\mu - \frac{ig}{\sqrt{2}} \left(T^+ W_\mu^+ +
  T^- W_\mu^- \right) - \frac{ig}{\cos \theta_W} \left(T^3 - Q \sin^2
  \theta_W \right) Z_\mu - ie Q A_\mu,
\end{equation*}
where $T^\pm \equiv \frac{1}{2} (T^1 \mp i T^2)$, and $e = g \sin
\theta_W$ is the electromagnetic coupling. In light of this derivative
one finds chiral \emph{charged currents}
\begin{equation}
  \mathcal{L}_{\cancel{SM}} ~\ni ~\frac{g}{\sqrt{2}} \left(\bar u_L^i
  \gamma^\mu V_{ckm}^{ij} d_L^j ~+~ \bar \nu_L^i
  \gamma^\mu e_L^i \right) W_\mu^+ ~+~ \mathrm{h.c};
\end{equation}
chiral \emph{neutral currents} 
\begin{equation}
  \mathcal{L}_{\cancel{SM}} ~\ni \sum\limits_{f_{L\!,\!R}}
  ~\frac{g}{\cos \theta_W} \bar f^i \gamma^\mu \left(T^3 - Q_f \sin^2
  \theta_W \right) f^i Z_\mu,
\end{equation}
where the sum is over both chiralities of all four flavors of fermion
excluding $\nu_R$; and the electromagnetic currents, coupling to Dirac
spinors,
\begin{equation}
  \mathcal{L}_{\cancel{SM}} ~\ni ~\left(\frac{2}{3}\, \bar u^i
  \gamma^\mu u^i ~-~ \frac{1}{3}\, \bar d^i \gamma^\mu d^i ~-~\bar e^i
  \gamma^\mu e^i \right) e A_\mu^+.
  \label{eq:Lqed}
\end{equation}
Recall that $T^3$ is $+1/2$ for $u_L$ and $\nu_L$, $-1/2$ for $d_L$
and $e_L$, and zero otherwise.

Note the presence of the matrix $V_{\rm ckm}$ in the charged currents
of the quarks. Like the bosons, mass eigenstates for the quarks are
generally different than flavor eigenstates; for flavor eigenstates
$u'_i,d'_i$ and mass eigenstates $u_i,d_i$, the mixing is given
by the transformations
\begin{equation*}
  u_i = U^u_{ij} u'_j, \qquad d_i = U^d_{ij} d'_j,
\end{equation*}
where $U^{u,d}_{ij}$ are $3\times3$ unitary matrices.
Inserting these transformations into a neutral current, one finds that
the factors cancel with each other due to Hermitian conjugation; in
the charged current, however, the new factors differ in flavor, and the
resulting contribution
\begin{equation}
  V_{\rm ckm} \equiv U^{u \dagger}_R U^d_L
\end{equation}
does not vanish in general. In fact, experiments have found that
$V_{\rm ckm}$ is slightly off diagonal, implying that its presence in
nature is physical. The matrix is parametrized by three mixing angles
(one for each pair of generations) and a single imaginary
phase,\footnote{Note that a general $3\times3$ unitary matrix has six
phases, but here, five of them can be absorbed into field
redefinitions.} which induces $CP$-violation in the model

The same phenomenon does not occur with leptons in the model due to
the masslessness of the neutrino; the single rotation matrix coming
from the charged leptons can be absorbed into a field redefinition.
That said, we know that neutrinos do in fact have differing flavor and
mass eigenstates, as their oscillation between mass eigenstates has
been measured by experiments \cite{kajita, *superk-atm,*sno}.  The
corresponding transformation
\begin{equation*}
  \nu_i = U_\nu^{ij} \nu'_j \equiv V_{\rm pmns}^{ij} \nu'_j
\end{equation*}
again consists of three angles, but generally may have two additional
phases, for a total of three, due to the suspected Majorana nature of
the neutrino. The mixing among generations is quite large in general,
and even approximately maximal for $\theta_{23} \sim 45^\circ$. In
fact, the largest (by far) angle of the CKM matrix, $\theta^{12}_{\rm
ckm} \sim 12^\circ$ is only about 50\% larger than the \emph{smallest}
angle in the PMNS, $\theta^{13}_{\rm pmns} \sim 9^\circ$. The phases
of the PMNS matrix are yet to be precisely measured, so the nature of
$CP$- violation there is not yet known.

Returning to the substitution of the redefined Higgs + vev into
eq.\,(\ref{eq:LHiggs}), one also finds that the Higgs boson itself
acquires a mass term (with the proper sign) $m_h = 2v
\sqrt{\lambda}$. Note that if I had \emph{not} made gauge
transformations to remove the additional components of $\phi$, we
would see that they show up as massless scalars in the new Lagrangian.
These components are known as \emph{Nambu-Goldstone bosons} and are a
general feature of spontaneously-broken field theories. Upon closer
inspection, one would find terms like
\begin{align}
  \mathcal{L}_{\cancel{SM}} ~\ni ~\frac{i}{2} gv
  \left(W_\mu^+ \partial^\mu \phi^- - W_\mu^- \partial_\mu \phi^+
  \right) ~- ~\frac{v}{2} \sqrt{g^2 + g'^2}\, Z_\mu \partial^\mu \eta,
\label{eq:LNGB}
\end{align}
where $\eta$ is the imaginary part of $h^0$; these rather bizarre
terms imply the gauge bosons can ``convert'' into the Goldstone bosons
through two-particle, momentum-dependent interactions. Further terms
show that in the interactions of the Goldstones with fermions, the
bosons ``imitate'' the gauge bosons in terms of the configurations of
fields with which they interact. These features led to the
interpretation that the Goldstones are ``eaten'' by the gauge bosons,
effectively becoming the longitudinal degrees of freedom absent in the
massless states. Any other gauge choice or interpretation of the
Goldstone bosons further confirm that the states are otherwise
unphysical.

\section{Measurement and The Success of the Standard Model} At
this point, I have introduced the basic structure of the model and the
interactions that arise from it. Application of the model to
real-world measurements is traditionally built upon Hamiltonian
formalism. In particular, if one defines from the Lagrangian a
\emph{Hamiltonian}
\begin{equation*}
  H = \int \mathbf{d}^3 \mathbf{x}\; \mathcal{H} \quad \mathrm{where}
  \quad \mathcal{H} \equiv \frac{\partial {\cal L}}{\partial
  \dot{\Psi}_i} \dot{\Psi}_i - \mathcal{L} \quad \propto~ \hat
  a_i^\dagger \hat a_i
\end{equation*}
for any field $\Psi_i$ in the model, then using any term
$\mathcal{H}_\mathrm{int} \in \mathcal{H}$ describing an interaction
of $\Psi_i$ with other fields $\Psi_j$, one can define the
\emph{S-matrix element} $\bra{\mathbf{p}_k \mathbf{p}_l}\, S \,
\ket{\mathbf{p}_i \mathbf{p}_j}$ for an interaction $\Psi_i \Psi_j
\rightarrow \Psi_k \Psi_l$ via the operator
\begin{equation*}
  S \equiv \lim_{{t,t_0}\to\pm\infty} \mathcal{T} \left[
\exp \left(-i \int_{t_0}^{t}\!\! dt'\, H_\mathrm{int}(t') \right)
\right] = \mathcal{T} \left[ \exp \left(-i \int_{-\infty}^{\infty}\!\!
d^4 x\; \mathcal{H}_\mathrm{int}(t) \right) \right].
\end{equation*}
This seemingly simple expression hides a great deal of complexity;
first note that
\begin{equation*}
  H_\mathrm{int}(t) = \mathrm{e}^{i H_0 (t - t_0)}
  H_\mathrm{int}\, \mathrm{e}^{-i H_0 (t - t_0)},
\end{equation*}
where $H_0$ is the free part of the Hamiltonian; furthermore,
considering the series expansion of the exponential, the
$n\mathrm{th}$ term in the series is
\begin{align*}
  S^{(n)} = (-i)^n \int_{t_0}^{t}\!\! dt_1 \int_{t_0}^{t_1}\!\! dt_2
  \dots \int_{t_0}^{t_{n\!-\!1}}\!\! dt_n\, H_\mathrm{int}(t_1)\dots
  H_\mathrm{int}(t_n) \\ = \frac{(-i)^n}{n!} \int_{t_0}^{t}\!\! dt_1
  \int_{t_0}^{t}\!\! dt_2 \dots \int_{t_0}^{t}\!\! dt_n\,\mathcal{T}
  \left[ H_\mathrm{int}(t_1)\dots H_\mathrm{int}(t_n) \right],
\end{align*}
where $\mathcal{T}$ implies one must take the \emph{time ordered
product} of the $H$ operators. If $H_\mathrm{int}$ is proportional to
some small coupling constant $g \ll 1$, as is the case for QED and
electroweak processes at low energies, then each term in the series
will be much smaller than the previous, so that one can treat the
calculation of $\bra{f}S\ket{i}$ perturbatively. This is an especially
crucial point because, despite of the asymptotic shrinking of the
terms, the full series is typically divergent; because of this,
entirely different methods are needed in cases of strong coupling $g
\sim 1$.

\begin{figure}[t]
  \begin{center}
  \includegraphics{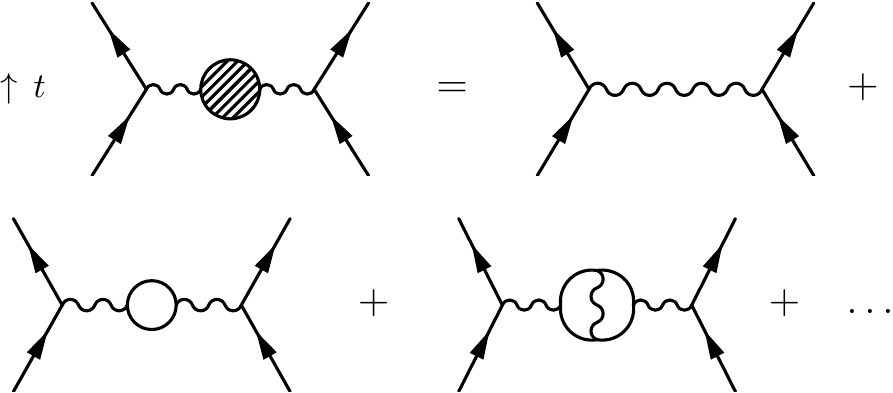} \caption[Expansion of $S$-matrix
  element for electron scattering.]{Feynman diagram expansion to third
  order of the $S$-matrix element for scattering of electrons by a
  photon.}
  \label{fig:1pi}
  \end{center}
\end{figure}

To further probe the $S$-matrix formalism, consider as an example the
simple QED scattering process $e^- e^- \rightarrow e^- e^-$; in this
case, $\mathcal{L} = i\bar\psi \gamma^\mu D_\mu \psi$, or
equivalently, $\mathcal{H}_\mathrm{int} = -Q e \bar\psi \gamma^\mu
\psi A_\mu$, such as for any term from eq.\,(\ref{eq:Lqed}). Figure
\ref{fig:1pi} shows the expansion of the scattering process in terms
of \emph{Feynman diagrams}, which are in one-to-one correspondence
with non-trivial terms in the $S$-operator expansion. The first such
term of the series, known as the \emph{tree-level} diagram, is
typically straightforward to calculate; for some processes, it may
also be a sufficient approximation to some low-energy measurement of
the matrix element. Note that in this case, the tree-level diagram
corresponds to the $n=2$ term in the series. Consider the pair of
$\mathcal{H}_{\rm int}$ operators in that term; each of the two fields
$\psi \sim \hat a$ act on the two initial electron states to
annihilate the incoming particles, each of the two fields $\bar\psi
\sim \hat a^\dagger$ act on the two final electron states to create
the outgoing particles, and the photon fields $A_\mu$ are \emph{Wick
contracted} with each other to create the propagator.

The second term in the expansion in Figure \ref{fig:1pi}
(corresponding to the $n=4$ term in the series) reveals a deeper
mathematical complication with $S$-matrix formalism. The loop in the
diagram, composed of two fermionic electron propagators, carries an
arbitrary momentum $\ell$, corresponding to an $\int\! d^4 \ell$ in
the calculation, which must be taken over all possible values of
$\ell$ $(-\infty, \infty)$. Fermionic propagators are $\sim i/\slashed
p$, so dimensional analysis suggests the integral is quadratically
divergent; these seemingly problematic loop factors are a general
feature of ``radiative corrections'' in a quantum field theory,
\emph{i.e.}, the quantum corrections to tree-level interactions 
arising from higher-order terms in the $S$-matrix. The apparent
intractability can be handled using a clever and intricate technique
called \emph{renormalization} \cite{feynman-reg,*schwinger,*tomonaga},
which uses a \emph{cut-off} energy scale or other \emph{regulator} to
quarantine the infinite part of the integral, then cancels that
infinite part against counter-terms associated to each of the
\emph{bare} parameters of the theory, namely the masses, coupling
constants, and wave-function normalization factors as they appear in
the original Lagrangian. In doing a complete analysis of the
renormalization of a particular theory, one finds not only that the
cut-off (ultimately $\rightarrow \infty$) is unphysical, but also that
the physical values of the parameters of the theory generally vary
with the overall energy scale of a measurement, and this variation is
determined by the finite parts of the higher-order loop diagrams in
the series expansion. The formalism describing this \emph{running} of
parameters with scale has a rich, group-like mathematical structure of
its own \cite{stueckelberg,wilson}.

With confidence that, despite its superficial complications,
$S$-matrix theory is mathematically valid, I can return to its use for
calculating measurable features of the SM. The non-trivial part of the
$S$ operator can be extracted explicitly by writing $S =
\boldsymbol{1} + iT$; furthermore, $T$ is related to the \emph{Feynman
amplitude} $\mathscr{M}$, generically known as the ``matrix element'',
by
\begin{equation*}
  \bra{\mathbf{p}_k \mathbf{p}_l}\, iT \, \ket{\mathbf{p}_i
  \mathbf{p}_j} = (2\pi)^4\, \delta^4(\Sigma p)\, i \mathscr{M},
\end{equation*}
where $\delta^4(\Sigma p) = \delta^4(p_i + p_j - p_k - p_l)$ gives the
total four-momentum conservation for the process. Since the
Hamiltonian, whose eigenvalues are energy, is a Hermitian operator,
$S$ is a unitary operator; consequently, the absolute square of a
$T$-matrix element gives the probability for the occurrence of the
corresponding interaction if the following conditions are satisfied:
(a) the free incoming particles are present at $t \rightarrow -\infty,
\mathbf{x} \rightarrow \infty$, (b) the system undergoes eternal time
evolution via the operator $\exp\left(-iHt\right)$, and (c) the free
outgoing particles are present at $t \rightarrow \infty, \mathbf{x}
\rightarrow \infty$. Using this prescription and the above definition
for $\bra{f} iT \ket{i}$, one can calculate the \emph{scattering cross
section} $\sigma$ of the interaction $\Psi_i \Psi_j \rightarrow \Psi_k
\Psi_l$:
\begin{equation*}
  \sigma = \frac{1}{4 E_i E_j v} \int \frac{\mathbf{d}^3
  \mathbf{p}_k}{(2\pi)^3\, 2 E_k} \int \frac{\mathbf{d}^3
  \mathbf{p}_l}{(2\pi)^3\, 2 E_l}\, (2\pi)^4\, \delta^4(\Sigma p) \;
  \lvert \mathscr{M} \rvert^{\,2},
\end{equation*}
where $v$ is the relative velocity of the incoming particles. A
similar expression can be written for the \emph{decay width} of a
massive particle. One can
make explicit measurements of a cross section or a decay width,
represented by some $S$-matrix element, by observing the output of
particle beams incident upon each other, so long as ($a$) the
interaction occurs in relative isolation, at a ``large'' distance from
the detectors, and ($b$) the output is observed a very large number
of times, so as to simulate the eternality of the probabilities.

Indeed, precisely such measurements have been made for decades, at
particle accelerator experiments such as the Tevatron, LEP, and now
the LHC; every probability associated with an interaction predicted by
the SM agrees with the experimental data to truly remarkable and
unprecedented levels of precision. Furthermore, several of the
particles of the SM were predicted to exist by the completed framework
\emph{prior to being observed}; the mass of each particle was
accurately predicted as well. This was the case for the heavy quarks,
the $W$ and $Z$ bosons, and, most recently, the Higgs boson $h^0$,
which was not seen until 2012. The mass of the Higgs was perhaps a bit
higher than originally expected, and so its observation had to wait
for the construction of CERN's Large Hadron Collider; yet, due to the
extremely thorough record of prior successes of the model, physicists
remained confident throughout the years that the Higgs boson would be
seen.

The model also makes similarly remarkable predictions involving the
precision of measured values related to the hydrogen atom, the
magnetic moment of the electron, and other low-energy or atomic
phenomena. These values were previously calculated in the context
of non-relativistic quantum mechanics or classical electromagnetism
and showed unexplained discrepancies with measurements; the
discrepancies are largely eliminated when the analogous calculations
are performed in the context of the SM.

\section{The Limitations of the Model and a Need for New Physics}
Despite the extreme robustness and precision of the Standard Model, it
is at the same time a manifestly incomplete theory, and it leaves some
number of mysteries unsolved. Some of the most obvious aspects of its
incompleteness are:
\begin{itemize}
  \item The model relies on the presence of roughly 19 parameters,
    including masses, coupling constants, and generational mixing
    parameters, whose values are known through measurement and are
    otherwise completely arbitrary; in some cases, the observed values
    are arguably fine-tuned. Such tunings include the more conceptual
    concern of the presence of the three generations of
    otherwise-identical fermions with different masses, where a unique
    and unexplained hierarchical mass spectrum exists for each flavor.
  \item The model predicts that neutrinos are massless, while
    there is ample experimental evidence otherwise. Freely propagating
    neutrinos are known to oscillate from one generation to another;
    the only known mechanism for such a process is through CKM-like
    mixing among flavor and mass eigenstates. Hence, neutrinos seem to
    have mass after all, however small those masses may be. 
  \item The model makes no mention whatsoever of gravity; furthermore,
    it consequently gives no explanation for the presence of dark
    energy and no realistic explanation for the presence of dark
    matter.
\end{itemize}
In addition to these omissions, there are few more subtle
peculiarities that suggest theoretical incompleteness:
\begin{itemize}
  \item Like the parameters of the model, the internal gauge symmetry
    group of the SM is \emph{ad hoc}, as it was originally determined
    primarily through phenomenological arguments.
  \item The negative scalar mass parameter and therefore the entirety
    of electroweak breaking is similarly arbitrary from the
    theoretical perspective; the Higgs mechanism was devised to solve
    the problem of giving mass to the particles and is not motivated
    by any aspect of the mathematical structure of the model.
  \item Radiative corrections to the Higgs propagator are
    quadratically dependent on the energy scale of the measurement;
    these strongly divergent contributions, which are unique to scalar
    fields, severely renormalize the mass of the particle. Naively,
    one would expect this to lead to arbitrarily large corrections to
    the mass, pushing it all the way up to the \emph{Planck scale},
    where gravitational effects become significant, $M_\mathrm{Pl}
    \sim 10^{18}$\,GeV. Yet, we see the Higgs boson to have a
    comparably minuscule mass of 126\,GeV; the SM offers no
    explanation for this truly enormous discrepancy. This puzzle is
    known as the \emph{hierarchy problem}.
\end{itemize}
These unsolved questions have led physicists to pursue a great number
of ideas for the extension of the standard model, to varying degrees
of success. So far, very little has been ``officially'' added to the
theory, as no definitive experimental evidence has been observed in
support of any hypothesis.

Soon after the completion of the SM framework in the early 1970s, a
new class of models emerged from attempts to extend the notion of
electroweak unification to more fundamental levels. It seemed that if
electromagnetism and weak interactions were unified earlier in the
universe, then perhaps \emph{that} era followed from the breaking of
\emph{yet another} unification of the electroweak force with the
strong force. This concept, known as \emph{Grand Unified theory},
offered some relief to the arbitrariness of the SM gauge group. The
first models were developed by Pati and Salam \cite{pati-salam} and
then Georgi and Glashow \cite{su5} in 1974. Further extensions of
these models in turn led to the development of $SO(10)$ unification,
which will be a primary topic for the remainder of this work.

Taking a closer look at the Higgs mass corrections, one will notice
that they arise from both bosonic and fermionic loops; furthermore,
these contributions come with opposite signs. This subtlety led some
physicists in the 1970s to propose a practical application of an
otherwise-esoteric idea known as \emph{supersymmetry}, which relates
bosons to fermions through a subtle extension of spacetime itself.  I
will introduce this concept in more detail in the next chapter.

%
%

\chapter{Supersymmetry}\label{3-susy}

\begin{figure}[t]
\begin{center}
        \includegraphics{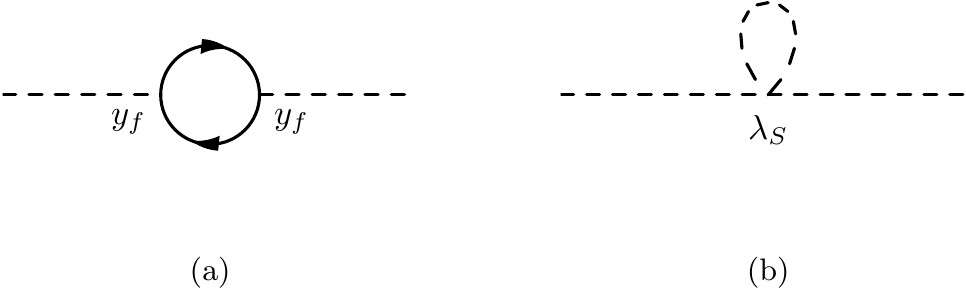} \caption[One-loop
        corrections from (a) a fermion and (b) a scalar to the Higgs
        mass squared $m_h^2$.]{One-loop diagrams from (a) a fermion
        $f$ and (b) a scalar $S$ for the Higgs propagator, which give
        corrections to the bare mass squared parameter $m_h^2$.}
	\label{fig:bfloops}
\end{center}
\end{figure}

Consider the diagrams for the one-loop corrections to the Higgs boson
mass squared parameter $m_h^2$ seen in Figure \ref{fig:bfloops};
the correction from a generic fermion $f$ in (a) can be written as
\vspace{-4mm}
\begin{equation}
  \Delta m_h^2 = - \frac{y_f^2}{8 \pi^2} \Lambda^2_{\mathrm{UV}}
  + \dots,
  \label{hloopf}
\end{equation}
where $\Lambda_{\mathrm{UV}}$ is the cutoff energy scale used to
regulate the loop integral for renormalization; the analogous
contribution from a generic scalar $S$, seen in Figure
\ref{fig:bfloops}(b) is
\begin{equation}
  \Delta m_h^2 = \frac{\lambda_S}{16 \pi^2} \Lambda^2_{\mathrm{UV}}
  + \dots
  \label{hloops}.
\end{equation}
The terms in ``\dots'' are at most logarithmically dependent on
$\Lambda_{\mathrm{UV}}$. Assuming no additional physics aside from
gravity, the cutoff is at the Planck scale, and these corrections are
at least 25 orders of magnitude larger than the physical value of $(126
\,\mathrm{GeV})^2$, depending on the size of the coupling constants.
Naively, this suggests a staggeringly large cancellation between the
bare Higgs mass $m_h$ and these corrections. Note that the
contributions from the log-divergent terms are a much more natural
$\mathcal{O}(m_h^2)$.

If instead one requires the $\Lambda^2_{\mathrm{UV}}$ corrections to
be similarly natural, then one fines a need for $\Lambda_{\mathrm{UV}}
\lsim \mathcal{O}(1)$\,TeV, which naively suggests a need for new
physics at that scale.

There is, however, a more creative solution one might consider. Since
the correction from the fermion is negative but the correction from
the scalar is positive, under the restriction that $y_f^2 =
\lambda_S$, then a theory with two such bosons for each fermion would
have a cancellation of these problematic terms against each other; in
fact the cancellation would persist to all orders. Since these are
interactions with the Higgs field, the above restriction on the
couplings corresponds to the scalar and the fermion having identical
masses.

In turns out that such a theory does exist. \emph{Supersymmetry}
employs fermionic operators to enable transformation of bosons into
fermions, and vice versa, through a subtle extension of spacetime
itself. The formalism was discovered in the early 1970s and was
explored for mainly novel reasons until the realization of its
application to the hierarchy problem discussed above. This chapter
will introduce the basic structure of supersymmetry (SUSY) and give
the form of a realistic extension of the standard model that utilizes
the concept to address not only the hierarchy problem, but also
several other aspects of the puzzles of the SM.

\section{Basic Supersymmetry Formalism} Consider a bosonic state
$\ket{b}$ and a fermionic state $\ket{f}$.  The generator of
supersymmetry is a fermionic operator $\hat Q$ such that $\hat Q
\ket{b} \sim \ket{f}$ and $\hat Q \ket{f} \sim \ket{b}$. In general, a
proper supersymmetric transformation trades a bosonic degree of
freedom for a fermionic one in a one-to-one manner. More
realistically, one can understand the Weyl spinor operators $\hat
Q^\alpha$ and $\hat Q^\dagger_{\dot{\alpha}}$ as a peculiar extension
of the Poincar\'e algebra such that
\begin{align}
  \left\{ \hat Q_\alpha, \hat Q^\dagger_{\dot{\alpha}} \right\} = -2
  \sigma^\mu_{\alpha {\dot{\alpha}}} \hat P_\mu,
  \label{eq:qq-p}
\end{align}
and
\begin{align}
  \left\{ \hat Q_\alpha, \hat Q_\beta \right\} = 0\,; \qquad \left\{
  \hat Q^\dagger_{\dot{\alpha}}, \hat Q^\dagger_{\dot{\beta}} \right\}
  = 0\,,
\end{align}
where $\hat P_\mu = i \partial_\mu$ is the generator of momentum and
$\sigma^\mu_{\alpha \dot{\alpha}}$ is the usual extension of the Pauli
matrices $(\mathbb{I}, \vec{\sigma})$, except I have written the
$SL(2,\mathbb{C})$ spinor space indices explicitly. The indices of
$Q^\alpha$ (and $\sigma^\mu_{\alpha {\dot{\alpha}}}$) are raised and
lowered using the Levi-Civita tensor $\epsilon^{\alpha \beta}$, with
$\epsilon^{12} = - \epsilon_{12} = 1$. Note also that
\begin{align}
  \left[\, \hat Q_\alpha, \hat P_\mu\, \right] = 0\,; \qquad \left[\,
  \hat Q^\dagger_{\dot{\alpha}}, \hat P_\mu\, \right] = 0\,
\end{align}
\emph{i.e.}, supersymmetric transformations commute with all
translations, implying that a boson and a fermion transforming into
one another under SUSY will have the same mass. The above relations
comprise a closed extension of the Poincar\'e algebra, forming what is
known as a \emph{graded algebra} or a superalgebra. This
supersymmetric loophole is the only exception to the Coleman-Mandula
``no-go'' theorem, which implies that the only symmetry group of the
$S$-matrix consistent with QFT is a direct product of Poincar\'e and
some internal compact Lie group.

Since we have not seen \emph{superpartner} particles for the light SM
particles in nature, it would seem that SUSY is broken symmetry at low
energies; however, in order to preserve the perfect cancellations in
the Higgs mass corrections, which requires that $y_f^2 = \lambda_S$
still holds in the broken theory, the breaking of SUSY must be
isolated from the dynamics. This prescription is known as \emph{soft
breaking} of the theory, and it is realized mainly through (positive)
mass terms for the superpartners, which may be the result of some
``hidden sector'' physics, cut off from the low energy physics, but
are otherwise free parameters. I will discuss this concept and its
implications in more detail shortly.

\subsection{Constructing a Supersymmetric Model} \footnote{This
discussion largely follows that of ref.\,\cite{martin}; please see that
work for further detail.} \label{susymodel} The most basic non-trivial
SUSY model one can construct involves a free single Weyl fermion $\psi
= \psi_\alpha$ and its two free scalar superpartners, which are
conventionally treated as one complex field $\phi = (A +
iB)/\sqrt{2}$. Note that for a realistic model with both matter
fermions and scalar bosons, each type of field will have the other
type as its superpartner; hence, I will keep this discussion very
general so it can apply to either case. The supersymmetric
transformation of a field is defined as
\begin{equation}
  -i \sqrt{2} \,\delta (\varepsilon) X \equiv \left[\, \varepsilon \hat
  Q + \varepsilon^\dagger \hat Q^\dagger, X\, \right]
  \label{eq:suptrans}
\end{equation}
for any field $X$ and infinitesimal parameter $\varepsilon_\alpha$,
which is a constant Grassmann (anti-commuting) spinor; the contraction
$\varepsilon \hat Q \equiv \epsilon^{\alpha \beta} \varepsilon_\alpha
\hat Q_\beta$, and $\varepsilon^\dagger \hat Q^\dagger$ is analogous.
One may expect that the corresponding supersymmetric Lagrangian is
simply
\begin{equation}
  \mathcal{L} = \partial^\mu \phi^* \partial_\mu \phi + i \psi^\dagger
  \bar \sigma^\mu \partial_\mu \psi,
  \label{eq:wz0}
\end{equation}
where $\psi^\dagger \bar \sigma^\mu \psi \equiv \psi^\dagger_{\dot
\alpha} (\bar \sigma^\mu)^{\dot \alpha \alpha} \psi_\alpha$. At first
glance, this will seem correct: the transformations of the fields are
\begin{align}
  \delta (\varepsilon) \phi &= \varepsilon \psi, & \delta
  (\varepsilon) \phi^* &= \varepsilon^\dagger \psi^\dagger, \nonumber\\ 
  \delta (\varepsilon) \psi_\alpha &= i \left( \sigma^\mu 
  \varepsilon^\dagger \right)_\alpha \partial_\mu \phi, & \delta 
  (\varepsilon) \psi^\dagger_{\dot\alpha} &= -i \left( \varepsilon 
  \sigma^\mu \right)_{\dot\alpha} \partial_\mu \phi^*,
  \label{eq:dbf}
\end{align}
where, \emph{e.g.}, $\left( \sigma^\mu \varepsilon^\dagger
\right)_\alpha \equiv \sigma^\mu_{\alpha \dot\alpha}
\varepsilon^{\dagger\,\dot\alpha}$; utilizing these transformations in
eq.\,(\ref{eq:wz0}), one finds that
\begin{align}
  \delta \mathcal{L}_\phi &= \varepsilon \partial^\mu \psi \partial_\mu
  \phi^* + \varepsilon^\dagger \partial^\mu \psi^\dagger \partial_\mu
  \phi, \nonumber \\
  \delta \mathcal{L}_\psi &= -\varepsilon \partial^\mu \psi
  \partial_\mu \phi^* - \varepsilon^\dagger \partial^\mu \psi^\dagger
  \partial_\mu \phi \,+ \,\mathrm{total~derivatives};
  \label{eq:dwz0}
\end{align}
since the total derivative vanishes in the action, $\mathcal{L}$ is in
fact invariant under a SUSY transformation.

Still, though, one must address the closure of the superalgebra.
Considering successive transformations $[\, \delta (\varepsilon_2),
\delta (\varepsilon_1)\, ] X$, one sees that
\begin{align}
  \left[\, \delta (\varepsilon_2), \delta (\varepsilon_1)\, \right] 
  \phi =& \,i \left( \varepsilon_2 \sigma^\mu \varepsilon_1^\dagger -
  \varepsilon_1 \sigma^\mu \varepsilon_2^\dagger \right) \partial_\mu
  \phi, \nonumber \\ 
  \left[\, \delta (\varepsilon_2), \delta (\varepsilon_1)\, \right]
  \psi_\alpha =& \,i \left( \varepsilon_2 \sigma^\mu
  \varepsilon_1^\dagger - \varepsilon_1 \sigma^\mu
  \varepsilon_2^\dagger \right) \partial_\mu \psi_\alpha \nonumber \\
  &+ \,i \varepsilon_{1\,\alpha} \varepsilon_2^\dagger \bar \sigma^\mu
  \partial_\mu \psi - i \varepsilon_{2\,\alpha} \varepsilon_1^\dagger
  \bar \sigma^\mu \partial_\mu \psi. 
  \label{eq:ddwz0}
\end{align}
For the scalar field, a product of SUSY transformations returns a
derivative of the field, as suggested by eq.\,(\ref{eq:qq-p}). The
fermion case is similar once one notes that the two extra terms in the
transformation will vanish on-shell, when the classical equation of
motion $\bar \sigma^\mu \partial_\mu \psi = 0$ holds. This is
something, but it is not enough to build a truly consistent
supersymmetric quantum model.

This problem can be resolved by introducing an \emph{auxiliary field}
into the system with the right properties.  The field $F$ will be a
complex scalar with $[F] = 2$, and the contribution to the Lagrangian
is
\begin{equation}
  \mathcal{L}_F = -F^* F;
  \label{eq:wzf}
\end{equation}
the field has a non-dynamical, algebraic equation of motion, and so
should be treated as unphysical. The field transforms under SUSY as 
\begin{equation}
  \delta (\varepsilon) F = i \varepsilon^\dagger \bar \sigma^\mu 
   \partial_\mu \psi, \qquad \delta (\varepsilon) F^* = -i
   \partial_\mu \psi^\dagger \bar \sigma^\mu \varepsilon; 
\end{equation}
combining this with an augmentation of the fermion transformations,
\begin{equation}
  \delta (\varepsilon) \psi_\alpha = i \left( \sigma^\mu 
  \varepsilon^\dagger \right)_\alpha \partial_\mu \phi +
  \varepsilon_\alpha F, \qquad \delta 
  (\varepsilon) \psi^\dagger_{\dot\alpha} = -i \left( \varepsilon 
  \sigma^\mu \right)_{\dot\alpha} \partial_\mu \phi^* +
  \varepsilon^\dagger_{\dot\alpha} F^*,
\end{equation}
gives the desired off-shell closure of the complete system.

Therefore, eq.\,(\ref{eq:wz0}) together with eq.\,(\ref{eq:wzf}) give
a complete supersymmetric Lagrangian for a free scalar, its fermionic
superpartner, and the corresponding auxiliary field, which is known as
the \emph{Wess-Zumino} model of supersymmetry; it will be the basis
for building a realistic model of SUSY-invariant interactions.

\subsubsection{Yukawa Interactions and the Superpotential} To
introduce interactions in the model, I first define the
\emph{superpotential} $W$:
\begin{equation}
  W \equiv \frac{1}{2} M^{ij} \phi_i \phi_j + \frac{1}{6} y^{ijk}
  \phi_i \phi_j \phi_k,
  \label{eq:genW}
\end{equation}
where the indices $i,j,k$ generically run over any flavor quantum
numbers.  Note that $W$ is \emph{holomorphic}, \emph{i.e.}, analytic
in $\phi$, and completely symmetric under exchange of indices. Now I
can write
\begin{equation}
  \mathcal{L}_{\mathrm{int}} = -\frac{1}{2} W^{ij} \psi_i \psi_j + W^i
  F_i + \mathrm{h.c.s},
  \label{eq:wzint0}
\end{equation}
where
\begin{align}
  W^{ij} \equiv \frac{\delta^2 W}{\delta \phi_i \delta \phi_j} &=
  M^{ij} + y^{ijk} \phi_k ~\mathrm{and} \nonumber \\
  W^i \equiv \frac{\delta W}{\delta \phi_i} &=
  M^{ij} \phi_j + \frac{1}{2} y^{ijk} \phi_j \phi_k. \nonumber \\
  \label{eq:Wijdefs}
\end{align}
The $F$-terms in the full Lagrangian lead to the algebraic equations
of motion
\begin{equation}
  F_i = -W^*_i \qquad \mathrm{and} \qquad F^{* i} = -W^i, \nonumber
\end{equation}
which I can utilize to rewrite the interaction Lagrangian as
\begin{equation}
  \mathcal{L}_{\mathrm{int}} = -\frac{1}{2} W^{ij} \psi_i \psi_j +
  \mathrm{h.c.} - V(\phi, \phi^*),
  \label{eq:wzint}
\end{equation}
where
\begin{equation}
  V(\phi, \phi^*) \equiv  W^i W^*_i = \Big\lvert \frac{\delta
  W}{\delta \phi_i} \Big\rvert^2 
\end{equation}
is the \emph{scalar potential} for the system, giving the usual mass,
cubic, and quartic terms for the scalar field(s)\,$\phi$; similarly,
the $W^{ij}$ term gives a (holomorphic) fermion mass term and Yukawa
coupling with the scalar parter $\phi$.

\subsubsection{Gauge Fields and Interactions} To expand a
Wess-Zumino-type model to include gauge interactions, I will first
need to consider the supersymmetric transformation of gauge bosons.
Like a scalar field, spin-1 fields will also have fermionic spin-1/2
superpartners. For a gauge field $A^a_\mu$, I will denote the
``gaugino'' superpartner as $\lambda^a_\alpha$.\footnote{Note that in
four-component bispinor notation, the gaugino is a \emph{Majorana}
fermion, meaning $\psi^\mathcal{C} = \psi$} The Lagrangian for the
gauge sector is then
\begin{align}
  \mathcal{L}_{\mathrm{g}} = ~-\frac{1}{4} A_{a}^{\mu\nu} A^{a}_{\mu\nu}
  - i \lambda^{\dagger a} \bar \sigma^\mu D_\mu \lambda^a +
  \frac{1}{2} D^a D^a;
\label{eq:wzg}
\end{align}
$D^a$ is, like $F$, an auxiliary field that allows the superalgebra to
close off-shell; unlike $F$, however, it is a real field (since the
on-shell boson has only one additional degree of freedom).
$F^a_{\mu\nu}$ is defined in the usual manner (\emph{e.g.}, as seen in
the previous chapter), and the covariant derivative acts on the
gaugino as $D_\mu \lambda^a = \partial_\mu \lambda^a + g f^{abc}
A^b_\mu \lambda^c$. Both $D^a$ and $\lambda^a$ transform in the
adjoint representation of the gauge group. The supersymmetric
transformations of the fields are
\begin{align}
  \delta (\varepsilon) A^a_\mu &= -\frac{1}{\sqrt{2}}
  \left(\varepsilon^\dagger \bar \sigma^\mu \lambda^a +
  \mathrm{h.c.}\right), \nonumber \\ \delta (\varepsilon)
  \lambda^a_\alpha &= \frac{i}{2 \sqrt{2}}\left(\sigma^\mu \bar
  \sigma^\nu \varepsilon \right)_\alpha A^{a}_{\mu\nu} +
  \frac{1}{\sqrt{2}}\, \varepsilon_\alpha D^{a},\nonumber \\ \delta
  (\varepsilon) D^{a} &= \frac{i}{\sqrt{2}} \left( \varepsilon^\dagger
  \bar \sigma^\mu D_\mu \lambda^a - D_\mu \lambda^{\dagger a} \bar
  \sigma^\mu \varepsilon \right).
  \label{eq:dg}
\end{align}

One couples the fermions $\psi$ and scalars $\phi$ to $A^a_\mu$
through the usual promotion of the derivative $\partial_\mu
\rightarrow D_\mu$ in the Lagrangian eq.\,(\ref{eq:wz0}); however, one
must also consider allowed fermion-boson-gaugino interactions, which
are of the form
\begin{equation}
  \mathcal{L}_{\mathrm{g,int}} = -g \sqrt{2} \left(\phi^* t^a
  \psi\, \lambda^a + \mathrm{h.c.}\right) + g\, \phi^* t^a \phi\, D^a,
  \label{eq:gint}
\end{equation}
where $t^a$ are the generators of the gauge group. As with $F$, one
can again use the algebraic equation of motion for the auxiliary field
$D^a = - g \phi^* t^a \phi$ to eliminate it from the Lagrangian. This
also results in an additional contribution to the scalar potential,
\begin{equation}
  V(\phi, \phi^*) \equiv  W^i W^*_i + \frac{1}{2}\, g^2\! \left( \phi^*
  t^a \phi \right)^2.
\end{equation}
Note this can also be written as $V = \lvert F \rvert^2 + \frac{1}{2}
D^2$, which gives rise to the common nomenclature ``F-term'' and
``D-term'' when referring to the two scalar potential contributions.
Note that in the presence of multiple gauge groups (as in the SM), one
finds a simple sum of contributions from each.

To guarantee invariance of the entire interacting model under SUSY
transformations, one must replace the derivatives in the
transformations $\delta \psi$ and $\delta F$ with gauge covariant
derivatives, and augment the transformation of $F$ by the inclusion of
a term involving the gaugino
\begin{equation}
  \delta (\varepsilon) F_i = i \varepsilon^\dagger \bar \sigma^\mu 
   D_\mu \psi_i - g \sqrt{2} \left(t^a \phi \right)_i
   \varepsilon^\dagger \lambda^{\dagger a}
\end{equation}
and similar for $F^{* i}$. Now the entire system is invariant (up to total
derivatives) under the transformations given by eqs.\,(\ref{eq:dg}), the
gauge covariant versions of (\ref{eq:dbf}), and the above
transformation for $F$.

\subsubsection{Soft Supersymmetry Breaking} As mentioned previously,
the absence of superpartners in nature suggests that SUSY is a broken
symmetry. One would like to find that the symmetry is broken
spontaneously, like that of electroweak theory; early on, the
possibilities of taking $\langle F \rangle \neq 0$ \cite{oraif} or
$\langle D \rangle \neq 0$ \cite{fayet-iliop} were explored
thoroughly; both options can be implemented in general SUSY models to
break the symmetry, but in the context of supersymmetric extension of
the standard model, both methods fail to give a realistic mass
spectrum for the superpartners. In the end, one is left to consider
\emph{soft breaking} of SUSY through terms with couplings of
explicitly positive mass dimension.

Soft breaking terms allowed in the general interacting model described
above are
\begin{equation}
  \mathcal{L}_{\mathrm{soft}} = - \frac{1}{2} M_g \lambda^a \lambda^a
  - \frac{1}{2} b^{ij} \phi_i \phi_j - \frac{1}{6} a^{ijk} \phi_i
  \phi_j \phi_k + \mathrm{c.c.s} - {(m^2)^i}_j\, \phi^{* j} \phi_i.
  \label{eq:Lsoft}
\end{equation}
We will see more about the consequences of these terms in the context
of the Minimally Supersymmetric SM in the next section. I will also
discuss briefly some mechanisms that could dynamically give rise to
these terms.

\subsection{Superfields}

In order to make supersymmetry manifest in a field theory, one needs
to consider \emph{superfields}, or multiplets containing a field and
its superpartner. In order to accommodate the fundamental spacetime
differences between bosons and fermions in the same object, one needs
to expand the spacetime itself to include four new fermionic
coordinates $x^\mu \rightarrow (x^\mu, \theta^\alpha,
\theta^\dagger_{\dot \alpha})$. These new coordinates of dimension
$[\theta] = -\frac{1}{2}$ commute with $x^\mu$ but anti-commute with
themselves and each other. Products or contractions of thetas are
generally the same as those for any Weyl fermions, but note also that
$\theta^\alpha \theta^\beta = -\frac{1}{2} \epsilon^{\alpha \beta}
\theta \theta$ for identical spinors. 

The Grassmann nature of the thetas has the peculiar implication that
the square of any individual component vanishes, $(\theta_1)^2 =
(\theta_2)^2 = 0$. As a result, any general function of $\theta$ and
$\theta^\dagger$ can be written as a terminating series. Therefore,
the most general superfield $\mathcal{S}$ one can write has the form
\begin{align}
  \mathcal{S}(x^\mu, \theta, \theta^\dagger) =\, &a + \theta \chi +
  \theta^\dagger \xi^\dagger  + \theta^2 b + (\theta^\dagger)^2 c +
  \theta^\dagger \bar \sigma^\mu \theta\, v_\mu \nonumber \\ &~+
  (\theta^\dagger)^2 \theta \eta + \theta^2 \theta^\dagger
  \zeta^\dagger  + (\theta^\dagger)^2 (\theta)^2 d,
  \label{eq:gensfield}
\end{align}
where all component fields are functions of spacetime. When comparing
to the fields in the previous section, one can determine that $a$ is
scalar-like, $\chi,\xi$ is fermion-like, $\eta,\zeta$ gaugino-like,
and $b,c,d$ auxiliary-field-like. The complex scalar component fields
$a,b,c,d$ give eight real bosonic degrees of freedom, $v_\mu$ gives
eight more as a complex vector field, and the (always complex) Weyl
fermion components $\chi,\xi^\dagger, \eta, \zeta^\dagger$ give
sixteen fermionic degrees of freedom. $\mathcal{S}$ transforms under
general SUSY transformations as a translation in superspace,
\begin{align*}
  \mathcal{S}(x^\mu, \theta, \theta^\dagger) \rightarrow \,&\exp
  \left[\, i\! \left( \varepsilon Q + \varepsilon^\dagger Q^\dagger
  \right) \right] \mathcal{S}(x^\mu, \theta, \theta^\dagger) \\ &=
  \mathcal{S} \left(x^\mu - i \varepsilon \sigma^\mu \theta^\dagger + i
  \theta \sigma^\mu \varepsilon^\dagger,~ \theta + \varepsilon,~
  \theta^\dagger + \varepsilon^\dagger \right);
\end{align*}
note that superfields are closed under multiplication, which is a
crucial factor for constructing Lagrangians.

We can write the SUSY generators as differential operators in
superspace:
\begin{equation}
  Q_\alpha = -i\, \frac{\partial}{\partial \theta^\alpha} - \left(
  \sigma^\mu \theta^\dagger \right)_\alpha \partial_\mu; \qquad
  Q^\dagger_{\dot\alpha} = i\, \frac{\partial}{\partial
  (\theta^\dagger)^{\dot\alpha}} + \left( \theta
  \sigma^\mu \right)_{\dot\alpha} \partial_\mu;
\end{equation}
Using these operators, one can show that supersymmetric
transformations written in terms of these differential operators are
equivalent to the transformations in terms of the quantum operators as
seen in eq.\,(\ref{eq:suptrans}):
\begin{equation*}
  \left[\, \varepsilon \hat Q + \varepsilon^\dagger \hat Q^\dagger,
  X\, \right] = \left( \varepsilon Q + \varepsilon^\dagger Q^\dagger
  \right) X 
\end{equation*}
for any superfield component $X$. One can also define the \emph{chiral
covariant derivatives}
\begin{equation}
  D_\alpha = \frac{\partial}{\partial \theta^\alpha} + i \left(
  \sigma^\mu \theta^\dagger \right)_\alpha \partial_\mu; \qquad
  D^\dagger_{\dot\alpha} = - \frac{\partial}{\partial
  (\theta^\dagger)^{\dot\alpha}} - i \left( \theta
  \sigma^\mu \right)_{\dot\alpha} \partial_\mu,
\end{equation}
such that $\delta (\varepsilon) (D_\alpha \mathcal{S}) = D_\alpha
(\delta (\varepsilon) \mathcal{S})$, and similar for
$D^\dagger_{\dot\alpha}$. Note that these operators satisfy the same
superalgebra as, and also anti-commute with, $Q$ and $Q^\dagger$.

\subsubsection{Irreducible Supermultiplets} The general superfield
$\mathcal{S}$ is a reducible representation in the superalgebra space.
This is perhaps evident in light of the independent supersymmetric
closure of \emph{each} of the sets of fields $\{\phi, \psi, F \}$ and
$\{A, \lambda, D\}$, as seen in the previous section. One can obtain
the desired irreducible multiplets by constraining $\mathcal{S}$ in
specific ways.

The \emph{chiral} or \emph{left-chiral superfield} $\Phi_L$, which
generically corresponds to an irreducible supermultiplet containing a
matter fermion or scalar boson, arises from the constraint equation
\begin{equation}
  D^\dagger_{\dot\alpha} \Phi_L = 0.
\end{equation}
Using the convenient change of variables $y^\mu \equiv x^\mu + i
\theta \sigma^\mu \theta^\dagger$, one can write a general chiral
superfield as
\begin{equation}
  \Phi_L (y, \theta) = \phi (y) + \sqrt{2} \theta \psi (y) +
  \theta^2 F (y);
  \label{eq:chisfield}
\end{equation}
where the component fields $\{\phi, \psi, F \}$ correspond to those
from the previous section. Note one can quickly determine that a chiral
superfield has $[\Phi] = 1$.

Similarly, the \emph{anti-chiral} or \emph{right-chiral superfield}
$\Phi^*_R$ is the complex conjugate of $\Phi_L$ and arises from the
constraint equation
\begin{equation}
  D_\alpha \Phi^*_R = 0;
\end{equation}
Using the corresponding change of variables $y^{\mu *} \equiv x^\mu -
i \theta \sigma^\mu \theta^\dagger$, one can write a general
anti-chiral superfield as
\begin{equation}
  \Phi^*_R (y^*, \theta^\dagger) = \phi^* (y^*) + \sqrt{2}
  \theta^\dagger \psi^\dagger
  (y^*) + (\theta^\dagger)^2 F^* (y^*).
  \label{eq:antichisfield}
\end{equation}

Finally, the \emph{vector superfield} $\mathcal{A}$, which is the
irreducible supermultiplet containing a gauge boson field, is obtained
by demanding the superfield is real, \emph{i.e.}, by imposing the
condition $\mathcal{S} = \mathcal{S}^*$. Comparing with
eq.\,(\ref{eq:gensfield}), this implies
\begin{equation*}
  a = a^*, \quad \chi = \xi, \quad b = c^*, \quad v_\mu =
  v_\mu^*, \quad \eta = \zeta, \quad d = d^*.
\end{equation*}
Note that the combinations of chiral/anti-chiral superfields $\Phi^*
\Phi$, $\Phi + \Phi^*$, and $i(\Phi^* - \Phi)$ are also real and hence
are vector superfields.

We can write the generalization of an infinitesimal gauge
transformation to supersymmetric form as 
\begin{equation}
  \mathcal{A}^a \rightarrow \mathcal{A}^a + i(\Omega^{* a} - \Omega^a)
  + g f_{abc} \mathcal{A}^b (\Omega^{* c} + \Omega^c)
  \label{eq:supergauge}
\end{equation}
for some chiral superfield gauge transformation parameter $\Omega$;
the expression simplifies in the usual manner for Abelian symmetry.
Such a transformation will yield the proper form for a gauge
transformation of the gauge boson field, as well as the proper
transformations for the gaugino $\lambda^a$ and auxiliary field $D^a$
for non-Abelian cases. Using a convenient supergauge choice $\Omega^*
= -\Omega$, known as the \emph{Wess-Zumino gauge}, one can write a
vector superfield in the form
\begin{equation}
  \mathcal{A}^a(x^\mu, \theta, \theta^\dagger) = \theta^\dagger \bar
  \sigma^\mu \theta\, A^a_\mu + (\theta^\dagger)^2 \theta \lambda^a +
  \theta^2 \theta^\dagger \lambda^{\dagger a} + \frac{1}{2}
  (\theta^\dagger)^2 \theta^2 D^a,
  \label{eq:vecsfield}
\end{equation}
where the component fields $\{A, \lambda, D \}$ correspond to those
for a supersymmetric gauge model from the previous section. In this
form, it is apparent that $[\mathcal{A}] = 0$.

All three types of superfields discussed above close independently
under multiplication.

\subsubsection{A Complete Superfield Lagrangian} Using the
superfield notation from the previous subsection and the details
introduced in Section \ref{susymodel}, one can write a complete
supersymmetric action in terms of integrals of superfields in
superspace. One might see the final form as rather unexpected, in that
it relies on several unusual intermediate results.

First I need to discuss how one performs Grassmann integration. Using
these basic rules,
\begin{equation*}
  \int d \theta^\alpha = 0, \qquad \int d \theta^\alpha \theta^\beta
  = \delta^{\alpha \beta},
\end{equation*}
and noting that $d^2 \theta = -\frac{1}{4} \epsilon_{\alpha \beta} d
\theta^\alpha d \theta^\beta$, one can see that the integration of a
function $f(\theta, \theta^\dagger)$ over some measure in superspace
picks out the coefficient in $f$ of the term with theta dependence
matching that of the signature; \emph{e.g.},
\begin{align*}
  &\int d^{\,2} \theta \,\mathcal{S} = b +
  \theta^\dagger \zeta^\dagger + (\theta^\dagger)^2 d, \\
  &\int d^{\,2} \theta d^{\,2} \theta^\dagger \,\mathcal{S} = d,
  \qquad \mathrm{etc.}
\end{align*}

Now, I can use the above principle to build my superfield Lagrangian
by integrating certain products of superfields over certain portions
of superspace. For instance, in the expansion of the superfield
product $\Phi^*_R \Phi_L$, one will find that the ``D-term'' $\sim
(\theta^\dagger)^2 \theta^2$ precisely gives the free Wess-Zumino
Lagrangian seen in eqs.\,(\ref{eq:wz0}) and (\ref{eq:wzf}):
\begin{equation}
  \left[\, \Phi^* \Phi\, \right]_D \equiv \int d^{\,2} \theta d^{\,2}
  \theta^\dagger \,\Phi^* \Phi = \partial^\mu \phi^* \partial_\mu \phi
  + i \psi^\dagger \bar \sigma^\mu \partial_\mu \psi -F^* F +
  \partial_\mu (\dots);
  \label{eq:dterm}
\end{equation}
similarly, if I reconsider the concept of the Wess-Zumino
superpotential $W(\phi)$ in the context of superfields, \emph{i.e.},
\begin{equation}
  W (\Phi) \equiv \frac{1}{2} M^{ij} \Phi_i \Phi_j + \frac{1}{6} y^{ijk}
  \Phi_i \Phi_j \Phi_k,
  \label{eq:Wsusy}
\end{equation}
one finds that the ``F-terms'' $\sim \theta^2$ for $W(\Phi)$ and
$W(\Phi^*)$ together give 
\begin{align}
  \left[\, W(\Phi)\, \right]_F + \left[\, W(\Phi^*)\, \right]_F 
  &\equiv \int d^{\,2} \theta \,W(\Phi) + \int d^{\,2} \theta^\dagger 
  \,W(\Phi^*) \nonumber \\
  &= -\frac{1}{2} W^{ij} \psi_i \psi_j + W^i F_i + \mathrm{h.c.s},
\end{align}
as seen in eq.\,(\ref{eq:wzint0}), which give the Yukawa interactions
between $\psi$ and $\phi$, holomorphic fermion mass terms, and the
usual self-interaction terms for $\phi$. Therefore, the complete
interacting Wess-Zumino Lagrangian can be written as
\begin{equation}
  \mathcal{L}_{\mathrm{WZ}} = \left[\, \Phi^* \Phi\, \right]_D +
  \left[\, W(\Phi)\, \right]_F + \left[\, W(\Phi^*)\, \right]_F.
\end{equation}

To expand the model to include a gauge sector, first note that chiral
superfields transform under supergauge transformations as
\begin{equation}
  \Phi \rightarrow \mathrm{e}^{2ig\Omega^a t^a} \Phi, \qquad
  \Phi^* \rightarrow \Phi^* \mathrm{e}^{-2ig\Omega^{* a} t^a}.
\end{equation}
Additionally, eq.\,(\ref{eq:supergauge}) implies that
\begin{equation}
  \mathrm{e}^{2 g \mathcal{A}^a t^a} \rightarrow
  \mathrm{e}^{2ig\Omega^{* a} t^a} \mathrm{e}^{2 g \mathcal{A}^a t^a}
  \mathrm{e}^{-2ig\Omega^a t^a}.
\end{equation}
Therefore, the product $\Phi^* \mathrm{e}^{2 g \mathcal{A}^a t^a}
\Phi$ is a supergauge-invariant vector superfield. Furthermore, the
D-term of this expression gives the terms in eq.\,(\ref{eq:gint}) as
well as the gauge covariant version of eq.\,(\ref{eq:dterm})
\begin{align}
  \left[\, \Phi^* \mathrm{e}^{2 g \mathcal{A}^a t^a} \Phi\, \right]_D 
  =& \;D^\mu \phi^* D_\mu \phi + i \psi^\dagger \bar \sigma^\mu D_\mu
  \psi -F^* F \nonumber \\ &-g \sqrt{2} \left(\phi^* t^a \psi
  \lambda^a + \mathrm{h.c.}\right) + g\, \phi^* t^a \phi D^a.
  \label{eq:gaugedterm}
\end{align}

To complete the model, I need a superfield formulation for the gauge
kinetic terms. One can achieve this by defining the chiral field
strength superfield as
\begin{equation}
  2g\, t^a \mathcal{F}^a_\alpha \equiv -\frac{1}{4} D^\dagger D^\dagger
  \left( \mathrm{e}^{-2 g \mathcal{A}^a t^a} D_\alpha \mathrm{e}^{2 g
  \mathcal{A}^a t^a} \right);
\end{equation}
in the Wess-Zumino gauge, this superfield has the form
\begin{equation}
  \mathcal{F}^a_\alpha = i \lambda^a_\alpha -\frac{i}{2} (\sigma^\mu
  \bar\sigma^\nu \theta)_\alpha A^{a}_{\mu\nu}
  + \theta^2 (\sigma^\mu D_\mu \lambda^{\dagger a})_\alpha
  + \theta_\alpha D^a,
\end{equation}
and similar for $\mathcal{F}^{\dagger \dot\alpha}_a$. Now one can
see that the desired Lagrangian arises from the F-term of the square
of $\mathcal{F}$,
\begin{equation}
  \frac{1}{2} \left[\, \mathcal{F}^a_\alpha \mathcal{F}_a^\alpha\,
  \right]_F = -\frac{1}{4} A_{a}^{\mu\nu} A^{a}_{\mu\nu} - i
  \lambda^{\dagger a} \bar \sigma^\mu D_\mu \lambda^a + \frac{1}{2}
  D^a D^a + \frac{i}{8} A_{a}^{\mu\nu} \tilde A^{a}_{\mu\nu},
\end{equation}
where the final term, with $\tilde A^{a}_{\mu\nu} \equiv \epsilon_{\mu
\nu \rho \sigma} A_{a}^{\rho\sigma}$, which contributes to $CP$-violation
but is known experimentally to be highly suppressed, can be recast as
a total derivative.

Finally, I can write the full Lagrangian for a gauge superfield
theory:
\begin{align}
  \mathcal{L} \,=& \int d^{\,2} \theta d^{\,2} \theta^\dagger \,\Phi^*
  \mathrm{e}^{2 g \mathcal{A}^a t^a} \Phi + \int d^{\,2} \theta \left(
  W(\Phi) + \frac{1}{4} \mathcal{F}^a_\alpha \mathcal{F}_a^\alpha
  \right) \nonumber \\ &+ \int d^{\,2} \theta^\dagger \left( W(\Phi^*)
  + \frac{1}{4} \mathcal{F}^{\dagger \dot\alpha}_a
  \mathcal{F}_{\dot\alpha}^{\dagger a} \right), 
  \label{eq:sfL}
\end{align}
which describes a complete interacting theory for matter fermions
and scalar and gauge bosons, as one sees in the SM, as well as the
interactions of their superpartners.

\section{The Minimally Supersymmetric Standard Model}

In order to implement supersymmetry as part of the model of the
universe, the most straightforward approach one can take is to assume
that each field of the Standard Model has a superpartner with which it
forms a superfield multiplet. The result of this extension is the
\emph{Minimally Supersymmetric Standard Model} (MSSM). In the MSSM,
each matter fermion has a scalar superpartner called a ``sfermion''
(slepton, squark, stop, etc.), and each gauge boson has a fermionic
gaugino partner (Wino, Bino, gluino, etc.). In each case, the SM field
and its superpartner have the same quantum numbers, with the obvious
exception of spin.

The Higgs scalar field also has a fermionic ``Higgsino'' superpartner,
but some adjustments have to be made for its case, because (a) adding
a single fermion to the theory with non-zero weak isospin and
hypercharge would spoil gauge anomaly cancellation in the electroweak
sector, and (b), as I will show in detail shortly, the requirement
that the superpotential is analytic in $\Phi$ (or $\Phi^*$) forbids
the simultaneous use of $\Phi^*$ for up-type Yukawa terms and $\Phi$
for down-type terms, as would be analogous to the SM. As a result, the
MSSM must contain \emph{two} Higgs superfields, $H_u$ and $H_d$, to
give mass to matter superfields of both flavors. The fields are both
$SU(2)$ doublets, with weak hypercharges $Y_w = 1/2$ for $H_u$ and
$Y_w = -1/2$ for $H_d$. The explicit forms of the doublet superfields
are
\begin{equation}
  H_u = \left(\! \begin{array}{c} H_u^+\\ H_u^0 \end{array}
    \!\right), \qquad
  H_d = \left(\! \begin{array}{c} H_d^0\\ H_d^- \end{array}
    \!\right),
\label{eq:Hud}
\end{equation}
with analogous forms for the scalar bosons and Higgsino partners. As a
result of this structure, the Higgs particle spectrum is significantly
expanded when compared to the SM.

I will denote superfields for matter fermions as the capital letters
of their SM counterparts ($Q, U, D, L, E$), while I will denote the
superfields of gauge bosons with their usual letters but in the
calligraphic font (${\cal W}, {\cal B}, {\cal G}$). Superpartners for
all fields will be denoted with tildes over the SM names
($\tilde{q},\tilde{e},\widetilde{W}$, etc.). This notation will stand
for the remainder of the thesis. A summary of the particle content of
the MSSM is given in Table \ref{table:MSSMsfields}.

\subsection{The MSSM Lagrangian and SUSY Breaking}

\subsubsection{The MSSM Superpotential} The superpotential of the MSSM
is highly constrained by SM gauge invariance; starting from the
general form in eq.\,(\ref{eq:Wsusy}), out of all possible $\Phi_i
\Phi_j$ and $\Phi_i \Phi_j \Phi_k$ combinations of the fields given in
Table \ref{table:MSSMsfields}, only four terms survive. Its complete
form is
\begin{equation}
  W_{\rm MSSM} = \epsilon_{ab} \left(-y_u^{ij} U^{\cal C}_i Q_j^a
  H_u^b + y_d^{ij} D_i^{\cal C} Q_j^a H_d^b + y_e^{ij} E_i^{\cal C}
  L_j^a H_d^b - \mu H_u^a H_d^b \right),
\label{eq:Wmssm}
\end{equation}
where $i=1,2,3$ is the generation index, $a=1,2$ is the $SU(2)$ index,
and color indices, which are simply contracted on the two quark
fields, are not shown. The F-term of this superpotential will give
rise to the following interactions:
\begin{itemize}
  \item the SM-like mass-inducing Yukawa couplings of matter fermions
    $\{u,d,e\}$ to the Higgs scalars $h_{u,d}^0$, of coupling strength
    $y_f$ ($f=u,d,e$), analogous to those seen in
    eq.\,(\ref{eq:LHiggs});
  \item couplings of fermions (up-type to down-type) to the charged
    Higgs scalar fields $h_{u,d}^\pm$, again of strength $y_f$;
  \item cubic scalar couplings of two sfermions
    $\{\tilde{u},\tilde{d},\tilde{e}\}$ to a Higgs scalar of strength
    $\mu^* y_f$;
  \item quartic scalar couplings of two sfermions
    to two Higgs scalars (\emph{e.g.}, $\tilde{u} \tilde{u} h_u h_u$)
    of strength $y_f^2$;
  \item Higgsino-fermion-sfermion interactions (\emph{e.g.},
    $u \tilde{u} \tilde{h}_u$), also of strength $y_f$;
  \item quartic four-sfermion couplings of strength $y_f^2$.
  \item Higgs scalar mass terms for $h_{u,d}$ with mass $\mu^2$;
  \item Higgsino mass terms $\mu (\tilde{h}_u^+ \tilde{h}_d^- -
    \tilde{h}_u^0 \tilde{h}_d^0)$ + h.c.
\end{itemize}

\begin{table}[t]
\begin{center}
\begin{tabular}{||c|c|c|c|c|c||} \hline\hline
	Superfield & SM field & partner & $SU(3)$ & $SU(2)$ & $Y_w$\\
          \hline
        $Q_i$ & $q_i$ & $\tilde{q}_i$ & {\bf 3} & {\bf 2} & 1/6\\
	$U_i$ & $u_i^{\cal C}$ & $\tilde{u}_i^{\cal C}$ & {\bf 3} & 
        {\bf 1} & 2/3\\
	$D_i$ & $d_i^{\cal C}$ & $\tilde{d}^{\cal C}_i$ & {\bf 3} & 
        {\bf 1} & -1/3\\
        $L_i$ & $\ell_i$ & $\tilde{\ell}_i$ & {\bf 1} & {\bf 2} &-1/2\\
        $E_i$ & $e^{\cal C}_i$ & $\tilde{e}^{\cal C}_i$ & {\bf 1} & 
        {\bf 1} & -1\\ \hline
        ${\cal B}$ & $B_\mu$ & $\widetilde{B}$ & {\bf 1} & {\bf 1} 
          & 0\\
        ${\cal W}^a$ & $W_\mu^a$ & $\widetilde{W}^a$ & {\bf 1} & 
          {\bf 3} & 0\\
        ${\cal G}^{a'}$ & $G_\mu^{a'}$ & $\widetilde{G}^{a'}$ & 
          \bf{8} & {\bf 1} & 0\\ \hline
        $H_u$ & $\phi_u$ & $\tilde{\phi}_u$ & {\bf 1} & {\bf 2} 
          & 1/2\\
        $H_d$ & $\phi_d$ & $\tilde{\phi}_d$ & {\bf 1} & {\bf 2} 
          & -1/2\\ \hline\hline
\end{tabular}
\caption[Superfields of the MSSM, their components, and their
representations and charges under the gauge symmetries of the
model.]{Superfields of the MSSM, their components, and their
representations and charges under the gauge symmetries of the model.}
\label{table:MSSMsfields}
\end{center}
\end{table}

There are actually a few additional terms one could add to the
superpotential that are allowed by gauge invariance, but which do not
conserve either \emph{baryon number} $B$ or \emph{lepton number} $L$;
these global quantum numbers, which are automatically conserved in the
SM, are assigned as $B = \pm \frac{1}{3}$ for quarks and anti-quarks,
respectively, and $L = \pm 1$ for leptons and anti-leptons,
respectively (each is zero otherwise). These values, like other
quantum numbers, are present at the superfield level as well. If one
were to allow terms in the superpotential which violate baryon or
lepton number by one unit, \emph{i.e.}, $\Delta B = 1$ or $\Delta L =
1$, then the following terms arise:
\begin{align}
  W_{\Delta L = 1} &= \epsilon_{ab} \left(\lambda_1^{ijk} L_i^a L_j^b
  E^{\cal C}_k + \lambda_2^{ijk} L_i^a Q_j^b D^{\cal C}_k + \mu'_i
  L_i^a H_u^b \right) \\
  W_{\Delta B = 1} &= \lambda_3^{ijk} U^{\cal C}_i D^{\cal C}_j
  D^{\cal C}_k
  \label{eq:BLbreak}
\end{align}
We can be sure that these terms are somehow absent or extremely
suppressed, because if they were present, and the couplings were
${\cal O}(1)$, tree level proton decay would arise at ordinary
energies, which is wildly inconsistent with experiment, and even with
the existence of stable matter.

One way to ensure the absence of the $B$- and $L$-violating terms is
to enforce the discrete symmetry $R$-\emph{parity}, which is
defined as
\begin{equation*}
  R = (-1)^{3(B-L)+2s},
\end{equation*}
where $s$ is spin. One can determine that all SM matter fermions and
Higgs bosons have $R = 1$, while all SUSY particles have $R = -1$.
Enforcement of $R$-parity means every interaction vertex has $R = 1$
overall, which has several important implications: (a) any vertex will
contain an even number of SUSY fields, and SUSY particles will always
be produced in even numbers, (b) the product of any SUSY particle
decay will contain an odd number of new SUSY fields, and (c) the
lightest SUSY particle (LSP) is stable and will be present at the end
of any SUSY decay process. The stability of the LSP, if taken with
the cosmologically-motivated requirement that it be electrically and
color neutral, suggests that it is an excellent candidate for the
composition of non-baryonic dark matter.

While $R$-parity may seem ad-hoc despite empirical motivations for its
existence, it actually has theoretical motivation as well in
the context of grand unified theory and some SO(10) models, in
particular, due to its relationship to $B-L$ symmetry, which is
typically gauged at high energies in SO(10) and is central to the
seesaw mechanism for neutrino masses. I will discuss these
topics further in the next chapter.

\subsubsection{Soft SUSY Breaking in the MSSM} The soft SUSY breaking
terms of the MSSM are those of the forms in eq.\,(\ref{eq:Lsoft}) that
are consistent with gauge invariance and $R$-parity. They are
\begin{align}
  {\cal L}_{\rm soft} = -\frac{1}{2} \left( M_1 \widetilde{B}
  \widetilde{B} + M_2 \,\widetilde{W}^a \widetilde{W}^a + M_3
  \,\widetilde{G}^{a'} \widetilde{G}^{a'} + {\rm h.c.} 
  \right) \nonumber \\
  + \,\epsilon_{ab} \left(-a_u^{ij} \tilde{u}^{\cal C}_i
  \tilde{q}_j^a H_u^b + a_d^{ij} \tilde{d}_i^{\cal C} \tilde{q}_j^a
  H_d^b + a_e^{ij} \tilde{e}_i^{\cal C} \tilde{\ell}_j^a H_d^b + 
  {\rm h.c.} \right) \nonumber \\[2mm]
  -(m^2_{\tilde{q}})^{ij} \tilde{q}_i^\dagger \tilde{q}_j - 
  (m^2_{\tilde{\ell}})^{ij} \tilde{\ell}_i^{\dagger} \tilde{\ell}_j 
  - (m^2_{\tilde{u}})^{ij} \tilde{u}_i^{\cal C} \tilde{u}_j^{{\cal C} *} 
  - (m^2_{\tilde{d}})^{ij} \tilde{d}_i^{\cal C} \tilde{d}_j^{{\cal C} *} 
  - (m^2_{\tilde{e}})^{ij} \tilde{e}_i^{\cal C} \tilde{e}_j^{{\cal C} *} 
  \nonumber \\[2mm] - m^2_{h_u} h_u^\dagger h_u - m^2_{h_d} h_d^\dagger 
  h_d - b \epsilon_{ab} (h_u^{* a} h_d^b + {\rm h.c.});
  \label{eq:MSSMsoft}
\end{align}
the summation over $a,a'$ for the gauginos runs over the generators,
while the $\epsilon$ contraction in the $a$-terms and $b$ Higgs term
is over $SU(2)$ indices as it was in (\ref{eq:Wmssm}). The daggers on
the scalars in the mass squared terms indicate complex conjugate of
the scalar but transpose in $SU(2)$ space.
Note that unlike the Yukawa couplings $y_f$, the $a_f$ couplings have
mass dimension. Since all the fields here acquire masses after EWSB
from the couplings in $W_{\rm MSSM}$, one expects physical masses to
be generated by a mixing of all relevant terms.

The soft breaking terms introduce 105 new parameters to the theory,
including numerous mixing angles and phases in addition to the masses
themselves. This fact is quite disconcerting without further context;
however, several important experimental considerations lead to
substantial constraints on the full parameter space. For instance, the
absence of evidence for substantial $CP$ violation in the universe
requires that phases are small or zero. Both the $a_e$ and
$m_{\tilde{e}}^2$ terms contribute to \emph{lepton flavor violation}
(LFV), which is the breaking of global lepton flavor number symmetries
present in the SM; this phenomenon occurs in processes such as $\mu
\rightarrow e \gamma$ and must be at least highly suppressed to agree
with experimental limits \cite{lfv}. The presence of arbitrary mass
matrices $m_{\tilde{f}}^2$ would also disrupt the suppression of
\emph{flavor changing neutral currents} (FCNC), which are exactly zero
at tree level in the SM and suppressed even at loop level through
cancellation.  Experimental limits on processes such as $K^0
\rightarrow \bar K^0$, \emph{i.e.}, $d \bar s \rightarrow s \bar d$,
strongly constrain the squark mass differences \cite{KKbar}.

These considerations motivate an extreme simplification of the
soft breaking parameter space, built on the following assumptions:
\begin{equation}
  a_f^{ij} \simeq A_f y_f^{ij}; \quad (m^2_{\tilde{f}})^{ij} \simeq
  m^2_{\tilde{f}} \,\delta^{ij}; \quad {\rm Im}\{A_f,M_i\} \simeq 0;
  \label{eq:lowuniv}
\end{equation}
These simplifications are the SUSY-scale realization of a high-energy
prescription known as \emph{universality}, which I will discuss in
more detail below.

There are several feasible mechanisms for dynamically generating the
soft breaking terms; each involves a \emph{hidden sector}, which
couples very weakly or not at all to the ``visible'' sector of SM
superpartners, and a \emph{messenger sector}, which \emph{mediates}
the hidden sector physics, \emph{i.e.} ``relays'' it to the visible
sector, creating the soft terms seen in (\ref{eq:MSSMsoft}). Popular
mechanisms for SUSY breaking are \emph{gravity-mediated breaking}, in
which a hidden sector auxiliary vev $\langle F \rangle$ is
communicated to the MSSM fields through gravitational effects, and
\emph{gauge-mediated breaking}, in which a similar vev is coupled to
messenger fields charged under the SM gauge group, so that soft terms
arise through multi-loop order interactions between the messenger
fields and MSSM fields via the SM bosons. Since the gauge bosons are
blind to generation and, in some cases, flavor in general, the
conditions in (\ref{eq:lowuniv}) may be naturally present. Other
possible mediators include anomalies and extra-dimensions. There is
little agreement on which mediator is ``most'' appropriate or
promising, as every prescription faces a list of at least minor
phenomenological issues.

Gravity and gauge mediation can also be readily explored in
\emph{supergravity}, which arises automatically when one considers
local supersymmetry transformations, \emph{i.e.}, gauged
supersymmetry. The gauging of supersymmetry unifies global SUSY with
the spin-2 field theory of the graviton. In this theory, the fermionic
Goldstone mode associated with the broken SUSY generator is eaten by
the spin-3/2 graviton superpartner, the \emph{gravitino}. Depending on
the mediator, the gravitino may have cosmological or even TeV scale
consequences. Additionally, an appropriately ``minimal'' supergravity
model gives rise to flavor \emph{universality}, mentioned above, where
at the GUT scale $M_U$,
\begin{align}
  A_u = A_d = A_e \equiv A_0,& \quad m^2_{\tilde{f}} = m^2_{h_u} =
  m^2_{h_d} \equiv m^2_0 ~~\forall\, f, \nonumber \\ b = B_0 \mu,&
  \quad M_1 = M_2 = M_3 \equiv m_{1/2},
  \label{eq:univ}
\end{align}
where the parameters $A_0, B_0, m_0, m_{1/2}$ are all determined by
the theory in terms of $\langle F \rangle$ and $M_{\rm Pl}$. The
weaker conditions seen in (\ref{eq:lowuniv}) arise through the running
of the parameters down from $M_{\rm U}$ to the soft breaking scale
$M_{\rm SUSY}$. As I will discuss shortly, taking universality at the
GUT scale means that it coincides with unification of the standard
model gauge couplings $g_s,g,g'$ in the MSSM, which will be a key
factor in motivating the synthesis of SUSY with $SO(10)$ grand
unification. I will assume universality throughout the remainder of
this work.

\subsubsection{The Complete MSSM Lagrangian and EWSB} With $W_{\rm
MSSM}$ and ${\cal L}_{\rm soft}$ defined, I can write the complete
MSSM Lagrangian, in terms of superfields, as
\begin{align}
  \mathcal{L}_{\rm MSSM} = \int d^{\,2} \theta d^{\,2} \theta^\dagger
  \Big\{\, Q_i^* \exp \left(2 g_s {\cal G}^{a'} \lambda^{a'} + 2 g
  \mathcal{W}^a T^a + g' \mathcal{B}/3 \right) Q_i ~+ \nonumber \\
  U_i^{\mathcal{C} *} \exp \left(2 g_s {\cal G}^{a'} \lambda^{a'} + 4
  g' \mathcal{B}/3 \right) U_i^{\cal C} + D_i^{\mathcal{C} *} \exp
  \left(2 g_s {\cal G}^{a'} \lambda^{a'} - 2 g' \mathcal{B}/3 \right)
  D_i^{\cal C} ~+ \nonumber \\[1mm] L_i^* \exp \left(2 g \mathcal{W}^a
  T^a - g' \mathcal{B} \right) L_i + E_i^{\mathcal{C} *} \exp \left(-2
  g' \mathcal{B} \right) E_i^{\cal C} \nonumber \\[1mm] +\; H_u^* \exp
  \left(2 g \mathcal{W}^a T^a + g' \mathcal{B} \right) H_u + H_d^*
  \exp \left(2 g \mathcal{W}^a T^a - g' \mathcal{B} \right) H_d
  \,\Big\} \nonumber \\ +\; \int d^{\,2}
  \theta\, \Big( W_{\rm MSSM} + \frac{1}{4} \mathcal{G}^{a'}_\alpha
  \mathcal{G}^\alpha_{a'} + \frac{1}{4} \mathcal{W}^a_\alpha
  \mathcal{W}^\alpha_a + \frac{1}{4} \mathcal{B}_\alpha
  \mathcal{B}^\alpha \,\Big) + {\rm c.c.} + {\cal L}_{\rm soft}. 
  \label{eq:Lmssm}
\end{align}
The D-terms for the chiral superfields in this Lagrangian will give
rise to the following interactions:
\begin{itemize}
  \item the SM kinetic terms and gauge boson interactions for the
    fermions $\{u,d,e\}$ and Higgs bosons $\{h_u,h_d\}$;
  \item the kinetic terms and gauge boson interactions of the SM
    superpartners $\{\tilde{u},\tilde{d},\tilde{e},\tilde{h_u},
    \tilde{h}_d\}$, which include cubic sfermion-sfermion-boson terms
    (\emph{e.g.}, $\tilde{f} \tilde{f} W $) of coupling strength $g$,
    quartic terms involving two sfermions and two gauge bosons
    (\emph{e.g.}, $\tilde{f} \tilde{f} W W$) of strength $g^2$, and
    cubic higgsino-higgsino-boson terms (\emph{e.g.}, $\tilde{h}
    \tilde{h} W$) of strength $g$;
  \item cubic fermion-sfermion-gaugino (\emph{e.g.}, $f \tilde{f}\,
    \widetilde{W}$) terms of coupling strength $g$; 
  \item quartic four-sfermion and four-Higgs boson terms of strength
    $g^2$.
\end{itemize}
The F-terms of the gauge field strength terms in this Lagrangian will give
rise to the following interactions:
\begin{itemize}
  \item the SM kinetic terms and self-interaction terms for the
    gauge bosons $\{G^a,W^a,B\}$;
  \item the kinetic terms for the gaugino superpartners
    $\{\widetilde{G}^a,\widetilde{W}^a,\widetilde{B}\}$ and their
    cubic gaugino-gaugino-boson self interactions of strength $g$.
\end{itemize}

\vspace{2mm}
The neutral Higgs scalar potential for the model is
\begin{align}
  V_h =&\, (|\mu|^2 + m_{h_u}^2) \lvert h^0_u \rvert^2 + (|\mu|^2 +
  m_{h_d}^2) \lvert h^0_d \rvert^2 - (B_0 \mu\, h^0_u h^0_d ~+ ~{\rm
  c.c.}) \nonumber \\ &+\; \frac{1}{8}(g^2 + g'^2)(\lvert h^0_u
  \rvert^2 - \lvert h^0_d \rvert^2)^2,
\end{align}
where I've set $h_u^+ = h_d^- = 0$ at the minimum (without loss of
generality) to avoid disturbing electromagnetism. Both $h^0_u$ and
$h^0_d$ acquire vevs to break EW symmetry. The values of $B_0$,
$\langle h_u^0 \rangle$, and $\langle h_d^0 \rangle$ can all be chosen
and real and positive through field redefinition and $U(1)_Y$ gauge
transformation. I'll define $\langle h_u^0 \rangle \equiv v_u$ and
$\langle h_d^0 \rangle \equiv v_d$; the two vevs relate to the SM vev
as $v_u^2 + v_d^2 = v^2$, where $v = 174$\,GeV (or $246\,{\rm GeV}/
\sqrt{2}$, as an alternate convention). It's customary to define
\begin{equation*}
  \tan \beta = \frac{v_u}{v_d}, \quad v_u < v_d,
\end{equation*}
so that $v_u = v \sin \beta$ and $v_d = v \cos \beta$.

Of the eight real scalar degrees of freedom in the two complex Higgs
doublets, three become the Goldstone bosons, eaten by the massive
gauge bosons after EWSB, which leaves five physical Higgs scalars in
the model. There are two charged bosons $h^\pm$, two neutral,
$CP$-even bosons $h^0$ and $H^0$, and one neutral, $CP$-odd
pseudo-scalar $A$; the lighter of the neutral scalars corresponds to
the Higgs of the standard model. The tree-level masses of the neutral
bosons can be written as
\begin{equation}
  m^2_{H,h} = \frac{1}{2} \left\{m^2_A + M_Z^2 \pm \sqrt{(m^2_A +
  M_Z^2)^2 - 4 m^2_A M^2_Z \cos^2 2\beta}\, \right\},
\end{equation}
where
\begin{equation*}
  m_A^2 = 2 \lvert \mu \rvert^2 + m^2_{h_u} + m^2_{h_d}.
\end{equation*}
One might notice that the lighter SM scalar mass is less than $M_Z$,
at least at tree level and for $m_A > M_Z$. If one includes the
largest loop correction, coming from the top and stop couplings, one
can obtain $m_h$ of up to about 135\;GeV or, which puts the
observed Higgs mass near the upper end of the comfortably consistent
parameter space of the MSSM.

In a manner similar to the mixing of the gauge bosons seen in the SM,
there is additional mixing among like-charged superpartners in the
MSSM. In particular, the like-charged Winos $\widetilde{W}^\pm$ and
Higgsinos $\tilde{h}_{u,d}^\pm$ mix to give the physical
\emph{charginos} $\chi^\pm$, and the two neutral gauginos
$\widetilde{B}, \widetilde{W}^0$ and Higgsinos $\tilde{h}^0_{u,d}$ mix
to give the four \emph{neutralinos} $\chi^0_i$. Since $SU(3)_C$ is
unbroken in the model, the \emph{gluinos} $\tilde{g}$, which would be
massless in the absence SUSY breaking, degenerately share the
soft-breaking Majorana mass $M_3$.

The particle and anti-particle fermion superpartners will also
generally mix with one another. The two physical scalar partners are
typically denoted simply by $\tilde{f}_{1,2}$.

\subsection{Gauge Coupling Unification} \label{g-unify}

In addition to solving the hierarchy problem, one of the more curious
and inviting features of the MSSM is the rather precise unification of
the three SM gauge couplings at high energies. To understand the
meaning of this statement, recall that, as mentioned briefly in the
previous chapter, the physical parameters of a gauge field theory
actually change with the energy scale of interaction due to
renormalization effects. The evolution of a gauge coupling $g$ is
governed by the \emph{beta function} \cite{callan,*symanzik}
\begin{equation}
  M \frac{\partial g}{\partial M} = \beta(g) ,
  \label{eq:beta}
\end{equation}
where $M$ is the energy scale in question, referred to as simply the
renormalization scale. The derivative here is often seen written as
$\partial / \partial (\ln M)$ or $\partial / \partial t$, with $t
\equiv \ln M$, for simplicity. Taking the above expression as an
equation of evolution, one can see that the running with energy of $g$
is a function of $g$ itself; furthermore, $\beta(g)$ will be a smooth
function such that the evolution can be viewed as a continuous,
group-like transformation for $M \rightarrow M + \delta M$. As a
result, eq.\,(\ref{eq:beta}) is known as the \emph{renormalization
group equation} (RGE) for $g$. For a general gauge theory, the beta
function due to single-loop-level corrections is
\begin{equation}
  \beta(g) = \frac{b g^3}{16 \pi^2} \equiv \frac{g^3}{16 \pi^2}
  \left(-\frac{11}{3} C_2 (G) + \frac{4}{3} n_f C(r) \right),
  \label{eq:betaSM}
\end{equation}
where $n_f$ is the number of fermions charged under the group in the
theory, and $C_2(G)$ and $C(r)$ are group theory factors. For an
$SU(N)$ theory, $C_2(G) = N$, while $C_2(G) = 0$ for an abelian group;
In the SM, $C(r)$ is normalized to 1/2 for $SU(2)_L$ and $SU(3)_C$ and
to $3Y^2/5$ for $U(1)_Y$. This unusual normalization for $U(1)_Y$ is
chosen to match the redefinition of the gauge coupling $g'$ used in
$SU(5)$ and $SO(10)$ grand unification, which I will discuss in more
detail in the next chapter. For a semi-simple theory of multiple
gauge groups such as the SM, one can consider a separate, independent
RGE for each coupling in the theory:
\begin{equation}
  M \frac{\partial g_i}{\partial M} = \frac{b_i g_i^3}{16 \pi^2},
  \label{eq:rge}
\end{equation}
for multiple couplings $g_i$. Notice that, given the beta function for
an $SU(N)$ coupling, the beta function will be negative for
sufficiently small $n_f$, which implies that the strength of the
coupling diminishes with increasing energy.  As a result, the coupling
strength should vanish at some high energy.  This property, known as
\emph{asymptotic freedom}, is a feature of both non-Abelian symmetries
of the SM.

\begin{figure}[t]
\begin{center}
        \includegraphics{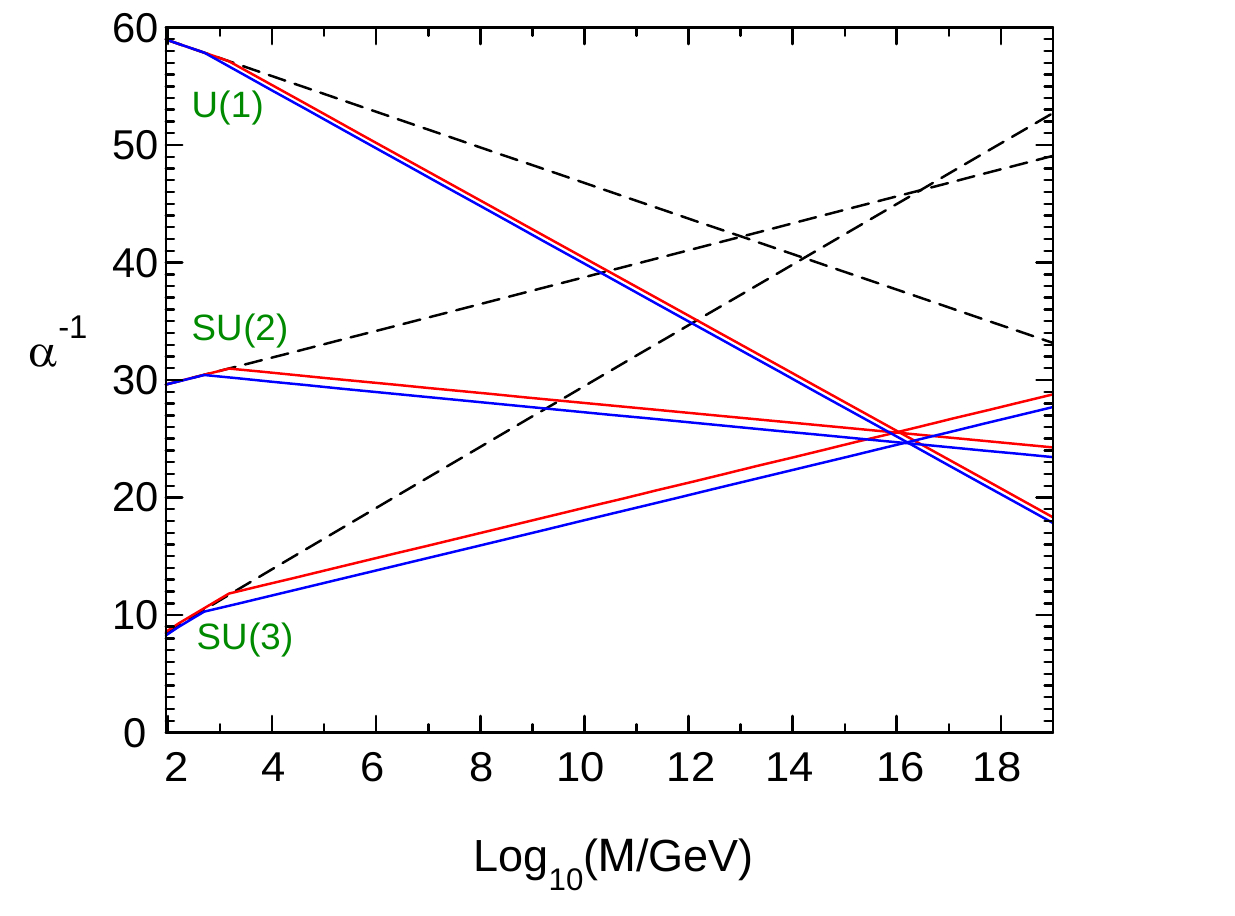} \caption[Renormalization
        group evolution of the inverse gauge couplings $\alpha_i^{-1}$
        for the SM and the MSSM.]{Renormalization group evolution of
        the inverse gauge couplings $\alpha_i^{-1}$ for the SM (dashed
        lines) and the MSSM (solid colored lines) \cite{martin}; for
        the MSSM case, the red vs.\,blue colored lines give bounds
        under variation of the superpartner masses.}
	\label{fig:running}
\end{center}
\end{figure}

For the standard model, careful counting of fields reveals that
\begin{equation}
  b_i = \left( \frac{41}{10}, -\frac{19}{6}, -7 \right), 
\end{equation}
where I've made the identifications $g_3 = g_s$, $g_2 = g$, and $g_1 =
\sqrt{\frac{5}{3}} g'$; Again, the change in normalization for $g'$ is
made for compatibility with $SU(5)$ grand unification. Conveniently,
if one writes the RGEs above in terms of the parameters $\alpha_i =
g_i^2/ 4\pi$, the resulting equations (still at one-loop order) are
linear in $\alpha_i^{-1}$:
\begin{equation}
  M \frac{~\partial \alpha_i^{-1}}{\partial M} = \frac{b_i}{2 \pi}.
  \label{eq:rgealpha}
\end{equation}
As a result, the running of the couplings will be straight lines on a
plot of coupling strength vs.\;$\log M$. That plot is given for the
three SM couplings in Figure \ref{fig:running}, shown as the black
dashed lines in the plot. Perhaps unexpectedly, the values of the
three couplings show signs of attempting to merge in the vicinity of
$10^{13}$\,GeV; this is a very tantalizing concept\dots could it be
that at very high energies, and hence in the very early universe, the
strong and electroweak forces were just different components of a
single interaction? This is of course similar to what we see in
electroweak unification; before EWSB, massless $W^a$ and $B$ bosons
would have mediated a single and perhaps long-range electroweak force,
resulting in a presumably unrecognizable universe. In the end, it
seems reasonable or even wise to assume that the merging of forces
continues as one moves back in time, and up in energy, toward the big
bang.

Yet, this vague trend in the SM is only the beginning of the story. In
the MSSM, due to the additional fields of varying species, the beta
function becomes
\begin{equation}
  \beta_{\rm S}(g) = \frac{g^3}{16 \pi^2} \left(-3 C_2 (G) +
  \sum\limits_\phi C(r(\phi)) \right),
  \label{eq:betaMSSM}
\end{equation}
where the sum over fields $\phi$ includes all the matter and Higgs
fields in the theory and their superpartners. The values of the
coefficients are
\begin{equation}
  b^{\rm S}_i = \left( \frac{33}{5}, 1, -3 \right).
\end{equation}
Note that the beta function for $SU(2)_L$ has changed signs. Looking
again at Figure \ref{fig:running}, the solid colored lines show the
running of $\alpha_i^{-1}$ in the MSSM; the red and blue lines for
each coupling give variation for a range of superpartner masses
0.5-1.5\,TeV. The merging of the coupling strengths has improved
dramatically, with a nearly exact agreement between the three coupling
values at an energy scale of $\sim 2 \times 10^{16}$\,GeV. This
behavior, known as \emph{gauge coupling unification}, seems almost too
good to be true, but does in fact arise for reasonable or even
preferred values for the parameters of the theory. Now perhaps one can
see why the prospect of combining theories of SUSY with those of grand
unification became so popular: this feature of the MSSM compels us to
explore the possibility that this merger is no accident. Adding
unification to the hierarchy problem solution and prospects for dark
matter, the lucrative nature of the MSSM is clear, and one might
understand why it created so much excitement for BSM physics, and why
its presence in BSM theories persists to this today, even despite an
increasingly long list of phenomenological difficulties.

Note though that I have still made no further mention of neutrino
masses, which, again, are strongly suggested by empirical data. Adding
neutrino masses to the MSSM is quite analogous to adding them in the
standard model, although the allowed soft breaking terms contribute
further to lepton flavor violation and the other phenomenological
complications discussed previously in the context of the charged
fermions. Even if one avoids those issues as before, it remains that
extending the MSSM to accommodate neutrino mass phenomenology is
starkly \emph{ad hoc}. In the context of grand unification, however,
this is not the case. A rather attractive mechanism for describing
neutrino masses goes hand-in-hand with $SO(10)$ grand unification,
which will be the topic of the next chapter.


%
%

\chapter{Grand Unification and Neutrino Mass}\label{4-so10}

Once the theory of electroweak unification and its spontaneous
breakdown via the Higgs mechanism were fully understood, grand
unification was perhaps an easy target for physicists looking to go
beyond the standard model. If the acquisition of a vev by a scalar
boson could break $SU(2)_L \times U(1)_Y$ down to $U(1)_{\rm em}$ and
a short-range weak force via massive vector bosons, then perhaps there
could be more such scalars, of even larger mass, governing additional
spontaneous breakdowns of higher dimensional groups to $SU(3)_C \times
SU(2)_L \times U(1)_Y$. Such a breakdown process would correspond to
the physical notion that the original symmetry of our universe was
quite a simple one (which can be taken literally in the context of
group theory), forced into a more elaborate configuration by the
nontrivial internal landscape of the quantum vacuum as spacetime
expanded and average energy density fell.

Yet, as previously mentioned, there are many reasons beyond aesthetics
to pursue unification. In addition to the highly suggestive nature of gauge
coupling unification discussed at the end of the previous chapter, GUT
models explain the seemingly arbitrary values for hypercharge in the
SM and consequently offer some basis for charge quantization; they
often restore parity symmetry in the gauge group; and they
may provide a framework more conducive to giving neutrinos mass.
Furthermore, specifically in the case of $SO(10)$, the right-handed
neutrino appears automatically, and neutrino masses arise quite
naturally, in connection to unification-scale breaking of $B-L$.

\section{Earlier Models of Unification}

\subsection{Pati-Salam and Left-Right Symmetry} \label{ps-lr} J.C.
Pati and A. Salam proposed the first model of partial unification in
1974 \cite{pati-salam}, based on the gauge group $SU(2)_L \times
SU(2)_R \times SU(4)_C$. The model treated lepton number as the
fourth color, and the resulting multiplets predictably contained new
fields with ``lepto-quark'' characteristics.

Left-right symmetric models restore the maximal breaking of parity
seen in the SM gauge group. These models were first developed by R.N.
Mohapatra, G. Senjanovic, and Pati \cite{rabi-pati-LR,*rabi-senj-LR},
also during 1974.\footnote{Right-handed currents had first been
proposed in the context of the SM by Mohapatra in 1972, as a possible
source of $CP$ violation \cite{rabi-RH}.} The simplest $L$-$R$ model is
based on the gauge group $SU(2)_L \times SU(2)_R \times U(1)_{B-L}$,
where the couplings are $g_{2L} = g_{2R}$ and $g'$. Such
models are really extensions of the SM model rather than unification
models, since no SM model multiplets are merged into larger
representations. With the addition of the $SU(2)_R$ gauge group and
the presence of $U(1)_{B-L}$, one can define electric charge as
\cite{rabi-marsh1,*rabi-marsh2}
\begin{equation*}
  Q = T^3_L + T^3_R + \frac{1}{2}(B-L);
\end{equation*}
this definition provides explanations for not only the seemingly
arbitrary values for hypercharge seen in the SM, but also for the
quantization of electric charge.

Since $SU(4) \supseteq SU(3) \times U(1)$, the left-right model can be
naturally embedded into Pati-Salam.

Left-right symmetry adds right-handed $W$ and $Z$ bosons to the SM and
collects the $SU(2)_L$-singlet fermions into doublets of their own:
\begin{equation}
  q_R \equiv \left(\!\begin{array}{c}u_R\\d_R\end{array}\!\right),
    \qquad \ell_R \equiv
    \left(\!\begin{array}{c}\nu_R\\e_R\end{array}\!\right);
\end{equation}
here, finally, one sees the addition of the right handed neutrino to
the model. Since right-handed neutrinos are not observed in our
low-energy world, the model will need some way to understand this. The
most popular solution utilizes the Majorana character of neutrinos as
follows. Consider the following scalar fields with $SU(2)_L \times
SU(2)_R \times U(1)_{B-L}$ representations \cite{rabi-seesaw}:
\begin{equation*}
  \Delta_L: (\boldsymbol{3}, \boldsymbol{1}, 2), \qquad 
  \Delta_R: (\boldsymbol{1}, \boldsymbol{3}, 2), \qquad
  \phi: (\boldsymbol{2}, \boldsymbol{2}, 0).
\end{equation*}
I can write interactions between these Higgs fields and the leptons
(for one generation) as
\begin{align}
  \mathcal{L}_{\rm Yuk} \ni h\, \bar\ell_L \phi\, \ell_R + \tilde h\,
  \bar\ell_L \tilde \phi\, \ell_R + if\! \left(\,\ell^T_L C^{-1}
  \sigma_2 \sigma_a \Delta^a_L \ell_L + \ell^T_R C^{-1}
  \sigma_2 \sigma_a \Delta^a_R \ell_R\, \right) + {\rm h.c.s}, \nonumber
  \\
  \label{eq:LRlepmaj}
\end{align}
where $\psi^T C^{-1} \psi$ is the Lorentz scalar for Majorana
fermions, and where $\tilde\phi = \sigma_2 \phi^* \sigma_2$. The
chiral Majorana interactions here violate lepton number conservation
by 2 units but conserve $B-L$. The $SU(2)$ structure of these terms
couples the neutrino to the neutral component of $\Delta$ for both the
left and right cases; hence, if either field acquires a vev, the
neutrinos will receive Majorana contributions to their masses. A vev
for $\phi$ will play the role of breaking EWSB and giving masses to
all of the fermions, including contributions to the neutrinos.
However, if $\langle \Delta_R \rangle \gg \langle \phi \rangle,
\langle \Delta_L \rangle$, then the right handed neutrinos will
acquire masses much heavier than the rest of the fields, which would
explain their absence in nature. The vev $\langle \Delta_R \rangle$
will also serve to break $SU(2)_R \times U(1)_{B-L} \longrightarrow
U(1)_Y$ if parity is broken in conjunction.

A closer look at the full neutrino mass matrix will reveal that the
left-handed neutrinos are $m_\nu \sim \langle \phi \rangle^2 / \langle
\Delta_R \rangle$, and are thus suppressed by the heavy scale.
Furthermore, if the vev $\langle \phi \rangle$ is inversely
hierarchical, then the solutions to the scalar potential give $\langle
\Delta_L \rangle \sim 0$, resulting in extremely small masses for the
left-handed neutrinos, also in agreement with observation. This
prescription, known as the \emph{seesaw mechanism}, has held as the
most phenomenologically viable explanation for neutrino mass for 35
years. It is also quite compatible with $SO(10)$ unification. I will
discuss the mechanism in more detail shortly.

\subsection[$SU(5)$ Grand Unified Theory]{$\boldsymbol{SU(5)}$ Grand
Unified Theory} \label{su5} Georgi and Glashow introduced the first
model of complete grand unification \cite{su5} in the same year as
Pati-Salam, based on the gauge group $SU(5)$. The SM gauge group has
rank $r=4$, where the rank of a Lie group is given by the dimension of
its \emph{maximal Cartan sub-algebra}, \emph{i.e.}, by the number of
diagonal generators in the algebra. A group can only be embedded in a
larger group if $r_{\rm small} \leq r_{\,\rm large}$, and $SU(5)$ is
the smallest simple Lie group of rank-4; therefore, it is the smallest
simple group in which the SM group can be embedded, and $SU(3) \times
SU(2) \times U(1)$ is a maximal subgroup.

The 15 matter fields per generation in the SM can be embedded into
$SU(5)$ using the conjugate fundamental representation
$\boldsymbol{\bar 5} \ni \!\{\ell, d_{\bar\rho}^{\cal C}\}$ and the
completely antisymmetric two-index representation {\bf 10} $\ni
\!\{q_\rho, u_{\bar\rho}^{\cal C}, e^{\cal C}\}$; $\rho = 1,2,3$ is
the color index.  Their explicit forms are
\begin{equation}
  \psi_a \equiv \left(\!\begin{array}{c} d_{\bar r}^{\cal C}\\[1mm]
    d_{\bar g}^{\cal C}\\[1mm] d_{\bar b}^{\cal C}\\ e\\ \nu
  \end{array}\!\right)_L; \qquad \chi_{ab} \equiv
  \left(\!\begin{array}{ccccc} 0& u_{\bar b}^{\cal C}& -u_{\bar
    g}^{\cal C}& u_r& d_r\\[1mm] & 0& u_{\bar r}^{\cal C}& u_g&
    d_g\\[1mm] & & 0& u_b& d_b\\ & & & 0& e^{\cal C}\\ & & & & 0
    \end{array}\!\right)_L.
\end{equation}

The model has 24 generators, and thus 24 gauge bosons, which decompose
under the SM group as
\begin{equation*}
  \{24\} = G(\boldsymbol{8},\boldsymbol{1},0) \oplus
  W(\boldsymbol{1},\boldsymbol{3},0) \oplus
  Y_w(\boldsymbol{1},\boldsymbol{1},0) \oplus
  X_\rho^{u,d}(\boldsymbol{3},\boldsymbol{2},-\frac{5}{6}) \oplus 
  \bar X_{\bar\rho}^{u,d}(\boldsymbol{\bar 3},\boldsymbol{2},
  \frac{5}{6}),
\end{equation*}
where the first three components correspond to the gluons, $W$ bosons,
and hypercharge boson, respectively. The remaining two components
carry both color and weak isospin; these fields are understood as 12
new individual $SU(5)$ bosons, which allow quark-lepton interaction at
a single vertex. The coupling $g_5$ to all bosons is universal, as
$g_5 = g_3 = g_2 = g_1 = \sqrt{\frac{5}{3}} g'$ at the unification
scale $M_{\rm U}$.

Note that in order to write the diagonal hypercharge generator such
that it preserves $SU(3)_C$, one will find that the diagonal entires
are fully determined by a single parameter plus the overall
normalization, and hence the action of this generator on the various
component fields fixes the values of $Y_w$ for all the SM fermions
precisely as needed.  Quantization of electric charge follows as an
implication.

The Higgs sector of $SU(5)$ has a minimum content of a 24-dimensional
adjoint field $\Phi$ and a 5-dimensional fundamental field $H_5$.
Breaking $SU(5) \longrightarrow G_{\rm SM}$ occurs via a vev $\langle
\Phi \rangle_{24}$, aligned with the diagonal ($\sim$ hypercharge)
generator $\lambda_{24}$. The breaking gives masses to the $X$ bosons
$M^2_X \sim g_5^2 V^2$, where $\langle \Phi \rangle = V \lambda_{24}$.

The {\bf 5} Higgs is essentially $(H_C^\rho \oplus \phi_{\rm SM})$,
\emph{i.e.}, a color triplet Higgs field and the SM Higgs doublet in a
single multiplet. EWSB occurs through the vev $\langle H_5 \rangle =
(0,0,0,0,v)^T$, which gives mass to the fermions through the couplings
\begin{equation}
  {\cal L}_{\rm Yuk} = h_{ij} \bar \psi_a^i\, \chi^j_{ab}
  H^\dagger_b + h'_{ij} \epsilon^{abcde} \chi^{Ti}_{ab}\, C^{-1}
  \chi^j_{cd} H_e + {\rm h.c.s}.
\end{equation}
The down-type and charged lepton masses are both given by the first
Yukawa term in the expression; as a result $m^i_e = m^i_d$ for
$i=1,2,3$. While these relationships are given at the unification
scale, only third generation Yukawa runnings are substantial enough to
correct the experimental inaccuracy of this relationship at low
energies (because $m_b \sim m_\tau$). In order to give realistic mass
eigenvalues to all the down-type fields, one can introduce a {\bf
45}-dimensional Higgs field $H^c_{ab}$.

Expansion of the $X$ gauge boson couplings to the matter
multiplets gives interactions with the individual fields of the form
\begin{align}
  {\cal L}_{X} = -\frac{g_5}{\sqrt{2}} X^{u \rho}_\mu
  \left(\epsilon_{\rho \sigma \tau} \bar u^{{\cal C}\sigma}_L
  \gamma^\mu u^\tau_L + \bar d_{L \rho} \gamma^\mu e^{\cal C}_L + \bar
  d_{R \rho} \gamma^\mu e^{\cal C}_R \right) \nonumber \\
  -\frac{g_5}{\sqrt{2}} X^{d \rho}_\mu \left(\epsilon_{\rho \sigma \tau}
  \bar u^{{\cal C}\sigma}_L \gamma^\mu d^\tau_L - \bar u_{L \rho}
  \gamma^\mu e^{\cal C}_L + \bar d_{R \rho} \gamma^\mu \nu^{\cal C}_R
  \right) + {\rm h.c.s}.
\end{align}
Note that some vertices include quark-lepton mixing. As a result,
through the exchange of an $\bar X^u$ boson, the process
\begin{equation*}
  uu \rightarrow de^+
\end{equation*}
is possible. Similarly,
\begin{equation*}
  ud \rightarrow ue^+
\end{equation*}
can occur through the exchange of a $\bar X^d$. Either process may
therefore lead to the decay of a nucleon. In particular, one sees
\begin{equation*}
  \tau (p \rightarrow \pi^0 e^+) \approx \frac{M^4_X}{g_5^4 m_p^5}
\end{equation*}
When $SU(5)$ theory was new, limits on proton lifetime were in the
vicinity of $10^{28\hbox{-}30}$\,GeV \cite{learned}, which implied
$M_X \gsim 10^{14\hbox{-}15}$\,GeV. Since then, lifetime limits have
risen by several orders of magnitude, and consequently the basic
$SU(5)$ model has been virtually ruled out as a viable theory of
nature (a few niches in the parameter space do technically remain).
One can make extensions to the model to salvage its validity, although
most require severe tuning of free parameters.

Other shortcomings of the model exist as well. Like the SM, the
$SU(5)$ model suffers from a ``gauge hierarchy problem'', in that
there is no basis for the extreme difference of the EW and unification
scales. Additionally, as in the SM and the MSSM, extension of the
model to include neutrino mass is completely \emph{ad hoc}. However,
the $SU(5)$ model can be embedded into the larger group $SO(10)$, in
which neutrino masses arise naturally. In fact, specifically in the
SUSY case, all of the above concerns see at least partial resolution.

Before discussing $SO(10)$ models, I will discuss the seesaw
mechanism for neutrino mass in more detail.

\section{The Seesaw Mechanism and Neutrino Masses} \label{seesaw}

Looking back at section \ref{ps-lr}, one can take the form of the
Higgs fields in the left-right model as \cite{rabi-seesaw}
\begin{equation*}
  \Delta_{L,R} \equiv \sigma^a \Delta_{L,R}^a =
  \left(\!\begin{array}{cc} \delta^+/\sqrt2 & \delta^{++}\\ \delta^0 &
    -\delta^+/\sqrt2
  \end{array}\!\right)_{L,R}, \qquad \phi \equiv
  \left(\!\begin{array}{cc} \phi^0_1 & \phi^+_1\\ \phi^-_2 & \phi^0_2
  \end{array}\!\right);
\end{equation*}
if the neutral components of the fields acquire vevs, I can write them
without loss of generality as
\begin{equation}
  \langle \Delta_{L,R} \rangle = \left(\!\begin{array}{cc} 0 & 0\\
    v_{L,R} & 0\end{array}\!\right),
    \qquad \langle \phi \rangle = {\rm e}^{i \alpha} 
    \left(\!\begin{array}{cc} \kappa & 0\\ 0 &
      \kappa'\end{array}\!\right).
\end{equation}
Now if I expand eq.\,(\ref{eq:LRlepmaj}) into components of the
$SU(2)_{L,R}$ multiplets, one finds the following neutrino mass terms:
\begin{align}
  \mathcal{L}_{\rm Yuk} \ni h_\nu \,\bar\nu_L \nu_R (\kappa + \kappa')
  \,{\rm e}^{i \alpha} + f v_L \,\nu^T_L C^{-1} \nu_L + f v_R \,\nu^T_R
  C^{-1} \nu_R + {\rm h.c.s}.
  \label{eq:numass}
\end{align}
Looking at the resulting neutrino mass matrix, in terms of its Weyl
components, one sees that, neglecting the phase $\alpha$,
\begin{equation}
  {\cal M}_\nu = \left(\!\begin{array}{cc} f v_L & \frac{1}{2}(h\kappa
    + \tilde h \kappa')\\ \frac{1}{2}(h\kappa + \tilde h \kappa') 
    & f v_R \end{array}\!\right);
    \label{eq:LR-mnu}
\end{equation}
The scalar potential for $\Delta_{L,R}$ and $\phi$ is quite extensive,
but under the assumption that $\kappa' \ll \kappa$ as well as $\kappa
\ll v_R$, one finds that
\begin{equation}
  v_L \simeq \frac{r \kappa^2}{2 v_R} \ll 1,
\end{equation}
where $r$ is a combination of parameters from the potential and
is generally small. Hence, the vev $v_L$ will be highly suppressed,
and one finds the following eigenvalues for ${\cal M}_\nu$:
\begin{equation}
  m_\nu \simeq f v_L - \frac{h^2 \kappa^2}{2 f v_R}, \qquad
  M_N \simeq 2 f v_R,
\end{equation}
where $N$ is the heavy $\sim$right-handed neutrino; the mass
eigenstates are generally linear combinations of $\nu_{L,R}$, but the
extremely hierarchical nature of the mass matrix leads to large
suppression of the mixing for the single-generation case.

This ``seesaw'' mechanism can be explored outside of the context of
left-right symmetry as well. In fact, one may consider simply adding
the right-handed neutrino to the SM under the assumptions that it must
be \emph{sterile}, \emph{i.e.}, a singlet under the full gauge group,
and that it is Majorana and heavy. Then the model is extended through
the inclusion of the terms
\begin{equation}
  \mathcal{L}_{SM} \ni y_\nu^{ij} \epsilon_{\alpha\beta} \,
  \bar\ell_{L i}^{\,\alpha} \phi^{* \beta}\, \nu_{R j} + \frac{1}{2}
  M^i_N \,\nu^T_{R i} C^{-1} \nu_{R i} + {\rm h.c.s};
  \label{eq:SMnuY}
\end{equation}
after EWSB, one finds a neutrino mass matrix similar in form to
(\ref{eq:LR-mnu}):
\begin{equation}
  {\cal M}_\nu = \left(\!\begin{array}{cc} 0 & y_\nu^{ij} v\\
    y_\nu^{ji} v & \delta_{ij} M_N^j \end{array}\!\right),
    \label{eq:type1-mnu}
\end{equation}
which will give left-handed eigenvalues of the form
\begin{equation}
  m_\nu \simeq -\frac{y_\nu^2 v^2}{M_N}.
\end{equation}
This form for neutrino mass, involving a Majorana term for the heavy
right-handed neutrinos only, is known as the \emph{type-I seesaw}.
Integrating out the heavy neutrinos leads to an effective dimension-5
operator of the form
\begin{equation}
  \mathcal{L}_{SM,{\rm eff}} \simeq \frac{y_\nu^2}{M_N}\,
  \bar\ell_\alpha \phi^\alpha \ell^\beta \phi^*_\beta,
  \label{eq:SMeffnu}
\end{equation}
first proposed by Weinberg in \cite{weinberg-nu5}. Note that to
obtain light neutrino masses of $m_\nu \ll 1$\,eV, the right-handed
mass scale will need to be $M_N \gsim 10^{14}$\,GeV, which is
surprisingly close to the scales of unification seen in
MSSM and $SU(5)$.

The alternative case for neutrino mass that includes the left-handed
Majorana term, as seen above in the left-right model case, and as will
be the case for $SO(10)$, is known as the \emph{type-II seesaw}. The
corresponding light neutrino masses for this case will generally be of
the form
\begin{equation}
  m_\nu \simeq f v_L - \frac{y_\nu^2 v^2}{f v_R},
\end{equation}
with $v_L \sim v^2 / v_R$. Note that generally the type-I term will be
present in the type-II case, although one may see dominance of either
term depending on the couplings and the scale of $v_R$. One can
implement type-II seesaw through extension of the SM as well, by for
instance adding a heavy triplet $\Delta_L$ with couplings of the form
$\ell^T \sigma_2 \Delta_L \ell$ and $\phi^T \sigma_2 \Delta_L \phi$,
which gives rise to an effective operator similar to that in
(\ref{eq:SMeffnu}). Other forms are plausible as well but typically
require more highly \emph{ad hoc} or tuned assumptions.

\section[$SO(10)$ Grand Unification]{$\boldsymbol{SO(10)}$ Grand
Unification}

\subsection[Representations of $SO(N)$ and $SO(2N)$]{Representations
of $\boldsymbol{SO(N)}$ and $\boldsymbol{SO(2N)}$} For the
N-dimensional fundamental representation of the group $SO(N)$, one can
define a basis in the conventional way,
\begin{equation*}
  \left(J^{ab} \right)_{mn} \equiv -i \delta^a_{\,[m}\delta^b_{\;n]} =
  -i \left( \delta^a_{\;m}\delta^b_{\;n} -
  \delta^a_{\;n}\delta^b_{\;m} \right) 
\end{equation*}
such that the Lie algebra bracket condition
\begin{equation}
  \left[\,J_{ab}, J_{cd}\, \right] = -i \left( \delta_{b[c} J_{ad]} +
  \delta_{a[d} J_{bc]} \right) 
  \label{eq:soNalg}
\end{equation}
is satisfied. These generators are of course analogous to the usual
angular momentum generators in $SO(3)$; thus, I can write the
orthogonal transformation (\emph{i.e.}, length-preserving rotation) of
an N-dimensional vector $V_m$ as
\begin{equation*}
  V_m \rightarrow O_{mn} V_n = \exp \left\{-\frac{i}{2} \theta_{ab}
  \left(J^{ab}\right)_{mn} \right\} V_n.
\end{equation*}
Tensor representations of larger dimensions can be constructed in the
usual way
\begin{equation*}
  T_{mn\dots} = V_m \otimes W_n \otimes \dots
\end{equation*}

In addition to fundamental and tensor representations, $SO(2N)$ will
have a spinor representation\footnote{One can of course construct a
spinor representation for $SO(N)$ with $N$ odd as well, though it
requires a bit more consideration.} in its universal covering group
$Spin(2N)$, and the Lie algebras of the two groups will be isomorphic.
In Euclidean analogy to the Dirac algebra of the Lorentz group, the
objects $\Gamma_m$, with $m=1,\cdots,2N$, are $2^N \times 2^N$
matrices that satisfy the Clifford algebra condition
\begin{equation}
  \left\{\Gamma_m, \Gamma_n \right\} = 2 \delta_{mn} \mathbb{I}_{2^N}
  \label{eq:cliff}
\end{equation}
and act on $2^N$-dimensional spinors $\psi$. If I define
\begin{equation}
  \Sigma_{mn} \equiv -\frac{i}{4} \left[\,\Gamma_m, \Gamma_n\, \right],
\end{equation}
one finds that the $\Sigma_{mn}$ satisfy the $SO(2N)$ algebra
(\ref{eq:soNalg}) and are therefore a valid representation of the
group. I can write the transformation of a spinor $\psi_\alpha$ as
\begin{equation*}
  \psi_\alpha \rightarrow U_{\alpha\beta} V_\beta = \exp
  \left\{-\frac{i}{2} \theta_{mn}
  \left(\Sigma_{mn}\right)_{\alpha\beta} \right\} V_\beta.
\end{equation*}

In analogy with $\gamma_5$ of the Dirac algebra, the object
\begin{equation*}
  \Gamma_0 \equiv i^{2N} \Gamma_1 \Gamma_2 \dots \Gamma_{2N}
\end{equation*}
allows for projection of the $2^N$-dimensional spinor into two
$2^{N - 1}$-dimensional chiral components by
\begin{equation}
  \psi_{L,R} = \frac{1}{2} \left( 1 \pm \Gamma_0 \right) \psi.
\end{equation}

Also of interest is the $Spin(2N)$ basis as an extension of an $SU(N)$
basis. If one takes the complex operators $\chi_a$, for
$a=1,2,\dots,N$ satisfying
\begin{equation*}
  \left\{\chi_a, \chi_b^\dagger \right\} = \delta_{ab},
\end{equation*}
then the operators $T_{ab} = \chi_a^\dagger \chi_b$ satisfy the
$\mathfrak{su}(N)$ Lie algebra, while the operators
\begin{align}
  \Gamma_{2a} \equiv \left(\chi_a + \chi_a^\dagger \right) \nonumber \\
  \Gamma_{2a-1} \equiv -i \left(\chi_a - \chi_a^\dagger \right)
  \label{eq:gammas}
\end{align}
are 2N objects satisfying the Clifford algebra in (\ref{eq:cliff}),
and therefore form a valid representation for $\Gamma_m$.

\subsection[The Basics of $SO(10)$ as an Interacting
Gauge Theory]{The Basics of $\boldsymbol{SO(10)}$ as an Interacting
Gauge Theory}

Following the prescription above, the rank-5 simple group $SO(10)$ has
a {\bf 16}-dimensional Weyl-spinor representation in its covering
group $Spin(10)$;\,\footnote{In keeping with convention, I will often
refer to this representation as the ``$SO(10)$ spinor'' rep.} the {\bf
16} decomposes in $SU(5) \times U(1)$ as ${\bf 10}\, \oplus\, {\bf
\bar 5}\, \oplus\, {\bf 1}$; given the matter field content of the
$SU(5)$ representations, this decomposition is highly suggestive.
Taking the $SU(5)$ representations as usual and the right-handed
neutrino as the singlet, one sees that all matter fermions and
anti-fermions of a single generation and chirality fit exactly into
one chiral $SO(10)$ spinor, denoted by $\psi_{L,R}$.  Since the
anti-particle fields of some chirality correspond to the particle
fields of opposite chirality, one finds all of the left- and
right-handed fields in a single chiral spinor. Therefore, in building
an $SO(10)$ model, I have no need for the full 32-dimensional spinor,
and I will simply denote the chiral spinor by $\psi$, which I assume
left-handed by convention.

The explicit arrangement of the field content in $\psi$ depends on the
choice of basis for the generators $\Sigma_{mn}$, and hence the choice
of basis for $\Gamma_m$ ($m=1,2,\dots,10$), for which there are many.
The end result is quite tedious not of much use other than for explicit
calculation. The kinetic term for $\psi$, however, can nonetheless be
written in a familiar form:
\begin{equation}
  {\cal L}_{\rm U,kin} = \bar \psi i \slashed{D} \psi = \bar \psi
  \gamma_\mu \left(i \partial_\mu + \frac{g_{\rm U}}{2} \Sigma_{mn}
  W_\mu^{mn} \right) \psi; 
\end{equation}
the matrix $\left(\Sigma_{mn} W_\mu^{mn} \right)_{ab}$ is generally
$32 \times 32$ in spin space but will be block diagonal and redundant
for reps based on the {\bf 16} spinor. $W_\mu^{mn}$ are the 45 gauge
bosons of the model (\emph{i.e.}, ${10 \choose 2}$), which decompose
under the SM gauge group as
\begin{align*}
  \{45\} = ~&G(\boldsymbol{8},\boldsymbol{1},0) \oplus
  W_L(\boldsymbol{1},\boldsymbol{3},0) \oplus
  X_{B-L}(\boldsymbol{1},\boldsymbol{1},0) \\[1mm] \oplus
  ~&X^{u,d}_\rho(\boldsymbol{3},\boldsymbol{2},-\frac{5}{6}) \oplus 
  \bar X^{u,d}_{\bar\rho}(\boldsymbol{\bar 3},\boldsymbol{2},\frac{5}{6}) 
  \oplus Y^{u,d}_\rho(\boldsymbol{3},\boldsymbol{2},\frac{1}{6})\oplus
  \bar Y^{u,d}_{\bar\rho}(\boldsymbol{\bar 3},\boldsymbol{2},-\frac{1}{6}) 
  \\ \oplus ~&A_\rho(\boldsymbol{3},\boldsymbol{1},\frac{1}{3}) \oplus 
  \bar A_{\bar\rho}(\boldsymbol{\bar 3},\boldsymbol{1},-\frac{1}{3}) 
  \\ \oplus ~&W^+_R(\boldsymbol{1},\boldsymbol{1},\frac{1}{2}) \oplus
  W^-_R(\boldsymbol{1},\boldsymbol{1},-\frac{1}{2}) \oplus
  W^3_R(\boldsymbol{1},\boldsymbol{1},0); 
\end{align*}
when compared to $SU(5)$, one might notice that (a) the diagonal
hypercharge generator has been swapped for the $B-L$ generator and
that of the neutral right-handed $W_R^3$, thereby increasing the rank
of the group by one, as expected, and (b) another set of bosons $Y$
with both color and $\boldsymbol{T}_L$ weak isospin are present, in
addition to the $X$ bosons of $SU(5)$. In fact, both the $X$ and $Y$
bosons have $\boldsymbol{T}_R$ isospin as well here, and pair off
cross-wise under $SU(2)_R$, as $\left(Y^u, X^u \right)_\rho,
\,\left(\bar Y^d, \bar X^d \right)_{\bar\rho}$, etc. For a complete
analysis of the bosons, their corresponding generators, and their
decompositions in several bases and for several subgroups, see,
\emph{e.g.}, \cite{oezer}.

\subsection[Fermion Masses and Higgs Representations in
$SO(10)$]{Fermion Masses and Higgs Representations in
$\boldsymbol{SO(10)}$}

Because particles and anti-particles in $SO(10)$ are together in the
same chiral spinor, generating mass terms requires additional
complexity when compared to the familiar low-energy theory. In
particular, one sees non-trivial algebraic structure in the Yukawa
couplings.

The tensor product of two chiral spinors decomposes in the group as
$\boldsymbol{16} \otimes {\bf 16} = {\bf 10} \oplus {\bf 120} \oplus
\boldsymbol{\overline{126}}$; the {\bf 10} and {\bf 120} are the
fundamental rep and the 3-index totally anti-symmetric rep,
respectively, and the 5-index, totally anti-symmetric rep {\bf 252}
decomposes into ${\bf 126} \oplus \boldsymbol{\overline{126}}$.
Therefore one expects the Yukawa couplings of Higgs fields to matter
in the model to appear in one of the three above representations.

In the simplest case, an $SO(10)$ model has only a {\bf
10}-dimensional Higgs field ${\rm H}_m$; its coupling to $\psi\psi$
has the explicit form
\begin{equation}
  {\cal L}_{\rm U,Yuk} \ni h_{ij} \psi^T_i B C^{-1} \Gamma_m \psi_j
  {\rm H}_m,
  \label{eq:10yuk}
\end{equation}
where the Yukawa coupling $h_{ij}$ is symmetric in the generation
space. The matrix $B$ appearing here plays a role analogous to that
of $C$ but in the $Spin(10)$ space: under the spin group, the spinor
$\psi$ and its conjugate transform as
\begin{equation*}
  \delta \psi = i \omega_{mn} \Sigma_{mn} \psi \qquad
  \delta \psi^\dagger = -i \omega_{mn} \psi^\dagger \Sigma_{mn},
\end{equation*}
where I've used that the generators $\Sigma_{mn}$ are Hermitian;
however,
\begin{equation*}
  \delta \psi^T = i \omega_{mn} \psi^T \Sigma_{mn}
\end{equation*}
does not transform like a conjugate field. Therefore, one defines the
matrix $B$ such that
\begin{equation*}
  \delta \left(\psi^T B \right) = -i \omega_{mn} \left(\psi^T B 
  \right) \Sigma_{mn}.
\end{equation*}
Explicitly, $B$ can be given as $B \equiv \Gamma_1 \Gamma_3 \Gamma_5
\Gamma_7 \Gamma_9$, which further implies that
\begin{equation*}
  B^{-1} \Gamma_m B = - \Gamma_m.
\end{equation*}

As in $SU(5)$ and the SM, I want a vev for H to break $SU(2)_L$ in
order to give the fermions mass. Looking at eq.\,(\ref{eq:gammas}),
note that for the fields $\chi^a$, the components $a=1,2,3$ relate to
color, while $a=4,5$ relate to left isospin. I will take the vev to
correspond to $a=5$, which implies $\langle\, {\rm H}_9,{\rm H}_{10}
\rangle \neq 0$. If I take $\langle\, {\rm H}_9 \rangle = v_1$ and
$\langle\, {\rm H}_{10} \rangle = v_2$, then one finds the following
terms for fermion masses (considering a single generation for now):
\begin{equation*}
  {\cal L}_{ {\rm Yuk},\slashed{\rm H}} = h(v_2 -
  v_1)\left(\bar d_L d_R + \bar e_L e_R \right) + h(v_2 +
  v_1)\left(\bar u_L u_R + \bar\nu_L \nu_R \right) + {\rm
  h.c.s};
\end{equation*}
this result implies $m_e = m_d$ and $m_u = m_\nu$. Although this is a
GUT-scale result, it cannot be made to agree with low-energy
observations, even when running effects are taken into account. This
is even more strongly the case for second generation; hence, to build
a realistic model, one needs additional Higgs Yukawas.
 
The next available option for Higgs field is the {\bf 120}-dimensional
field $\Sigma_{mno}$, which couples to the fermions by
\begin{equation}
  {\cal L}_{\rm U,Yuk} \ni g_{ij} \psi^T_i B C^{-1} \Gamma_m \Gamma_n
  \Gamma_o \psi_j \Sigma_{mno};
\end{equation}
the Yukawa coupling matrix $g_{ij}$ is anti-symmetric in order to
preserve $SO(10)$ invariance; therefore, this Yukawa can only
contribute to mass mixing among generations.

There are several potential vevs that do not disturb color invariance.
If I choose $\langle \Sigma_{789},\Sigma_{780} \rangle \neq 0$ (I will
use ``0'' instead of ``10'' for multi-index fields to avoid
confusion), then the resulting mass relationships are
\begin{equation*}
  m^i_d = 3 m^{ij}_e, \qquad m^i_u = 3 m^{ij}_\nu;
\end{equation*}
\emph{i.e.}, the contribution to the $(ij)$-element of electron mass
matrix is proportional to the $i^{th}$ down mass, and similar for the
up-type particles. Clearly this Higgs field would need to be used in
conjunction with others to achieve a realistic mass spectrum.

The final choice for a Higgs is the $\boldsymbol{\overline{126}}$
field $\bar\Delta_{mnopq}$; its coupling to the fermions
is
\begin{equation}
  {\cal L}_{\rm U,Yuk} \ni f_{ij} \psi^T_i B C^{-1} \Gamma_m \Gamma_n
  \Gamma_o \Gamma_p \Gamma_q \psi_j \bar{\Delta}_{mnopq},
  \label{eq:126yuk}
\end{equation}
where $f_{ij}$ is symmetric. The following vevs preserve $SU(3)_C$:
\begin{equation*}
  \langle \bar\Delta_{1278m} \rangle = \langle
  \bar\Delta_{3478m} \rangle = \langle \bar\Delta_{5678m}
  \rangle \neq 0, \quad m = 9\;{\rm or}\;10,
\end{equation*}
which give the mass relations
\begin{equation*}
  m^{ij}_e = -3 m^{ij}_d, \qquad m^{ij}_\nu = -3 m^{ij}_u;
\end{equation*}
this result nicely predicts the observed $\frac{m_e}{m_\mu}:
\frac{m_d}{m_s}$ ratio, but does not agree with third generation
observations. A realistic mass spectrum can though be obtained through
a combination of H and $\bar\Delta$.

The $\boldsymbol{\overline{126}}$ Higgs may play another important
role in the fermion mass spectrum. Under decomposition to left-right
models, the field contains a right-handed triplet part. A vev for this
component breaks $B-L$, and it couples to $\nu_R \nu_R$ as in
eq.\,(\ref{eq:LRlepmaj}); furthermore, the field corresponds to
the $SU(5)$ singlet, so it does not disturb $SU(3)_C \times SU(2)_L$.
Hence, if this triplet acquires a vev around the GUT scale, it will
simultaneously explain the suppression of right-handed currents and
activate the type-I seesaw for neutrino mass.

\subsection[Spontaneous Symmetry Breaking in $SO(10)$]{Spontaneous
Symmetry Breaking in $\boldsymbol{SO(10)}$} $SO(10)$ has two maximal
subgroups of relevance to symmetry breaking:
\begin{equation*}
  SO(10) \supseteq SU(5) \times U(1), ~SO(6) \times SO(4);
\end{equation*}
The context of the former should be clear, as it has been mentioned
previously. To understand the significance of the latter
decomposition, note that
\begin{equation*}
  Spin(6) \cong SU(4), \qquad Spin(4) \cong SU(2)_- \times SU(2)_+;
\end{equation*}
hence, $Spin(6) \times Spin(4) \cong$ Pati-Salam (PS); more
specifically, the breakdown of $SO(10)$ to PS is
\begin{equation*}
  Spin(10) \longrightarrow SU(2)_L \times SU(2)_R \times SU(4)_C
  \times \mathbb{Z}_2;
\end{equation*}
in full $SO(10)$ representations, the $\mathbb{Z}_2$ symmetry is
manifested as \emph{D-parity} \cite{dparity}; the explicit form of a
$D$-parity transformation is 
\begin{align*}
  D(V_m) \equiv&\; \exp(-i \pi J_{23}) \exp(i \pi J_{67}) \\
  D(\psi) \equiv&\; \exp(-i \pi \Sigma_{23}) \exp(i \pi \Sigma_{67}) 
  = - \Gamma_2 \Gamma_3 \Gamma_6 \Gamma_7,
\end{align*}
which corresponds to a pair of $\pi$-rotations in the (23) and (67)
planes of the 10-dimensional vector space of the fundamental. Since
the matter field $\psi$ contains only fields of a single chirality,
there can be no well-defined notion of parity in $SO(10)$; $D$-parity
then plays a role to create to the possibility for the presence of $C$
and $P$ at lower energies. 

As I mentioned earlier, the matter spinor decomposes under $SU(5)
\times U(1)$ as ${\bf 16} = {\bf 10} \oplus {\bf \bar 5} \oplus {\bf
1}$; under Pati-Salam, the decomposition makes ``left-right''
splitting manifest: ${\bf 16} = ({\bf 2},{\bf 1},{\bf 4}) \oplus ({\bf
1},{\bf 2},\boldsymbol{\bar 4})$, but let me reiterate that
right-handed fields are still explicitly absent; for example, the
doublet one might be inclined to call ``$q_R$'' is actually $q_L^{\cal
C}$. In breaking $SO(10)$ to Pati-Salam, the $\mathbb{Z}_2$ coming
from conservation of $D$-parity corresponds to ${\bf 2}_L
\leftrightarrow {\bf 2}_R$ under charge conjugation symmetry. Hence
one finds Pati-Salam with ``left-right'' symmetry, in the sense that
$g_{2L} = g_{2R}$, but nonetheless defined with left-handed
antiparticle fields rather than right-handed particle fields.

For either class of breaking possibilities, one must of course
consider only vevs which leave $SU(3)_C \times U(1)_{\rm em}$
unbroken; furthermore, since one expects to find that group as a
consequence of breaking the usual SM gauge group, further restriction
to vevs which leave $SU(2)_L$ in tact is also needed. Note that in
general the Higgs fields with components that acquire vevs will not
be those that couple to matter; \emph{i.e.}, additional
representations of Higgs may be present in the scalar potential of
the $SO(10)$ model, coupled only to other Higgs fields.

\paragraph{$\boldsymbol{SO(10) \rightarrow SU(5)}$.} To induce the
breaking of $SO(10)$ to $SU(5)$, one simply gives a vev to the
$SU(5)$-singlet component of some appropriate Higgs, which usually
also breaks $B-L$.  Two such choices are the {\bf 1} of a ${\bf 16}_H$
or {\bf 126}. The 2-index, totally anti-symmetric {\bf 45} rep of
$SO(10)$ contains the {\bf 24} of $SU(5)$, so if one includes that
field, the breaking of $SU(5) \rightarrow$ SM proceeds as discussed in
section \ref{su5}.

Assuming $SO(10)$ breaks at the GUT scale, $M_{\rm U} \sim 2 \times
10^{16}$\,GeV and $SU(5)$ breaks at its canonical scale of $M_X \sim
10^{14\hbox{-}15}$, this model would be ruled out by proton decay
constraints; hence any applications of these breaking patterns would
need to be at higher scales in more elaborate models.

\paragraph{$\boldsymbol{SO(10) \rightarrow}$ PS \& Left-Right.}
Breaking $SO(10)$ to the Pati-Salam gauge group is a considerably more
fruitful choice, with not only many choices for path of breaking, but
also the possibility for robust intermediate scale physics, because
left-right symmetric models are phenomenologically eligible for
breaking at scales as low as 1\,TeV, although doing so sacrifices the
possibility for implementing the seesaw mechanism specifically as
described in section \ref{seesaw}.

Some of the most common vev choices for breaking to PS include the
$(\boldsymbol{1},\boldsymbol{1},\boldsymbol{1})$ component of the
2-index, traceless symmetric {\bf 54} rep and the $(\boldsymbol{1},
\boldsymbol{1},\boldsymbol{1})$ or $(\boldsymbol{1},\boldsymbol{1},
\boldsymbol{15})$ component of the 4-index, anti-symmetric {\bf 210}
rep. The {\bf 54} option preserves $D$-parity, while the {\bf 210}
choices do not. In the {\bf 54} case, one can further break to
$SU(3)_C \times SU(2)_L \times SU(2)_R \times U(1)_{B-L}$ through the
$(\boldsymbol{1}, \boldsymbol{1},\boldsymbol{15})$ component of {\bf
45}, which also breaks $D$-parity.

In all of the cases described above, breaking to the SM requires
$SU(2)_R \times U(1)_{B-L} \rightarrow U(1)_Y$; in the PS cases, one
must also break $SU(4)_C$ as well, but since $SU(4)_C \supseteq SU(3)_C
\times U(1)_{B-L}$, the breaking of $B-L$ will accomplish both
tasks.\footnote{One can instead break only $SU(2)_R \rightarrow
U(1)_R$ if looking to leave $SU(4)_C$ (and hence $B-L$) in tact.} The
most common approaches involve vevs for either the $(\boldsymbol{1},
\boldsymbol{3}, \boldsymbol{10})$ component of
$\boldsymbol{\overline{126}}$, denoted $\bar\Delta_R$, or the
singlet of $\boldsymbol{16}_H$.  The $\boldsymbol{\overline{126}}$
case has clear advantages over that of $\boldsymbol{16}_H$: 
\begin{itemize}
  \item One can see from the PS representation of $\bar\Delta_R$
    that it is a right-handed triplet, which is precisely the object
    present in the right-handed Majorana neutrino mass term in
    eq.\,(\ref{eq:LRlepmaj}). Hence the vev $\langle \,
    \bar\Delta_R \rangle \equiv v_{B-L} = v_R$, and
    implementation of the seesaw mechanism comes for free from the
    $B-L$ breaking; this attractive scenario of a single mechanism
    performing two crucial duties in the model is quite economical to
    say the least.  Furthermore, the $\boldsymbol{\overline{126}}$
    coupling $f_{ij}$ will be highly constrained by the mass spectrum
    of the charged fermions, and yet will be present in the Majorana
    neutrino terms also; so the economy of the model extends to its
    number of parameters as well.

    In contrast, one must include higher dimensional operators or
    singlet fields to obtain the $\nu^{\cal C}$ mass term in the case
    with $\boldsymbol{16}_H$.

  \item The $\bar\Delta_R$ breaks of $B-L$ by two units in the
    emergence of the $\nu^{\cal C} \nu^{\cal C}$ mass term. Note that
    for a supersymmetric model, this leaves $R$-parity, $R =
    (-1)^{3(B-L)+2s}$, conserved. This is of course attractive if one
    would like to suppress $R$-parity violating terms and retain the
    potential for an LSP dark matter candidate.
    
    The $\boldsymbol{16}_H$ field, however, corresponds to the
    $\nu^{\cal C}$ component and therefore breaks $B-L$ by a single
    unit, which is $R$-parity odd. As a result, one finds $R$-parity
    violating terms among the higher dimensional operators involving
    $\boldsymbol{16}_H$.
\end{itemize}
\begin{center}
  ***
\end{center}
The procedure for constructing a properly broken subgroup
at some scale requires several steps when considering larger groups
such as $SO(10)$, especially in the rank-reducing cases. First, one
must rescale all the generators for the ``before'' and ``after''
groups such that they share a common normalization. Next, for a
breaking of the form $G_1 \times G_2 \longrightarrow G_0$ at energy
scale $M$, where generators $T_1$ and $T_2$ will merge in the breaking
as
\begin{equation*}
  T_0 = a_1 T_1 + a_2 T_2,
\end{equation*}
then the corresponding gauge couplings $g_1,g_2,g_0$ must satisfy the
following boundary condition:
\begin{equation}
  \frac{1}{\alpha_0(M)} = \frac{a_1^2}{\alpha_1(M)} +
  \frac{a_2^2}{\alpha_2(M)},
\end{equation}
where $\alpha_i = g^2_i /4\pi$ is the fine structure constant for the
group $G_i$. Finally, one must consider the running of each coupling
between the various scales. In particular, the evolution of $\alpha_i$
between two mass scales $M_2 > M_1$ follows from the RGE for the
coupling:
\begin{equation}
  \frac{1}{\alpha_i(M_1)} = \frac{1}{\alpha_i(M_2)} - \frac{b_i}{2\pi}
  \ln \left(\frac{M_2}{M_1}\right),
  \label{eq:alphaGUTrun}
\end{equation}
where $b_i$ are model and group-specific beta function coefficients
discussed in section \ref{g-unify}. Note that in cases involving
multi-step breaking patterns and multiple couplings, these
relationships will be used iteratively. In this manner, one can
develop the precise relationships between low-scale measured
parameters and (heavy:light) mass scale ratios, which can be used to
experimentally test GUT models, set lower limits on heavy scales, etc.
One pertinent example is the ability to constrain GUTs using the
experimental limits on $\sin^2\theta_W = \alpha_{\rm em} /
\alpha_{2L}$ combined with the higher order corrections to its value
coming from the relationship in (\ref{eq:alphaGUTrun}).

\subsection[Supersymmetry and $SO(10)$]{Supersymmetry and
$\boldsymbol{SO(10)}$} Since some of the unresolved issues of the SM
are obviated by SUSY, and some others are successfully attended to by
$SO(10)$ unification, it would seem quite wise to consider the merging
of the two frameworks into a SUSY $SO(10)$ model of the universe. Most
clearly of importance is that non-SUSY GUT models face the problems
with quadratic divergences in loop corrections to Higgs masses. In
addition to the benefits coming from one framework or the other, a few
added benefits arise from the combination, including possible
restrictions of soft $CP$ phases in SUSY, similar constraint of the
strong $CP$ phase, and, as I mentioned in the previous section, the
possibility of automatic $R$-parity conservation.

The promotion of $SO(10)$ to a supersymmetric model follows quite
straightforwardly from the process for constructing the MSSM; in
particular, the SM fermion content is unchanged (other than the
addition of the right-handed neutrino, of course), and all of the same
formalism applies for new scalar and gauge boson superpartners,
auxiliary fields, etc.

One caveat does arise with respect to vevs for the various Higgs
fields: for any field with a vev that reduces the rank of the group,
one must include the barred partner for the field, so that their
D-terms in the scalar potential cancel with each other; this keeps
SUSY unbroken above the desired scale, which is thought to be ${\cal
O}({\rm TeV})$. In particular, the breaking $SU(2)_R \times U(1)_{B-L}
\rightarrow U(1)_Y$ will require $\boldsymbol{\overline{126}} + {\bf
126}$ or $\boldsymbol{16}_H + \boldsymbol{\overline{16}}_H$.

As an example, consider the well-known ``minimal'' SUSY $SO(10)$
model, which includes {\bf 10} and $\boldsymbol{\overline{126}}$ Higgs
fields coupling to matter plus a {\bf 210} field to initiate the GUT
scale breaking. Yukawa terms in the superpotential for can be written
by simply promoting the fermionic matter spinors and Higgs scalars in
eqs.\,(\ref{eq:10yuk}) and (\ref{eq:126yuk}) to superfields;
the remaining terms will be all quadratic or cubic superfield products
allowed by the $SO(10)$ invariance, of the form in eq.\,(\ref{eq:Wsusy}).  
The resulting superpotential for the this model, up to ${\cal O}(1)$
numerical factors, is
\begin{gather}
  W_{\rm U} = M_{210} \hat\Phi^2 + \lambda
  \hat\Phi_{lmno} \hat\Phi_{nopq} \hat\Phi_{pqlm} + M_{10}
  \hat{\bf H}^2 + M_{126} \hat\Delta \hat{\bar\Delta} \nonumber \\ +
  \eta\, \hat\Phi_{lmno} \hat\Delta_{lmpqr} \hat{\bar\Delta}_{nopqr} +
  \,\hat{\bf H}_l\, \hat\Phi_{mnop} \left( \gamma \hat\Delta_{lmnop}
  +\, \bar\gamma \hat{\bar\Delta}_{lmnop} \right) \nonumber \\ + h_{ij}
  \Psi_i B \Gamma \Psi_j \hat{\bf H} + f_{ij} \Psi_i B\, \Gamma \Gamma
  \Gamma \Gamma \Gamma \Psi_j \hat{\bar\Delta}, 
  \label{eq:Wmgut}
\end{gather}
where $i,j = 1,2,3$ are the generation indices, $l,m,n,\ldots =
1,\ldots,10$\,\;are\,\;$SO(10)$\,\;indices, and I have suppressed the
$SO(10)$ indices for straightforward contractions. Here I have used
hats in the denotations of the Higgs superfields to distinguish them
from their scalar components; otherwise, my notation conventions from
Chapter 3 for denoting superfields and their components will remain in
tact for the rest of this work. 

One more point of interest is that any Higgs superfield in the theory
in an $SU(2)_L \times SU(2)_R$ bi-doublet representation,
\emph{i.e.}, with PS quantum numbers ({\bf 2},{\bf 2},$x$), that
also breaks to an $SU(3)_C$ singlet will contribute to the
linear combinations which remain light and play the roles of $H_{u,d}$
at the electroweak scale. Contributions will generally come even from
components which do not couple to matter, through mixing with those
that do, once vevs are acquired. I will discuss this topic in more
detail in the next section, where I will give the details of the model
on which this work is based.

\section[A SUSY $SO(10)$ Model of Unification]{A SUSY
$\boldsymbol{SO(10)}$ Model of Unification}
\label{model} The SUSY $SO(10)$ model on which my proton decay
analysis is based has \textbf{10}, $\boldsymbol{\overline{126}}$, and
\textbf{120} Higgs superfields with Yukawa couplings contributing to
fermion masses; denotation of each is consistent with the previous
section. The superpotential for the model is given by
eq.\,(\ref{eq:Wmgut}) plus the following additional terms due to the
presence of the {\bf 120} field:
\begin{gather}
  W_{\rm U} \ni M_{120} \hat\Sigma^2 + \kappa\, \hat\Sigma_{mno}
  \hat{\bf H}_p \hat\Phi_{pmno} + \rho\, \hat\Sigma_{lmp}
  \hat\Sigma_{nop} \hat\Phi_{lmno} \nonumber \\ + \,\hat\Sigma_{lmn}\,
  \hat\Phi_{nopq} \left( \zeta \hat\Delta_{opqlm} +\, \bar\zeta
  \hat{\bar\Delta}_{opqlm} \right) + g_{ij} \Psi_i B\, \Gamma \Gamma
  \Gamma \Psi_j \hat\Sigma, 
  \label{eq:W120}
\end{gather}
where again $i,j = 1,2,3$ are the generation indices, and I have
suppressed the $SO(10)$ indices for total contractions. Here
$\Psi_i$ is the \textbf{16}-dimensional matter spinor containing
chiral superfields for all the SM fermions (of one generation) plus
the left-handed anti-neutrino.

\paragraph{Type-I Seesaw Breaking Pattern.} For the type-I seesaw
implementation, breaking of $SO(10)$ to MSSM proceeds as follows:
\begin{align*}
  &\langle \Phi(\boldsymbol{1},\boldsymbol{1}, \boldsymbol{1}) \rangle:
  SO(10) \longrightarrow SU(4)_C \times SU(2)_L \times
  SU(2)_R,~(\slashed{D}) \\
  &\langle \bar\Delta(\boldsymbol{1},\boldsymbol{3}, \boldsymbol{10})
  \rangle \equiv v_R: SU(4)_C \times SU(2)_L
  \times SU(2)_R \longrightarrow {\rm MSSM}.
\end{align*}
Note that $\langle \Delta(\boldsymbol{1},\boldsymbol{3},
\boldsymbol{\overline{10}}) \rangle = v_R$ is also present such that
D-term contributions will cancel. The value of $\langle \Phi \rangle$
is taken at the coupling unification scale $M_{\rm U} \sim 2 \times
10^{16}$\,GeV, and $v_R$ at $\sim\!10^{15}$\,GeV; hence any running
under PS is negligible. As discussed previously, the
$\hat{\bar\Delta}_R$ component superfield couples to the right-handed
neutrino ${\cal N}^{\cal C}$. Thus the acquisition of the vev $v_R$
will lead to the Majorana mass term
\begin{equation}
  W_{\cal N} \ni f_{ij} \hat{\bar\Delta}_R {\cal N}^{\cal C} {\cal
  N}^{\cal C} \xrightarrow{\text{$\langle \bar\Delta_R \rangle$}}
  f v_R \nu_R^T C^{-1} \nu_R;
\end{equation}
furthermore, this term will induce a type-I seesaw mass for $\nu_L$ after
EWSB:
\begin{equation}
  m_\nu = -\frac{y_\nu^2 v^2}{f v_R}.
\end{equation}

\paragraph{Type-II Seesaw Breaking Pattern.} The coupling to matter of
the left-handed PS (and SM) triplet $\hat{\bar\Delta}_L \equiv
\hat{\bar\Delta}(\boldsymbol{3},\boldsymbol{1},
\boldsymbol{\overline{10}})$ as seen in eq.\,(\ref{eq:LRlepmaj}) is
present in any model with a $\boldsymbol{\overline{126}}$ field;
hence, to give a type-II Majorana mass to the neutrino, one simply
must give a vev to the scalar $\langle \bar\Delta_L \rangle \equiv
v_L$. That said, the only motivation for giving such an extremely tiny
vev, ${\cal O}(10^{-2}\,{\rm eV})$, is strictly empirical. However, if
the vev for $v_L$ were instead inversely related to a heavy scale
already present in the theory, then its small value would be nicely
consistent. In order to create such a scenario, the most
straightforward option is to include a {\bf 54} multiplet $\hat{\bf
S}_{mn}$ in the Higgs spectrum.  This field adds the following
pertinent terms to the superpotential (among others not important
here):
\begin{equation}
  W_{\rm U}~ \ni ~\xi\, \hat{\bf S}_{mn} \hat{\bf H}_m \hat{\bf H}_n +
  \eta'\, \hat{\bf S}_{lm} \hat\Delta_{lnopq} \hat\Delta_{mnopq} +
  \bar\eta'\, \hat{\bf S}_{lm} \hat{\bar\Delta}_{lnopq}
  \hat{\bar\Delta}_{mnopq}.
  \label{eq:W54}
\end{equation}
The F-term of $\hat{\bf S}$ then gives rise to a scalar operator of
the form
\begin{equation*}
  \left[W(\hat{\bf S})\right]_F \ni \xi \bar\eta' H_u \bar\Delta_L
  H_u \bar\Delta_R,
\end{equation*}
which will consequently appear in the F-term for $\hat{\bar\Delta}_L$
as well, leading to the scalar potential
\begin{equation*}
  V(\bar\Delta_L) = M^2_{126} \lvert \bar\Delta_L \rvert^2 + \xi
  \bar\eta' H_u \bar\Delta_L H_u \bar\Delta_R;
\end{equation*}
now the vevs for $h^0_u$ and $\bar\Delta_R$ will induce a vev for
$\hat{\bar\Delta}_L$ of the form
\begin{equation*}
  v_L \equiv \langle \bar\Delta_L \rangle = \frac{\xi \bar\eta' v_u^2
  v_R}{M^2_{126}} \sim \frac{1}{M_{\rm U}}.
\end{equation*}
Thus the full low-scale neutrino mass matrix becomes
\begin{equation}
  m_\nu = f v_L -\frac{y_\nu v^2 y_\nu^T}{f v_R}.
  \label{eq:typeiimass0}
\end{equation}
However, an examination of this expression in light of the values for
the various parameters will reveal that the type-I and type-II
contributions in (\ref{eq:typeiimass0}) are generally comparable.
Hence this prescription is not enough on its own to give type-II
dominance. To induce a truly dominant type-II seesaw, one needs
additional structure to somehow decouple the mass of $\bar\Delta_L$
from that of $\bar\Delta$.

One particularly nice way to accomplish this, which was first
discussed in \cite{rabi-typeII}, goes as follows. One first breaks
$SO(10)$ together with $B-L$ by giving a vev to $\bar\Delta_R$ at a
scale $\gsim 10^{17}$\,GeV, resulting in $SU(5)$; here, the
left-handed triplet $\hat{\bar\Delta}_L$ is part of the two-index
symmetric {\bf 15} representation. Generally the {\bf 15} components
coming from the $\boldsymbol{\overline{126}}$, {\bf 126}, and {\bf
210}, will have comparable masses. The vev for $\hat{\bar\Delta}_L
\sim 1/M_{126}$, so for larger $v_L \sim {\cal O}({\rm eV})$, one
would like to lower the scale to $M_{126} \lsim 10^{13}$\,GeV;
however, the decomposition of $\boldsymbol{\overline{126}}$ gives rise
to additional $SU(5)$ reps such as {\bf 45} and $\overline{\bf 50}$,
which also have masses $\sim\!M_{126}$; if all such multiplets become
so light, gauge coupling unification will be irreparably damaged. The
day is saved, though, by the presence of the {\bf 54} Higgs $\hat{\bf
S}$, which decomposes under $SU(5)$ as ${\bf 15} \oplus
\boldsymbol{\overline{15}} \oplus {\bf 24}$, and thus contributes to
the {\bf 15} mass matrix but not those of {\bf 45} and {\bf 50}.  As a
result, the masses for {\bf 15} can be tuned to the required light
scale without other consequences, and the vev for $\bar\Delta_L$
\begin{equation*} v_L = \frac{f \xi \bar\eta'
  v_u^2}{M_{\bar\Delta_L}}
\end{equation*}
can be larger as needed for type-II dominance.

With the light mass for $\bar\Delta_L$ on hand, one breaks $SU(5)$ at
the usual coupling unification scale by $\Phi(\boldsymbol{24}) \in {\bf
210}$; hence, the $SO(10)$ breaking chain for type-II dominance is
\begin{align*}
  &\langle \bar\Delta(\boldsymbol{1}) \rangle: SO(10) \longrightarrow
  SU(5),~(\slashed{D}) \\ 
  &\langle \Phi(\boldsymbol{24}) \rangle:SU(5) \longrightarrow 
  {\rm MSSM};
\end{align*}
I've used notation for the $SU(5)$ reps here, but note that the two
components present correspond precisely to those acquiring vevs in the
type-I case.

\begin{center}
  ***
\end{center}
After breaking to MSSM, $SU(2)_L$ doublets with SM quantum numbers
($\left(\,{\bf 1},{\bf 2},-\frac{1}{2} \,\right)$ + c.c\,), which
have their origins in the PS bi-doublet components $\hat{\bf
H}(\boldsymbol{2},\boldsymbol{2}, \boldsymbol{1})$, 
$\hat{\bar\Delta}(\boldsymbol{2},\boldsymbol{2},\boldsymbol{15})$, and
$\hat\Sigma(\boldsymbol{2},\boldsymbol{2}, \boldsymbol{1}) +
\hat\Sigma(\boldsymbol{2},\boldsymbol{2}, \boldsymbol{15})$, have the
following couplings to matter superfields in the superpotential:
\begin{gather}
  W_{\rm Yuk} = h \epsilon_{ab} \left\{ \hat{\bf H}^b_u
  \left(Q^aU^{\cal C} + L^a{\cal N}^{\cal C} \right) + \hat{\bf H}^b_d
  \left(Q^aD^{\cal C} + L^aE^{\cal C} \right) \right\} \nonumber \\ +
  \frac{f \epsilon_{ab}}{\sqrt{3}} \left\{ \hat{\bar\Delta}^b_u
  \left(Q^aU^{\cal C} - 3L^a{\cal N}^{\cal C} \right) +
  \hat{\bar\Delta}^b_d \left(Q^aD^{\cal C} - 3L^aE^{\cal C} \right)
  \right\} \nonumber \\ + g \epsilon_{ab} \left\{ \hat\Sigma^{1 b}_u
  \left(Q^aU^{\cal C} + L^a{\cal N}^{\cal C} \right) + \hat\Sigma^{1
  b}_d \left(Q^aD^{\cal C} + L^aE^{\cal C} \right) \right\} \nonumber
  \\[2mm] + \frac{g \epsilon_{ab}}{\sqrt{3}}
  \left\{\hat\Sigma^{15 b}_u \left(Q^aU^{\cal C} - 3L^a{\cal
  N}^{\cal C} \right) + \hat\Sigma^{15 b}_d \left(Q^aD^{\cal C}
  - 3L^aE^{\cal C} \right) \right\}, 
\end{gather}
where I've suppressed generation and color indices. As one can see,
these doublets come in pairs with opposite hypercharge and so have the
form of the SUSY Higgs doublets $H_{u,d}$.  Furthermore, these fields
will mix with one another, and also with doublets from \textbf{126} and
\textbf{210}, to form mass eigenstates. If I take all such component
fields in the obvious basis as
\begin{equation*}
  \varphi_u \equiv \left( \hat{\bf H}_u, \hat\Sigma_u^1,
  \hat\Sigma_u^{15}, \hat\Delta_u, \hat{\bar\Delta}_u, \hat\Phi_u
  \right), 
\end{equation*}
and similar for $\varphi_d$, but with $\hat\Delta_u \rightarrow
\hat{\bar\Delta}_d$ and ``vice versa'', then the mass matrix
$\mathcal{M_D}$ is defined such that the mass states are given by
$\varphi_d^T \mathcal{M_D} \,\varphi_u$; the form of $\mathcal{M_D}$
can be seen in \cite{aulgarg}. The matrix is diagonalized by a
bi-unitary transformation\, ${\cal U} \mathcal{M_D} {\cal V}^T$,
giving the mass eigenstates for the doublet superfields as linear
combinations of the component fields. Note that this matrix is fully
determined by the couplings and vevs of the superpotential (although
the majority of those parameters are virtually unconstrained), and so
the fields are generally expected to be heavy; however, one doublet
pair must remain light in order to play the role of the MSSM Higgs
doublets $H_{u,d}$.  This point requires the imposing of the condition
$Det\,\mathcal{M_D}\sim0$ ({\it i.e.}, $M_{\rm SUSY} \sim 0$ when
compared to the GUT scale), which can be realized by fine-tuning one
of the parameters in the matrix, conventionally chosen to be the mass
of $\hat{\bf H}$, $M_{10}$. This choice will have implications for
proton decay analysis, which I will discuss in the next section.

In light of this establishment of the MSSM doublets, the effective
Dirac fermion mass matrices can be written as
\begin{align}
  {\cal M}_u &= \tilde{h}+r_2 \tilde{f}+r_3\tilde{g} \nonumber \\ 
  {\cal M}_d &= \frac{r_1}{\tan\beta}
  (\tilde{h}+\tilde{f}+\tilde{g}) \nonumber \\
  {\cal M}_e &= \frac{r_1}{\tan\beta}
  (\tilde{h}-3\tilde{f}+c_e\tilde{g}) \nonumber \\
  {\cal M}_{\nu_D} &= \tilde{h} - 3 r_2 \tilde{f} + c_\nu \tilde{g},
  \label{eq:mass}
\end{align}
where $1 / \tan \beta$ takes $v_u \rightarrow v_d$ for down-type
fields. The couplings with the tildes are given by \cite{olddmm}
\begin{gather}
  \tilde h \equiv {\cal V}_{11} h\, v_u; \quad \tilde f \equiv
  \frac{{\cal U}_{14} f v_u}{r_1 \sqrt{3}}; \quad \tilde g \equiv
  \frac{{\cal U}_{12} + {\cal U}_{13}/\sqrt{3}}{r_1} g\, v_u;
  \nonumber \\ r_1 \equiv \frac{{\cal U}_{11}}{{\cal V}_{11}}; \quad
  r_2 \equiv r_1 \frac{{\cal V}_{15}}{{\cal U}_{14}}; \quad r_3 \equiv
  r_1 \frac{{\cal V}_{12} - {\cal V}_{13}/\sqrt{3}}{{\cal U}_{12} +
  {\cal U}_{13}/\sqrt{3}}; \nonumber \\ c_e \equiv \frac{{\cal U}_{12}
  - {\cal U}_{13} \sqrt{3}}{{\cal U}_{12} + {\cal U}_{13}/\sqrt{3}};
  \quad c_\nu \equiv r_1 \frac{{\cal V}_{12} + {\cal V}_{13}
  \sqrt{3}}{{\cal U}_{12} + {\cal U}_{13}/\sqrt{3}};
\end{gather}
where ${\cal U}_{I\!J},\, {\cal V}_{I\!J}$ are the unitary matrices
that diagonalize $\mathcal{M_D}$. 

The light neutrino mass matrix is given in general by the type-II
seesaw mechanism as
\begin{equation}
  {\cal M}_\nu = f v_L - {\cal M}_{\nu_D}\left(f v_R\right)^{-1}\left(
  {\cal M}_{\nu_D}\right)^T;
\label{eq:neu}
\end{equation}
I will separately consider the cases of type-I and type-II dominance
as outlined previously. Note that the inverse dependence on $f$ in the
type-I term intimately connects the neutrino mass matrix to the
charged sector matrices, which makes the model quite predictive. Also
note that I will consider only normal mass hierarchy in this analysis.

The matrices $h$ and $f$ are real and symmetric, and $g$ is pure
imaginary and anti-symmetric; hence, the Dirac fermion Yukawa
couplings are Hermitian in general, and their most general forms can
be written as \smallskip
\begin{gather}
  \tilde{h} = \left(\begin{array}{ccc} 
    h_{11} & h_{12} & h_{13}\\
    h_{12} & h_{22} & h_{23}\\
    h_{13} & h_{23} & M \end{array}\right), \qquad
  \tilde{f} = \left(\begin{array}{ccc}
    f_{11} & f_{12} & f_{13}\\
    f_{12} & f_{22} & f_{23}\\
    f_{13} & f_{23} & f_{33}\end{array}\right), \nonumber \\[4mm]
  \tilde{g} = i \left(\begin{array}{ccc} 
    0 & g_{12} & g_{13} \\
    -g_{12} & 0 & g_{23} \\
    -g_{13} & -g_{23} & 0 \end{array}\right) \label{eq:y0}.
\end{gather}
$M \equiv h_{33} \sim m_t$ is singled out to stress its dominance over
all other elements. The three matrices as written have a total of 15
parameters; taken in combination with  ratios $r_i$ and $c_\ell$, the
model has a total of 21 parameters. Correspondingly, there are in
principle 22 measurable observables, including all masses, mixing
angles, and $CP$ violating phases, associated with the physical
fermions, although the three PMNS phases and one neutrino mass have yet
to be observed. Therefore one would prefer to have no more than
18 parameters in the model, and generally speaking fewer parameters
indicates greater predictability.

Furthermore, as I will discuss in more detail shortly, the
dimension-five effective operators that arise in proton decay go like
products of Yukawa coupling elements, $\sim \lambda_{ij}
\lambda^{'}_{kl}$ ($\lambda = h,f,g$); therefore, increasing the
number of $\lambda_{ij}$ elements that are small or zero will increase
the number of negligible or vanishing contributions to the decay
width.  This idea was given thorough consideration in \cite{dmm1302},
and the couplings suggested by the authors are as follows: 
\begin{gather}
  \tilde{h} = \left(\begin{array}{ccc} 
    0 & & \\
    & 0 & \\
    & & M \end{array}\right), \qquad
  \tilde{f} = \left(\begin{array}{ccc}
    \sim 0 & \sim 0 & f_{13}\\
    \sim 0 & f_{22} & f_{23}\\
    f_{13} & f_{23} & f_{33}\end{array}\right), \nonumber \\[4mm]
  \tilde{g} = i \left(\begin{array}{ccc} 
    0 & g_{12} & g_{13} \\
    -g_{12} & 0 & g_{23} \\
    -g_{13} & -g_{23} & 0 \end{array}\right) \label{eq:y}.
\end{gather}
Note that $\tilde{h}$ is an explicitly rank-1 matrix, with $M \sim
\mathcal{O}(1)$; thus, at leading order, the \textbf{10} Higgs H $\sim
m_t$ contributes to the third generation masses and nothing more. This
feature has been explored in models demonstrating a discrete flavor
symmetry in {\it e.g.} \cite{dmmflavor,ddms}, and may therefore be
dynamically motivated.  Taking $f_{12} \sim 0$ is equivalent to a
partial diagonalization of $\tilde{f}$, which can be done without loss
of generality in the presence of a rank-1 $\tilde h$; the restriction
on $f_{11}$ is clearly phenomenologically motivated by the smallness
of first-generation masses, in the same way the dominance of the
parameter $M$ corresponds to the largeness of third-generation masses.
As a result of these assumptions, the above Yukawa texture should give
rise to sufficient proton decay lifetimes without the need for the
usual extreme cancellations.

It is further preferred for proton decay that $f_{13},\; g_{12} \ll
1$, although $f_{13}$ plays a role in setting the size of the reactor
neutrino mixing angle $\theta_{13}$, so the above restriction may
create some tension in the fitting.

In carrying out the numerical minimization, I will allow $f_{11}$ and
$f_{12}$ to have small but non-vanishing values,
$\mathcal{O}(10^{-4})$, for the sake of giving accurate
first-generation masses without creating tension in other elements.
The results of that analysis will be discussed in section \ref{fit},
after I discuss the details of calculating proton decay. 


%
%

\chapter{The Details of Proton Decay}\label{5-pdecay}

In addition to the SM doublets present in each of the GUT Higgs
superfields, which contribute to the emergence of $H_{u,d}$ at the
SUSY scale, the heavy fields similarly contain SM-type $SU(3)$
\emph{color triplets} ($\left(\,{\bf 3},{\bf 1},-\frac{1}{3}
\,\right)$ + c.c\,) in their decompositions. These fields come from
the PS components $\hat{\bf H}(\boldsymbol{1},\boldsymbol{1},
\boldsymbol{6})$, $\hat{\bar\Delta}(\boldsymbol{1}, \boldsymbol{1},
\boldsymbol{6}) + \hat{\bar\Delta}_R$, and $\hat\Sigma(\boldsymbol{1},
\boldsymbol{3}, \boldsymbol{\bar6}) + \hat\Sigma(\boldsymbol{1},
\boldsymbol{1}, \boldsymbol{\overline{10}})$. Furthermore, there are
two more exotic types of triplets that also lead to $B$- or
$L$-violating vertices: $\left(\,{\bf 3},{\bf 1}, -\frac{4}{3}
\right)$ + c.c, which interact with two up-type or two down-type
$SU(2)_L$ singlet fermions, and $\left(\,{\bf 3},{\bf 3}, -\frac{1}{3}
\right)$ + c.c, which interact with a pair of $SU(2)_L$ doublets. The
above components have the following couplings to matter superfields in
the superpotential:
\begin{gather}
  W_{\rm \slashed{B}\slashed{L}} = h \left\{ \hat{\bf H}_{\cal T}
  \left(\frac{1}{2} \epsilon_{ab} Q^aQ^b + U^{\cal C}E^{\cal C}
  \right) + \hat{\bf H}_{\bar{\cal T}} \left(\epsilon_{ab} Q^aL^b +
  U^{\cal C}D^{\cal C} \right) \right\} \nonumber \\ + f \left\{
  \hat{\bar\Delta}_{\cal T} \left(\frac{1}{2} \epsilon_{ab} Q^aQ^b -
  U^{\cal C}E^{\cal C} \right) + \hat{\bar\Delta}_{\bar{\cal T}}
  \left(\epsilon_{ab} Q^aL^b - U^{\cal C}D^{\cal C} \right) \right\} +
  f \sqrt{2}\, \hat{\bar\Delta}^R_{\cal T}\, U^{\cal C}E^{\cal C}
  \nonumber \\ + g \sqrt{2}\, \left\{ \left( -\hat\Sigma^6_{\cal T} +
  \hat\Sigma^{10}_{\cal T} \right) U^{\cal C}E^{\cal C} +
  \hat\Sigma^6_{\bar{\cal T}}\, U^{\cal C}D^{\cal C} + \epsilon_{ab}\,
  \hat\Sigma^{10}_{\bar{\cal T}}\, Q^aL^b \right\} \nonumber \\ + 2f
  \hat{\bar\Delta}_C\, D^{\cal C}E^{\cal C} + 2g\, \hat\Sigma_C\,
  D^{\cal C}E^{\cal C} + 2g\, \hat\Sigma_{\bar C}\, U^{\cal C}U^{\cal
  C} \nonumber \\ - 4f\, Q\, i\sigma_2 \hat{\bar\Delta}_{\bar Q}\, L -
  2g\, Q\, i\sigma_2 \hat\Sigma_Q\,Q - 4g\, Q\, i\sigma_2
  \hat\Sigma_{\bar Q}\,L,
  \label{eq:trip-int}
\end{gather}
where I have again suppressed generation and color indices. Note that
all of the terms present violate baryon or lepton number. The terms in
the final two lines represent the exotic couplings.

Like the doublets, the ordinary color triplets will mix after the
GUT-scale breaking to form mass eigenstates; again, this mixing
includes triplets contained in the {\bf 210} and {\bf 126} fields not
contributing to fermion masses.  The resulting $7\times7$ triplet mass
matrix ${\cal M_T}$ is diagonalized by ${\cal X M_T Y}^T$ to give the
eigenstates. The exotic types will mix amongst themselves as well in
their own $2\times2$ matrices. These matrices are again fully
determined by the heavy vevs and the parameters of the $SO(10)$
superpotential. Since there is no light triplet analog to $H_{u,d}$
found in the low-scale particle spectrum, all of the fields can be
heavy, although the presence of the same parameters in both the
doublet and triplet matrices makes the decoupling of the
doublet-triplet behavior a substantial topic itself.

\begin{figure}[t]
\begin{center}
  \includegraphics{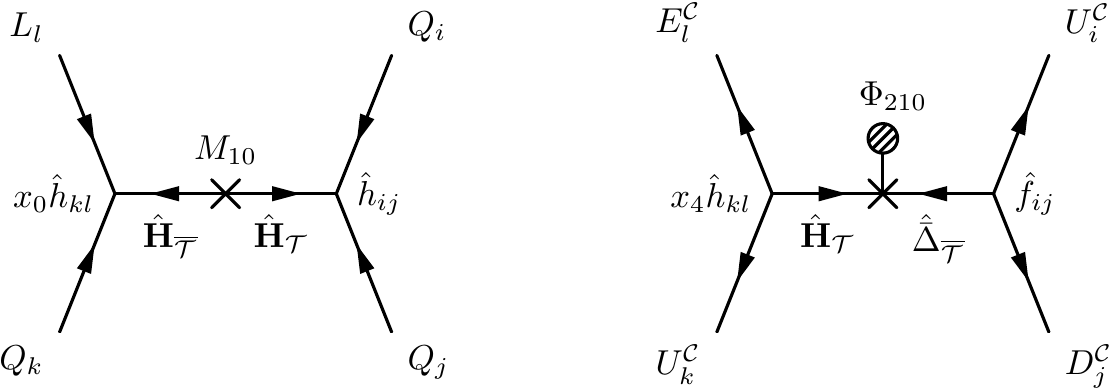} 
  \caption[Examples of superfield diagrams that can lead to proton
  decay.]{Examples of superfield diagrams that lead to proton decay in
  this model. The hats on the couplings indicate mass basis, and the
  parameters $x_i$ contain the triplet mixing information unique to
  the specific pairing of couplings present in each diagram (see
  below).} 
    \label{fig:susyfd}
\end{center}
\end{figure}
 
$T$-channel exchange of conjugate pairs of any of these triplets,
through a mass term or interaction with a heavy Higgs field such as
{\bf 54} or {\bf 210}, leads to operators that change two quarks into
a quark and a lepton; this is the numerically dominant mechanism
through which a proton can decay into a meson and a lepton;
corresponding $s$-channel decays through the scalar superpartners of
these triplets, as well as $s$-channel decays through the $SU(5)$-like
gauge bosons $X,Y$, are suppressed by an additional factor of $1/
M_{\rm U}$ and so are generally negligible in comparison.\footnote{The
dominant mode in $X$-boson exchange, $p \rightarrow \pi^0 e^+$, may be
comparable if the relevant threshold corrections are large.} Figure
\ref{fig:susyfd} shows Feynman diagrams for two examples of the
operators in question.

\section{The Effective Potential} At energies far below the GUT
scale, the triplet fields are integrated out, giving four-point
effective superfield operators, which give rise in turn to
four-fermion operators. The corresponding effective superpotential is
\begin{equation}
  {\cal W}_{\slashed{B}\slashed{L}} =
  \frac{\epsilon_{\rho\sigma\tau}}{M_\mathcal{T}} \left(
  \widehat{C}^L_{ijkl} Q^\rho_i Q^\sigma_j Q^\tau_k L_l +
  \widehat{C}^R_{[ijk]l} U^{\mathcal{C} \rho}_i D^{\mathcal{C}
  \sigma}_j U^{\mathcal{C} \tau}_k E^\mathcal{C}_l \right),
\label{eq:effW}
\end{equation}
where $i,j,k,l = 1,2,3$ are the generation indices and
$\rho,\sigma,\tau = 1,2,3$ are the color indices; $SU(2)$ doublets are
contracted pairwise.  This potential has $\Delta L = 1$ and $\Delta B
= 1$ and so also has $\Delta(B - L) = 0$. $M_\mathcal{T}$ is a generic
mass for the triplets, which I will take $\sim\! M_{\rm U}$. Note the
anti-symmetrization of $i,k$ in the $C_R$ operator; this is the
non-vanishing contribution in light of the contraction of the color
indices. The analogous anti-symmetry for the $L$ operator is ambiguous
in the current notation, but I will tend to the issue shortly.

The effective operator coefficients $C_{ijkl}$ are of the form
\begin{align}
  C^R_{ijkl} &= x_0 h_{ij} h_{kl} + x_1 f_{ij} f_{kl} +
  x_2 g_{ij} g_{kl} + x_3 h_{ij} f_{kl} +
  x_4 f_{ij} h_{kl} + x_5 f_{ij} g_{kl} \nonumber \\
  & + x_6 g_{ij} f_{kl} + x_7 h_{ij} g_{kl} +
  x_8 g_{ij} h_{kl} + x_9 f_{il} g_{jk} +
  x_{10} g_{il} g_{jk} \nonumber \\
  C^L_{ijkl} &= x_0 h_{ij} h_{kl} + x_1 f_{ij} f_{kl} -
  x_3 h_{ij} f_{kl} - x_4 f_{ij} h_{kl} + y_5 f_{ij} g_{kl} +
  y_7 h_{ij} g_{kl} \nonumber \\
  & + y_9 g_{ik} f_{jl} + y_{10} g_{ik} g_{jl}.
  \label{eq:Cs}
\end{align}
The couplings $h,f,g$ as written correspond to matter fields in the
flavor basis and undergo unitary rotations in the change to mass
basis, as indicated by the hats on $\widehat{C}^{L,R}$ in
eq.\,(\ref{eq:effW}) above; I will save the details of the change of
basis for later in the discussion. The parameters $x_i,y_i \sim {\cal
X}_{I\!J}, {\cal Y}_{I\!J}$ are elements of the unitary matrices that
diagonalize the triplet mass matrix $\mathcal{M_T}$, or the
corresponding matrices for the exotic triplets. Note that several
identifications have already been made here: $y_{0,1} = x_{0,1}$ and
$y_{3,4} = -x_{3,4}$; looking at eq.\,(\ref{eq:trip-int}), one can see
the would-be parameters $y_{2,6,8} = 0$. Also note that $x_0 \sim
M_{10}$ is the {\bf 10} mass parameter fixed by the tuning condition
for $M_\mathcal{D}$.  The parameters $x_{9,10}$ and $y_{9,10}$
correspond to the exotic triplets; the indices of those terms are
connected in unique ways as a result of the distinct contractions of
fields.

The left-handed term in eq.\,(\ref{eq:effW}) can be further expanded by
multiplying out the doublets as
\begin{equation}
  {\cal W}_{\slashed{B}\slashed{L}} \ni
  \frac{\epsilon_{\rho\sigma\tau}}{M_\mathcal{T}} \left(
  \widehat{C}^L_{[ijk]l} U^\rho_i D^\sigma_j U^\tau_k E_l -
  \widehat{C}^L_{i[jk]l} U^\rho_i D^\sigma_j D^\tau_k
  \mathcal{N}_l \right),
\label{eq:effWL}
\end{equation}
where $\mathcal{N}$ is the left-handed neutrino superfield.  Note that
the coefficients $C^L$ are anti-symmetrized in the indices of the
like-flavor quarks, again due to the anti-symmetry of color index
contraction, as discussed above for $C^R$. This anti-symmetry will be
crucial in restricting the number of contributing channels for decay.
 
\section{Dressing the Operators} 
\begin{figure}[t]
\begin{center}
  \includegraphics{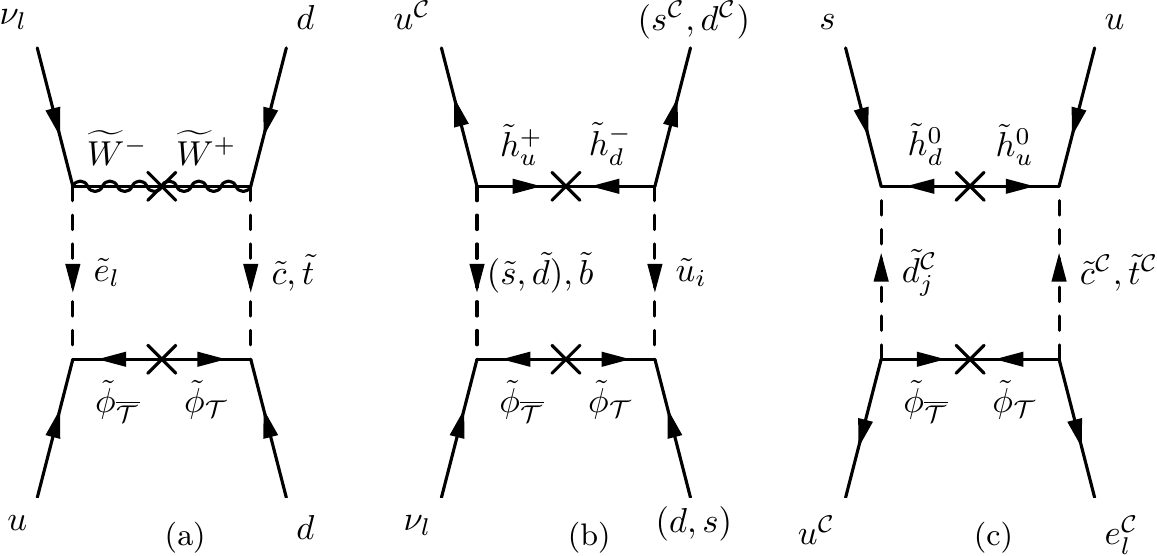} 
  \caption[Examples of dressed diagrams leading to proton
  decay.]{Examples of dressed diagrams leading to proton decay in the
  model. $\phi = {\rm H},\bar\Delta,\Sigma$. Diagram (a) shows a
  contribution to $p \rightarrow \pi^+ \bar{\nu}_l$; integrating out
  the triplets gives an effective operator of type $C^{L}udue$.
  Diagram (b) shows a $C^{L}udd\nu$-type operator contributing to $K^+
  \bar{\nu}_l$.  Diagram (c) shows a $C^{R} u^\mathcal{C}
  d^\mathcal{C} u^\mathcal{C} e^\mathcal{C}$-type operator
  contributing to $K^0 e_l^+$, for $l=1,2$. Note where more than one
  field is listed, each choice gives a separate contributing channel,
  except for the dependent exchange of $(s \leftrightarrow d)$ in
  (b).} 
  \label{fig:compfd}
\end{center}
\end{figure}

Holomorphism of the superpotential
forbids conjugate-mixing mass terms like $M_\mathcal{T}
\phi_{\mathcal{T}} \phi_{\overline{\cal T}}$ for $\phi = {\rm
H},\bar\Delta,\Sigma$ scalar boson components of the triplet
superfields; therefore, diagrams of the type in Figure
\ref{fig:susyfd} can only be realized at leading order through
conjugate pairs of {\it Higgsino} triplet mediators.  Thus, in
component notation, each vertex will be of the form $\lambda\,
\tilde{\phi}_\mathcal{T} q\,\tilde{q}$ or similar, with $\lambda =
h,f,g$ as appropriate. Therefore, the squarks and sleptons must be
``dressed'' with gaugino or (SUSY) Higgsino vertices to give $d=6$
effective operators of the four-fermion form needed for proton decay.
Depending on the sfermions present, diagrams may in principle be
dressed with gluinos, Winos, Binos, or Higgsinos. Examples of
appropriately-dressed component-field diagrams which give proton decay
are shown in Figure \ref{fig:compfd}.

In the following subsections, I will discuss the implications for each
type of dressing and determine which types will contribute leading
factors in the proton decay width. Note that I will give this
discussion in terms of $\widetilde{B}$, $\widetilde{W}^0$, and
$\tilde{h}_{u,d}^{\pm,0}$, rather than $\widetilde{A}$,
$\widetilde{Z}$, $\widetilde{\chi}^\pm_i$, and $\widetilde{\chi}^0_i$,
because $(a)$ I am assuming a universal mass spectrum for
superpartners to satisfy FCNC constraints, meaning the mass and flavor
eigenstates coincide for the gauge bosons, and $(b)$ the mixing of
Higgsinos, while not typically negligible, will result in chargino or
neutralino masses different from Higgsino mass parameter $\mu$ by
$\mathcal{O}(1)$ factors as long as gaugino soft masses are relatively
small compared to $M_{\rm SUSY}$; since precise values of such masses
are insofar unknown, and since so many of the SUSY and GUT parameter
values needed for the decay width calculations are similarly unknown,
I will take $m_{\tilde{h}^\pm} \sim m_{\tilde{h}^0} \sim \mu$ in order
to simplify the calculation, especially for computational purposes.

\subsection{Gluino Dressing} Two limitations are readily apparent
when considering dressing by gluinos. First, the lepton will have to
be a fermion leg in the triplet exchange operator, as in Figure
\ref{fig:compfd} (b) or (c), since a slepton cannot be dressed by a
gluino. Second, since $SU(3)_c$ interactions are
generation-independent, the gluino can only take $\tilde{u}
\rightarrow u$, $\tilde{s} \rightarrow s$, etc. The latter may seem a
fairly innocuous idea on its own, but consider that proton decay to a
kaon or pion will involve operators with one and zero
second-generation quarks as external legs, respectively, with all
others first-generation. Taking these two points together with the
generation-index anti-symmetry of the $C_{ijkl}$ operators, which
implies that $i\neq k$ for the $U_iD_jU_kE_l$ operators and $j\neq k$
for the $U_iD_jD_k\mathcal{N}_l$ operators, one can see by inspecting
a dressed diagram that only diagrams with exactly one each of $U,D,S$
in the triplet operator may be successfully dressed by the gluino.
This constraint implies that gluino dressing can contribute only to $p
\rightarrow K^+ \bar{\nu}$ decay mode; furthermore, the absence of
$UDUE$-type contributions implies no right-handed channels.

\begin{figure}[t]
\begin{center}
  \includegraphics{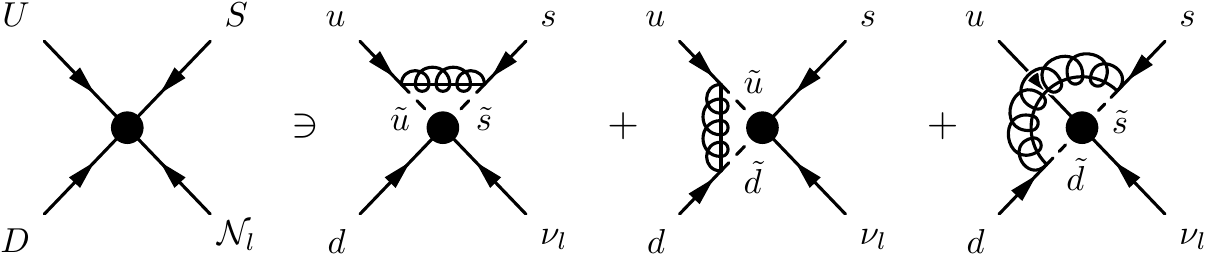}
  \caption[Gluino dressings of the $UDS\mathcal{N}$ operator that
  would contribute to $p \rightarrow K^+ \bar{\nu}_l$.]{ Gluino
  dressings of the $d=5$ operator $M_\mathcal{T}^{-1}
  \widehat{C}^L_{1[12]l} UDS\mathcal{N}$ that would contribute to $p
  \rightarrow K^+ \bar{\nu}_l$; in the limit of universal squark
  masses, the three diagrams sum to zero by a Fierz identity. NOTE:
  gluino mass insertions have been omitted from the diagrams for
  readability.} 
  \label{fig:gludress}
\end{center}
\end{figure}

Taking these constraints into account, and thus looking specifically
at variants of the $UDS\mathcal{N}$ operator, there are three
independent terms one can write \cite{belyaev}, which correspond to
the dressed diagrams shown in Figure
\ref{fig:gludress}:\,\footnote{Each term like ``$(u^\rho \nu_l)$'' is
actually $(u^\rho)^T C^{-1} \nu_l$; the details have been suppressed
simply for readability.}
\begin{equation}
  \epsilon_{\rho\sigma\tau} U^\rho D^\sigma S^\tau \mathcal{N}_l \ni
  \epsilon_{\rho\sigma\tau} \left\{(u^\rho \nu_l)(\tilde{d}^\sigma
  \tilde{s}^\tau) + (d^\sigma \nu_l)(\tilde{u}^\rho \tilde{s}^\tau)
  + (s^\tau \nu_l)(\tilde{u}^\rho \tilde{d}^\sigma)\right\}.
  \label{eq:udsn-q}
\end{equation}
Applying the gluino dressing to each term gives the following sum
of four-fermion effective operators:
\begin{equation}
  \overset{\tilde{g}}{\longrightarrow} \quad \epsilon_{\rho\sigma\tau}
  \left(\frac{\alpha_s}{4\pi}\right) \left\{\kappa_1 (u^\rho
  \nu_l)(d^\sigma s^\tau) + \kappa_2 (d^\sigma \nu_l)(u^\rho s^\tau)
  + \kappa_3 (s^\tau \nu_l)(u^\rho d^\sigma) \right\},
  \label{eq:gludr}
\end{equation}
where the parameters $\kappa_a$ contain factors from the scalar and
gluino propagators in the loop integral. The scalar propagators are
different in general; however, recall that I am assuming universality,
meaning that all sfermion masses are equal to leading order. In that
case, all $\kappa$s are equal and can be factored out of the brackets.
The sum left inside the brackets is zero by a Fierz identity for
fermion contractions \cite{goh}, and so the contribution from gluino
dressing to the $K^+ \bar{\nu}$ decay mode vanishes under the
universal mass assumption.

\subsection{Bino Dressing} As with $SU(3)_c$, $U(1)_Y$
interactions are also flavor-diagonal; thus, the same constraints
apply here as in the gluino case, and possible contributions are to
the $K^+ \bar{\nu}$ mode only.

Looking again at the $UDS\mathcal{N}$ operator, for terms in which the
neutrino is a fermion leg, the argument is analogous to that given for
the gluino dressing: the diagrams involved are identical to the three
in Figure \ref{fig:gludress} except with $\tilde{g} \rightarrow
\widetilde{B}$; starting again from expression (\ref{eq:udsn-q})
and applying the Bino dressing, one arrives at an expression similar
to (\ref{eq:gludr}) but containing hypercharge coefficients in
addition to the $\kappa_a$:
\begin{align} \label{eq:binodr-q}
  \overset{\widetilde{B}}{\longrightarrow} \quad
  \epsilon_{\rho\sigma\tau} \left(\frac{\alpha_1}{4\pi}\right) \{
  \kappa_1 Y_d Y_s & (u^\rho \nu_l)(d^\sigma s^\tau) + \kappa_2 Y_u
  Y_s (d^\sigma \nu_l)(u^\rho s^\tau) \\ & + \kappa_3 Y_u Y_d (s^\tau
  \nu_l)(u^\rho d^\sigma)\}; \nonumber
\end{align}
however, $u,d,s \in Q_i$ are all left-handed quarks with $Y =
\frac{1}{6}$, so the hypercharge products factor out, and again the
fermion sum vanishes by the Fierz identity.

\begin{figure}[t]
\begin{center}
  \includegraphics{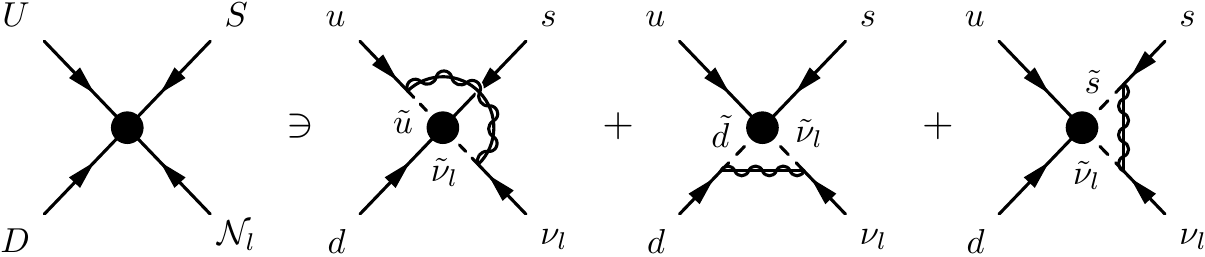}
  \caption[Bino dressings of the $UDS\mathcal{N}$ operator involving a
  sneutrino that would contribute to $p \rightarrow K^+
  \bar{\nu}_l$.]{Bino dressings of the $d=5$ operator
  $M_\mathcal{T}^{-1} \widehat{C}^L_{1[12]l} UDS\mathcal{N}$ involving
  a scalar neutrino that would contribute to $p \rightarrow K^+
  \bar{\nu}_l$; again, in the limit of universal squark masses, the
  three diagrams sum to zero by a Fierz identity.  NOTE: Bino mass
  insertions have been omitted from the diagrams for readability.}
  \label{fig:bdress}
\end{center}
\end{figure}

Because leptons carry hypercharge, there are three additional diagrams
one should include in Figure \ref{fig:gludress} if dressing instead by
the Bino, namely, those involving the scalar neutrino; these diagrams
are shown in Figure \ref{fig:bdress}, and the corresponding terms from
the triplet operator are
\begin{equation}
  \epsilon_{\rho\sigma\tau} U^\rho D^\sigma S^\tau \mathcal{N}_l \ni
  \epsilon_{\rho\sigma\tau} \left\{(d^\sigma s^\tau)(\tilde{u}^\rho
  \tilde{\nu}_l) + (u^\rho s^\tau) (\tilde{d}^\sigma \tilde{\nu}_l) +
  (u^\rho d^\sigma)(\tilde{s}^\tau \tilde{\nu}_l) \right\}.
  \label{eq:udsn-nu}
\end{equation}
Applying the Bino dressing to each of these terms gives another sum
of four-fermion effective operators involving hypercharge:
\begin{align} \label{eq:binodr-nu}
  \overset{\widetilde{B}}{\longrightarrow} \quad
  \kappa\,\epsilon_{\rho\sigma\tau} \left(\frac{\alpha_1}{4\pi}\right)
  \{Y_u Y_\nu & (d^\sigma s^\tau)(u^\rho \nu_l) + Y_d Y_\nu (u^\rho
  s^\tau)(d^\sigma \nu_l) \\ & + Y_s Y_\nu (u^\rho d^\sigma)(s^\tau
  \nu_l)\}; \nonumber
\end{align}
this group of terms has a different product of hypercharges from that
of (\ref{eq:binodr-q}), but it still has a single common product among
the three terms, so I can again factor it out, which results in yet
another vanishing contribution by the Fierz argument.  Hence, the
entire Bino dressing contribution to the $K^+ \bar{\nu}$ mode also
vanishes under the universal mass assumption.  

\subsection{Wino Dressing} As the flavor-diagonal restrictions of the
gluino and Bino also apply to the $\widetilde{W}^0$ but {\it not} to
the $\widetilde{W}^\pm$, the two cases must be considered separately.
That said, one additional restriction applicable in both cases is the
ability to interact with only left-handed particles; thus there will
be no contribution here from the $R$-type operators.

\paragraph{Neutral Wino.} As noted, dressing with the
$\widetilde{W}^0$ is also restricted to $UDS\mathcal{N}$ contributions
to the $K^+ \bar{\nu}$ mode. The terms to be dressed are the same as
those in the Bino case, given by expressions (\ref{eq:udsn-q}) and
(\ref{eq:udsn-nu}); however, in applying the dressing, one finds a kink
in the previous argument:
\begin{align}   
  \overset{\widetilde{W}^0}{\longrightarrow} \quad &
  \kappa\,\epsilon_{\rho\sigma\tau} \left(\frac{\alpha_2}{4\pi}\right)
  \{T^3_d T^3_s (u^\rho \nu_l)(d^\sigma s^\tau) + T^3_u T^3_s
  (d^\sigma \nu_l)(u^\rho s^\tau) + T^3_u T^3_d (s^\tau \nu_l)(u^\rho
  d^\sigma)\} \nonumber \\ =&
  \,\frac{\kappa\,\epsilon_{\rho\sigma\tau}}{4}
  \left(\frac{\alpha_2}{4\pi}\right) \{(u^\rho \nu_l)(d^\sigma s^\tau)
  - (d^\sigma \nu_l)(u^\rho s^\tau) - (s^\tau \nu_l)(u^\rho
  d^\sigma)\},
  \label{eq:winodr-q} \\ \nonumber \\
  \overset{\widetilde{W}^0}{\longrightarrow} \quad &
  \kappa\,\epsilon_{\rho\sigma\tau} \left(\frac{\alpha_2}{4\pi}\right)
  \{T^3_u T^3_\nu (d^\sigma s^\tau)(u^\rho \nu_l) + T^3_d T^3_\nu
  (u^\rho s^\tau)(d^\sigma \nu_l) + T^3_s T^3_\nu (u^\rho
  d^\sigma)(s^\tau \nu_l)\} \nonumber \\ =&
  \,\frac{\kappa\,\epsilon_{\rho\sigma\tau}}{4}
  \left(\frac{\alpha_2}{4\pi}\right) \{(d^\sigma s^\tau)(u^\rho \nu_l)
  - (u^\rho s^\tau)(d^\sigma \nu_l) - (u^\rho d^\sigma)(s^\tau
  \nu_l)\};
  \label{eq:winodr-nu} 
\end{align}
the negative weak isospin carried by the down-type fields prevents use
of the Fierz identity argument. Thus it seems I have finally
found a non-vanishing contribution to proton decay, albeit to only
this one mode.

There is something yet to be gained from the Fierz identity in this
case: the same zero sum seen in the previous cases tells one
that in each expression here, the sum of the two negative terms is
equal to the first term; furthermore, note that the final expressions
in (\ref{eq:winodr-q}) and (\ref{eq:winodr-nu}) are actually
identical. Therefore, I can collect the above contributions into one
expression:
\begin{align}
  \overset{\widetilde{W}^0}{\longrightarrow} \quad 2~\times~ &
  \frac{\kappa\,\epsilon_{\rho\sigma\tau}}{4}
  \left(\frac{\alpha_2}{4\pi}\right) (-2) \{(u^\rho s^\tau)(d^\sigma
  \nu_l) + (u^\rho d^\sigma)(s^\tau \nu_l)\} \nonumber \\ =&
  ~-\kappa\,\epsilon_{\rho\sigma\tau}
  \left(\frac{\alpha_2}{4\pi}\right) \{(u^\rho s^\tau)(d^\sigma \nu_l)
  + (u^\rho d^\sigma)(s^\tau \nu_l)\}.
  \label{eq:winodr-sum} 
\end{align}
Including the factors from the triplet operator, I can write an
operator for the entire neutral Wino contribution to $K^+ \bar{\nu}$:
\begin{equation}
  \mathscr{O}_{\widetilde{W}^0} = ~\kappa\,\epsilon_{\rho\sigma\tau}
  \left(\frac{\alpha_2}{4\pi}\right) M_\mathcal{T}^{-1}
  \widehat{C}^L_{1[12]l}\, \{(u^\rho s^\tau)(d^\sigma \nu_l) + (u^\rho
  d^\sigma)(s^\tau \nu_l)\},
  \label{eq:wino0-tot} 
\end{equation}
where the sign cancels with that from the $UDD\mathcal{N}$ term in
eq.\,(\ref{eq:effW}). The details of $\kappa$ will be discussed in the
next subsection. Note I could have instead written the above
expressions in terms of $(d^\sigma s^\tau)(u^\rho \nu_l)$ alone; I
choose this version simply because the up-up- and down-down-type
pairings in the latter expression are not found in Higgsino or charged
Wino modes and so are not otherwise used in calculation.

\paragraph{Charged Wino.} The assumption of universal mass means that
the sfermions are simultaneously flavor and mass eigenstates;
therefore, the would-be CKM-like unitary matrix for each is simply the
identity, $U^{\tilde{f}} \sim \mathbb{I}$. As a result, the unitary
matrix present in the fermion-sfermion-Wino couplings is not $V_{\rm
ckm}$ or $V_{\rm pmns}$, but rather the single unitary matrix
corresponding to the fermion rotation. Nonetheless, this rotation
allows for the mixing of generations at the dressing vertices, and the
limitations found on the neutral current dressings are not applicable.
This is quite crucial since it allows for contributions from diagrams
with any sfermion propagator not forbidden by the anti-symmetry of the
$C^L_{ijkl}$ operator.  Proton decay modes involving neutral kaons or
pions, which have $u\bar{u}$ or $d\bar{d}$ as external quarks, would
be intractable without generation mixing.  Such mixing will of course
come at the expense of suppression from an off-diagonal element in the
pertinent unitary matrix, which will typically be
$\mathcal{O}(10^{-2\mbox{-}3})$; hence, one can begin to see an
indication of why the $K^+ \bar{\nu}$ mode is so dominant in the full
proton decay width.

One additional constraint on charged Wino dressing involves the Wino
mass insertion. Unlike the gauginos discussed so far, $W^\pm$ are the
antiparticles of \emph{each other}, rather than either being its own
antiparticle. As a result, the Wino mass term is of the form
$M_{\widetilde{W}} \widetilde{W}^+ \widetilde{W}^-$; in order to
involve one $\widetilde{W}^+$ and one $\widetilde{W}^-$ in the
dressing, the two sfermions involved must be of opposite $SU(2)$
flavor. As a result, triplet operators of the form $u \tilde{d} u
\tilde{e}$, $\tilde{u} d \tilde{u} e$ (or the RH equivalents), $u
\tilde{d} \tilde{d} \nu$, and $\tilde{u} d d \tilde{\nu}$ do not
contribute.

Beyond these constraints, the generational freedom of the sfermions
leads to numerous contributions to each of the crucial decay modes,
$K^+ \bar{\nu}$, $K^0 \ell^+$, $\pi^+ \bar{\nu}$, and $\pi^0 \ell^+$,
where $\ell = e,\mu$. In particular the $UDUE$- and
$UDD\mathcal{N}$-type operators each contribute to {\it each} mode
through multiple channels. A list of all such contributions would
likely be overwhelming to the reader no matter how excellent my
choices of notation, but one can find the relevant diagrams in
Appendix \ref{fds}.

\subsection{Higgsino Dressing}
When compared to the others, Higgsino dressing is wildly
unconstrained. First, the low-scale Yukawa couplings governing the
fermion-sfermion-Higgsino interactions couple a left-handed field to a
right-handed one, so clearly the dressing can be applied to both
$C^L$- and $C^R$-type triplet operators. Also, since charged and
neutral Higgsinos couple through the same Yukawas, both types of
interactions can mix generations, meaning the generation-diagonal
constraints on the rest of the neutral-current dressings do not apply
to $\tilde{h}^0_{u,d}$. The only previously-mentioned restriction that
{\it does} apply is, like the charged Wino, the mass term for the SUSY
Higgs couples $H_u$ to $H_d$, so it therefore cannot contribute
through the triplet operators with sfermions of like $SU(2)$ flavor.
One remaining minor restriction is that one will not see the triplet
operator $\tilde{u}du\tilde{e}$ dressed by $\tilde{h}^\pm$ nor
$u\tilde{d}d\tilde{\nu}$ dressed by $\tilde{h}^0$ because each would
result in an outgoing left-handed anti-neutrino.

One can find cases in the literature ({\it e.g.} \cite{goh}) of
Higgsino-dressed contributions being counted as negligible when
compared to those from the Wino; this is usually because if one
exchanges the $g_2^2\,V_{\mathrm{Cabibbo}}$ found in a typical
dominant Wino contribution for a $y^u_{ii'}\, y^d_{kk'}\, \tan\beta$
found in a typical dominant Higgsino contribution, the resulting value
will be smaller by at least a factor of $\mathcal{O}(10)$. Of course
one makes several assumptions in such a comparison: $\mu \sim
M_{\widetilde{W}}$ for one, but additionally that $(a)$ $\tan\beta$ is
small or moderate, and $(b)$ the $C_{ijkl}$ coefficients are usually of
roughly the same magnitude for any combination of $i,j,k,l$ present.

For this analysis, though, neither assumption is valid: I have already
mentioned that I will consider large $\tan\beta$ for maximal
applicability; furthermore, due to the rank-1 texture of the $h$
coupling and the related sparse or hierarchical textures of $f$ and
$g$ as shown in eq.\,(\ref{eq:y}), many of the $C_{ijkl}$ are small or
zero, creating large disparities between the values from one
contribution to the next. This discrepancy from expectation is further
enhanced by the tendency for the unitary matrices $U^f$, which
give the off-diagonal suppressions at the dressing vertices in this
model, to individually deviate from the hierarchical structure of
$V_{\rm ckm}$.

To see the extent to which these two properties can lead to surprises
in numerical dominance, consider that, for example, I find $C^L_{1213}
\sim C^L_{3213}\,U^d_{31}$; one might expect that $U^d_{31} \sim
V_{ub}$ and $C^L_{1213} \sim C^L_{3213}$, so therefore the former term
is much larger than the latter, but in fact neither assumption is
accurate. 

As a result of these model characteristics, I find that the dominant
contributions from Higgsino-dressed diagrams are generally comparable
to those from Wino-dressed diagrams. This statement further applies to
contributions from {\it right-handed} operators as well. Thus I made
no {\it a priori} assumptions about which of the $C^L$- or $C^R$-type
Higgsino-dressed contributions might be excluded as negligible.

Because both the $U^\mathcal{C} D^\mathcal{C} U^\mathcal{C}
E^\mathcal{C}$ operators and the $\tilde{h}^0_{u,d}$ dressing
contribute to all of the pertinent decay modes, the complete list of
channels dressed by the Higgsino is considerably more plentiful than
that of the Wino and so would be even more overwhelming, but again one
can find all of the pertinent diagrams in Appendix \ref{fds}.

\section{Building the Partial Decay Width Formulae} As I discussed
above in the Higgsino dressing subsection, the Yukawa texture seen in
eq.\,(\ref{eq:y}) leads to $(a)$ unusually extreme variation in the
sizes of the $C_{ijkl}$ coefficients, depending strongly on the index
values present, and $(b)$ textures for the unitary matrices $U^f$
which deviate substantially from that of $V_{\rm ckm}$. The
repercussions of these features clearly extend beyond affecting the
relative size of Wino and Higgsino channel contributions. For one, the
off-diagonal suppressions $U^f_{kk'}$ present in most charged Wino
diagrams cannot be dependably approximated as $V^{\rm ckm}_{kk'}$;
fortunately, the GUT-scale $U^f$ are fixed by the fermion fitting, and
since the running of such unitary matrices is small, I can simply use
them at the $\widetilde{W}^\pm$ vertices as reasonable approximations
to their low-scale counterparts.

Another complication due the Yukawa texture is the disturbance of
typically useful assumptions about which channels dominate the
calculation. Such assumptions include dominance of Higgsino channels
with $\tilde{t},\tilde{b},\tilde{\tau}$ intermediate states or Wino
channels $\propto V_{ii}$ or $V_{\mathrm{Cabibbo}}$. In the absence of
the general validity of any such simplification, I am compelled to
presume that {\it any} channel might be a non-negligible contribution
to decay width.

Thus, I initially treated all possible channels as potentially
significant; however, in the interest of saving considerable
computational time, I chose an abridged set of contributions to
include in my numerical analysis through inspection of tentative
calculations, although my threshold for inclusion was quite
conservative. It seemed to me that conventional methods of keeping
only the most dominant terms for calculation might easily lead to
drastically underestimated decay widths, in that if I exclude ten
``negligible'' terms smaller than leading contributions by a factor of
ten, then I have evidently excluded the equivalent of a leading
contribution. To fully avoid such folly, I used a cutoff of roughly
1/50 for exclusion, and made cuts on a per-triplet-operator basis,
which translates to three or four significant figures of precision in
the decay widths.

The Feynman diagrams for all non-vanishing channels of proton decay
for the $K^+ \bar{\nu}_l$, $K^0 \ell^+$, $\pi^+ \bar{\nu}_l$, and
$\pi^0 \ell^+$ modes are catalogued in Appendix \ref{fds}.

Calculation of a proton partial decay width can be broken into three
distinct parts. The first part is the evaluation of the ``internal'',
$d=6$ dressed diagrams discussed in the previous subsection; each
diagram corresponds to an effective operator of the form $X\,qqq\ell$,
where $X \sim M^{-1}_\mathcal{T}\, C_{ijkl}\,$\dots~is a numerical
coefficient unique to each decay channel. Note that here each $q$ is a
single quark fermion, not a doublet. The second part is the evaluation
of a hadronic factor that quantifies the conversion of the three
external quarks of a dressed diagram--plus one spectator quark--into a
proton and a meson. The third and final part is the evaluation of the
``external'' effective diagram for $p \rightarrow \mathrm{M} \bar\ell$
giving the decay width of the proton. I will go through the details
of each stage before giving the resulting decay width expressions.

\subsection{Evaluating the Dressed Operators} The evaluation of one
such dressed $d=6$ box diagram involves calculating the loop integral
but no kinematics, because the physical particles carrying real
momenta here are the proton and the meson, not the quarks. The loop
factor is not divergent and is of the same general form for every
channel; furthermore, as the heavy triplets are common to all diagrams
and the sfermion masses are assumed to be equal, the only factors in
the loop that vary from one channel to the next are the couplings and
masses associated with either the Wino or Higgsino. The remaining
variation from one diagram to the next depends entirely on the
particle flavors, which is apparent in the external fermions and
encoded in the $C_{ijkl}$ coefficients and the unitary matrices
involved in rotation to mass basis. Thus, I can write the operator for
any pertinent diagram as a generic Wino- or Higgsino coefficient times
one of several flavor-specific ``sub-operators''; the forms of the
general operators are
\begin{equation}
  \mathscr{O}_{\widetilde{W}} = \left(\frac{\alpha_2}{4\pi}\right)
  \left(\frac{1}{M_\mathcal{T}}\right)\, I\left(M_{\widetilde{W}},
  m_{\tilde{q}}\right) \mathscr{C}_{\widetilde{W}}^{\mathcal{A}}
\end{equation}
and
\begin{equation}
  \mathscr{O}_{\tilde{h}} = \left(\frac{1}{16\pi^2}\right)
  \left(\frac{1}{M_\mathcal{T}}\right)\, I\left(\mu,
  m_{\tilde{q}}\right) \mathscr{C}_{\tilde{h}}^{\mathcal{A}},
\end{equation}
where\footnote{One might notice that this expression for $I(a,b)$
differs from what is usually given in the literature for analogous
proton decay expressions; the discrepancy is due to my inclusion of
the universal mass assumption prior to evaluating the loop integral.
\vspace{2mm}}
\begin{equation}
  I(a,b) = \frac{a}{b^2\!-\!a^2} \left\{\,1\, +\,
  \frac{a^2}{b^2\!-\!a^2} \log\left(\frac{a}{b}\right) \right\},
  \nonumber
\end{equation}
and the sub-operators $\mathscr{C}^{\mathcal{A}}$ are\footnote{I do
not list the neutral Wino operator again here, but looking back at
eq.\,(\ref{eq:wino0-tot}), one can see that $\kappa = I
\left(M_{\widetilde{W}}, m_{\tilde{q}}\right)$.}
\begin{align}
  \mathscr{C}_{\widetilde{W}}^I &= \frac{1}{2} (u^T\, C^{-1}\, d_j)\,
  \widehat{C}^L_{[ij1]l}\, U^d_{ii'}\, U^\nu_{ll'}\,
  (d^T_{i'}\, C^{-1}\, \nu_{l'}) \nonumber \\
  \mathscr{C}_{\widetilde{W}}^{I\!I} &= \frac{1}{2} (u^T\, C^{-1}\, 
  e_l)\, \widehat{C}^L_{[1jk]l}\, U^d_{kk'}\, U^u_{j1}\,
  (d^T_{k'}\, C^{-1}\, u) \nonumber \\
  \mathscr{C}_{\widetilde{W}}^{I\!I\!I} &= -\frac{1}{2} (u^T\, C^{-1}\, 
  d_{k})\, \widehat{C}^L_{1[jk]l}\, U^u_{j1}\, U^e_{ll'}\, (u^T\, C^{-1}\,
  e_{l'}) \nonumber \\
  \mathscr{C}_{\widetilde{W}}^{I\!V} &= -\frac{1}{2} (d_j^T\, C^{-1}\, 
  \nu_l)\, \widehat{C}^L_{i[jk]l}\, U^d_{ii'}\, U^u_{k1}\,
  (d^T_{i'}\, C^{-1}\, u)
  \label{eq:CWops}
\end{align}
for the (charged) Wino,
\begin{align}
  \mathscr{C}_{\tilde{h}^\pm}^{I} &= (u^T\, C^{-1}\, e_l)\,
  \widehat{C}^L_{[1jk]l}\: y^{d\, \dagger}_{kk'}\, y^{u\,
  \dagger}_{j1}\, (d^{\,\mathcal{C}\,T}_{k'}\, C^{-1}\,
  u^{\mathcal{C}}) \nonumber \\ 
  \mathscr{C}_{\tilde{h}^\pm}^{I\!I} &= -(u^T\, C^{-1}\, d_{k})\,
  \widehat{C}^L_{1[jk]l}\: y^{u\, \dagger}_{j1}\, y^{e\,
  \dagger}_{ll'}\, (u^{\mathcal{C}\,T}\, C^{-1}\,
  e^{\mathcal{C}}_{l'}) \nonumber \\
  \mathscr{C}_{\tilde{h}^\pm}^{I\!I\!I} &=
  -(d_j^T\, C^{-1}\, \nu_l)\, \widehat{C}^L_{i[jk]l}\: y^{d\,
  \dagger}_{ii'}\, y^{u\, \dagger}_{k1}\,
  (d^{\,\mathcal{C}\,T}_{i'}\, C^{-1}\, u^{\mathcal{C}}) \nonumber \\ 
  \mathscr{C}_{\tilde{h}^\pm}^{I\!V} &= (u^{\mathcal{C}\,T}\, C^{-1}\, 
  d^{\,\mathcal{C}}_j)\, \widehat{C}^R_{[ij1]l}\: y^u_{ii'}\, 
  y^e_{ll'}\, (d^T_{i'}\, C^{-1}\, \nu_{l'}) \nonumber \\
  \mathscr{C}_{\tilde{h}^\pm}^V &= (u^{\mathcal{C}\,T}\, C^{-1}\,
  e^{\mathcal{C}}_l)\, \widehat{C}^R_{[1jk]l}\: y^u_{kk'}\, y^d_{j1}\,
  (d^T_{k'}\, C^{-1}\, u)
  \label{eq:Chpops}
\end{align}
for the charged Higgsino, and
\begin{align}
  \mathscr{C}_{\tilde{h}^0}^{I} &= -(u^T\, C^{-1}\, d_{k})\,
  \widehat{C}^L_{[ij1]l}\: y^{u\, \dagger}_{i1}\, y^{e\,
  \dagger}_{ll'}\, (u^{\mathcal{C}\,T}\, C^{-1}\,
  e^{\mathcal{C}}_{l'}) \nonumber \\
  \mathscr{C}_{\tilde{h}^0}^{I\!I} &= -(u^T\, C^{-1}\, e_l)\,
  \widehat{C}^L_{[1jk]l}\: y^{d\, \dagger}_{kk'}\, y^{u\,
  \dagger}_{j1}\, (d^{\,\mathcal{C}\,T}_{k'}\, C^{-1}\,
  u^{\mathcal{C}}) \nonumber \\ 
  \mathscr{C}_{\tilde{h}^0}^{I\!I\!I} &=
  (d_j^T\, C^{-1}\, \nu_l)\, \widehat{C}^L_{i[jk]l}\: y^{u\, \dagger}_{i1}\,
  y^{d\, \dagger}_{kk'}\, (u^{\mathcal{C}\,T}\, C^{-1}\, 
  d^{\,\mathcal{C}}_{k'}) \nonumber \\ 
  \mathscr{C}_{\tilde{h}^0}^{I\!V} &= -(u^{\mathcal{C}\,T}\, C^{-1}\, 
  d^{\,\mathcal{C}}_j)\, \widehat{C}^R_{[ij1]l}\: y^u_{i1}\, y^e_{ll'}\,
  (u^T\, C^{-1}\, e_{l}) \nonumber \\
  \mathscr{C}_{\tilde{h}^0}^V &= -(u^{\mathcal{C}\,T}\, C^{-1}\,
  e^{\mathcal{C}}_l)\, \widehat{C}^R_{[1jk]l}\: y^u_{k1}\, y^d_{jj'}\,
  (u^T\, C^{-1}\, d_{j'}) 
  \label{eq:Ch0ops}
\end{align} 
\noindent for the neutral Higgsino, where I have suppressed the color
indices everywhere. Again the hats on $\widehat{C}^{L,R}$ indicate
$\hat{h},\hat{f},\hat{g}$ are rotated to the mass basis, which I will
discuss in detail shortly. Note that $UDUE$ and $UDD\mathcal{N}$
operators generally differ by a sign, as do diagrams dressed by
$\tilde{h}^\pm_{u,d}$ and $\tilde{h}^0_{u,d}$; the latter difference
arises from the $SU(2)$ contraction in the SUSY Higgs mass term. These
sign differences create the potential for natural cancellation within
the absolute squared sums of interfering diagrams, and even for
cancellation of entire diagrams with each other in some cases. Also
note that the Yukawa couplings are Hermitian in this model, hence the
distinction above between $y^f$ and $y^{f\,\dagger}$ is not relevant
for this work.

I utilized two additional observations to simplify the implementation
of the above operators. First, I took values for the superpartner
masses such that $\mu,M_{\widetilde{W}} \ll m_{\tilde{q}}$, which
implies $I(a,b) \simeq a/b^2$. Also, because I am only interested in
the combined contribution of the three neutrinos, and because the
total contribution is the same whether one sums over flavor states or
mass states, I made the replacement $U^\nu_{ll'} \rightarrow
\delta_{ll'}$ for $\mathscr{C}_{\widetilde{W}}^I$ and took $l = l'~
\Rightarrow~ y^e_{ll'} = m^e_l/v_d$ for
$\mathscr{C}_{\tilde{h}^\pm}^{I\!V}$.

Since the unitary matrices $U^f$ do not appear in the SM (+ neutrino
sector) Lagrangian except in the CKM and PMNS combinations, the
non-diagonal SUSY Yukawas $y^f$ present in the
$\mathscr{C}^{\mathcal{A}}$ are not physically determined.
Fortunately in our GUT model full {\it high}-scale Yukawas are defined
by the completely determined fermion sector. Furthermore, it is known
that unitary matrices such as the CKM matrix experience only small
effects due to SUSY renormalization. Thus, since the low-scale
masses are of course known, I can define good approximations to the
SUSY Yukawas needed by using the high-scale $U^f$ to rotate
the diagonal mass couplings at the proton scale, divided by
the appropriate vevs:
\begin{equation}
  y^{u} = \frac{1}{v_u}\: U_u\, \left(\mathcal{M}^{\rm wk}_u\right)^D\,
  U_u^\dagger, \nonumber
\end{equation}
where $v_u = v \sin \beta$, or, in component notation,
\begin{equation}
  y^{u}_{ij} = \frac{1}{v_u}\: \sum\limits_k\, m^u_k\, U^u_{ik}\,
  U^{u\,*}_{jk}.
\end{equation}
I can similarly write
\begin{align}
  y^{d}_{ij} = \frac{1}{v_d}\: \sum\limits_k\, m^d_k\, U^d_{ik}\,
  U^{d\,*}_{jk} \nonumber \\
  y^{e}_{ij} = \frac{1}{v_d}\: \sum\limits_k\, m^e_k\, U^e_{ik}\,
  U^{e\,*}_{jk}, \nonumber
\end{align}
where $v_d = v \cos \beta$. Mass values used were taken from the
current PDG \cite{pdg}; light masses are run to the 1-GeV scale, top
and bottom masses are taken on-shell. Note that since the Yukawa
factors always appear in pairs of opposite flavor in the Higgsino
operators, and since $\frac{1}{\sin\beta\cos\beta} \simeq \tan\beta$
for large $\beta$, the Higgsino contributions to proton decay are
$\sim \frac{\tan^2\beta}{v^4}$ for this model.

There are generally two distinct mass-basis rotations possible for
each of the $UDUE\,\mbox{-}$, $UDD\mathcal{N}$-, and $U^{\mathcal{C}}
D^{\mathcal{C}} U^{\mathcal{C}} E^{\mathcal{C}}$-type triplet
operators; the difference between the two depends on whether the
operator is ``oriented'' ({\it i.e.}, in the diagram) such that the
lepton is a scalar. For a given orientation, a unitary matrix
corresponding to the fermionic field at one vertex in the triplet
operator will rotate every coupling present in ${C}^{L,R}$ pertaining
to that vertex; an analogous rotation will happen for the other vertex
in the operator. For example, looking at the $\pi^+ \bar{\nu_l}$
channel in Figure \ref{fig:compfd}(a), every coupling $\lambda_{ij}$
($\lambda = h,f,g$) from $C^L_{ijkl}$ present at the
$\tilde{\phi}_{\mathcal{T}}$ vertex will be rotated by some form of
$U^d$; similarly all $\lambda'_{kl}$ present at the
$\tilde{\phi}_{\overline{\mathcal{T}}}$ vertex will be rotated by some
$U^u$. The down quark field shown is a mass eigenstate quark resulting
from the unitary rotation, which one can interpret as a linear
combination of flavor eigenstates: $d_j = U^d_{jm}\, d'_{m}$, with
$j=1$; applying the same thinking to the up quark, one also has
$u_k^T = u'^T_{p}\, U^{u\,T}_{pk}$, with $k=1$. To work out the
details of the rotations, I start with the $d=5$ operator written
in terms of flavor states\footnote{Recall the scalars are both mass
and flavor eigenstates under the universal mass assumption. Also note
``$\lambda'$'' is again my name for the second generic coupling, and
the prime has nothing to do with basis; I will continue to use hats to
indicate rotated couplings.}, $\sum\nolimits_a x_a (\tilde{u}_i\,
\lambda^a_{im}\, d'_m) (u'_p \lambda'^a_{pl}\, \tilde{e}_l)$, where I
have expanded $C^L_{impl}$ in terms of its component couplings and
chosen the indices with the malice of forethought; now I can write
\begin{align}
  &\sum\limits_a x_a (\tilde{u}^T_i\, C^{-1} \lambda^a_{im}\, d'_{m})
  (u'^{\,T}_p\, \lambda'^a_{pl}\, C^{-1}\, \tilde{e}_l) \nonumber \\
  = ~ &\sum\limits_a x_a (\tilde{u}^T_i\, C^{-1}
  \underbrace{\lambda^a_{im}\, U^{d\, \dagger}_{mj}}_{\displaystyle
  \equiv \hat{\lambda}^a_{ij}}\, \underbrace{U^d_{jn}\,
  d'_{n}}_{\displaystyle d_j}) (\underbrace{u'^{\,T}_p\,
  U^{u\,T}_{pk}}_{\displaystyle u_k^T}\, \underbrace{U^{u\,*}_{kq}\,
  \lambda'^a_{ql}}_{\displaystyle \equiv \hat{\lambda}'^a_{kl}}\,
  C^{-1}\, \tilde{e}_l). \nonumber
\end{align}
Using the new definitions for $\hat{\lambda}$, one can see that the
rotated coefficient $\widehat{C}^L$ corresponding to the expression in
eq.\,(\ref{eq:Cs}) has become
\begin{align}
  \widehat{C}^L_{ijkl} &= x_0 \hat{h}_{ij} \hat{h}_{kl} + x_1
  \hat{f}_{ij} \hat{f}_{kl} - x_3 \hat{h}_{ij} \hat{f}_{kl} + \dots
  \nonumber \\ 
  &= x_0 (h\, U_d^\dagger)_{ij} (U_u^* h)_{kl}\, +\, x_1
  (f\, U_d^\dagger)_{ij} (U_u^* f)_{kl}\, -\, x_3 (h\,
  U_d^\dagger)_{ij} (U_u^* f)_{kl}\, +\, \dots
\end{align}
Note that this version of $\widehat{C}^L$ is only valid for $\tilde
u_id_ju_k\tilde e_l$-type operators, with this particular orientation
in the diagram; there is an analogous pair of rotations for $u\tilde
d\tilde ue$, as well as two each for $UDD\mathcal{N}$ and
$U^{\mathcal{C}} D^{\mathcal{C}} U^{\mathcal{C}} E^{\mathcal{C}}$,
giving a total of six possible schemes.

\subsection{From Quarks to Hadrons} As mentioned above, the composite
hadrons $p$ and $K,\pi$ (in addition to the lepton) carry physical
momenta in the proton decay process, {\it not} the ``external'',
``physical'' quarks seen in the dressed operators above. Therefore
one is in need of calculating a factor like $\bra{\mathrm{M}}
(qq)q \ket{p}$, where M\:$= K, \pi$ is the final meson
state. More explicitly these objects will look like
\begin{align*}
  &\bra{K^+} \epsilon_{\rho\sigma\tau} (u^\tau s^\sigma)_L\, d^\rho_L
  \ket{p} \\ &\bra{K^0} \epsilon_{\rho\sigma\tau} (u^\rho s^\tau)_R\,
  u^\sigma_L \ket{p} \\ &\bra{\pi^0} \epsilon_{\rho\sigma\tau}
  (u^\sigma d^\tau)_L\, u^\rho_R \ket{p} \\ & \qquad \qquad \vdots
\end{align*}
Such matrix elements are calculated using either chiral Lagrangian
methods or a three-point function (for M, $p$, and the $(qq)q$
operator) on the lattice; in either case, the result is determined in
part by a scaling parameter $\beta_H$ defined by $\bra{0} (qq)q
\ket{p(s)} = \beta_H P_L u_p (s)$, where $P_L$ is the left-chiral
projection matrix and $u_p (s)$ is the Dirac spinor for an incoming
proton of spin $s$. In principle $\beta_H$ is not necessarily the same
for cases where the quarks have different chiralities, but the values
usually differ only in sign, which is irrelevant when the entire
factor is squared in the decay width expression.

While lattice methods have advanced significantly since the early
years of SUSY GUT theory, there is still a substantial amount of
uncertainty present in the calculation of both $\beta_H$ and the
matrix element factors; some groups have even obtained contradictory
results when applying the two methods in the same work
\cite{gavela-king}. Some more recent works ({\it
e.g.}\,\cite{fukugita})~using more advanced statistics and larger
lattices seem to be converging on trustworthy answers, but it is still
normal to see results vary by factors of (1/2 - 5) for a single decay
mode from one method to the next, where the values for the matrix
elements themselves are $\mathcal{O}(10) \times \beta_H$. Thus I will
simply take the admittedly favorable approach of using
$\bra{\mathrm{M}} (qq)q \ket{p(s)} \sim \beta_H P u_p$ for all modes.

It is not uncommon to see values as low as $\beta_H = 0.003$ used in
other works calculating proton decay \cite{donoghue}, but while
calculated values have indeed varied as much as (0.003 - 0.65) over
the years \cite{fukugita}, the value is now most commonly found in the
range (0.006 - 0.03) \cite{claudson}, with a tendency to prefer
$\beta_H \sim 0.015$, as seen in \cite{fukugita}. Again, I will take a
slightly optimistic approach and use $\beta_H = 0.008$.

\subsection{The $p \rightarrow \mathrm{M} \bar\ell$ Effective Diagram
and the Decay Width of the Proton} 
\begin{figure}[t]
\begin{center}
  \includegraphics{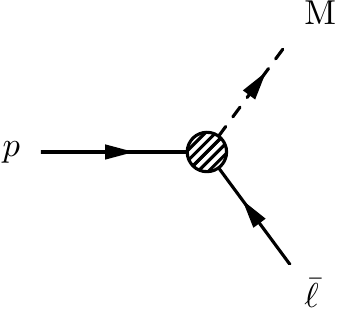} 
  \caption[Proton decay to a meson and an anti-lepton.]{Proton decay
  to a meson and an anti-lepton; the effective operator vertex
  contains hadronic and renormalization factors as well as the sum of
  all $d=6$ dressed operators contributing to the mode.} 
  \label{fig:pdecay}
\end{center}
\end{figure}

Ultimately it is a deceptively
simple two-body decay that I am calculating, as shown in Figure
\ref{fig:pdecay}. The corresponding decay width can be determined by
the usual phase-space integral expression:
\begin{equation}
  \Gamma = \frac{1}{2 M_p} \int \frac{\mathbf{d}^3
  \mathbf{p}}{(2\pi)^3\, 2 E_{\mathrm{M}}} \int \frac{\mathbf{d}^3
  \mathbf{p}}{(2\pi)^3\, 2 E_\ell}\, (2\pi)^4\; \delta^4(p_p -
  p_{\mathrm{M}} - p_\ell)\; \frac{1}{2}\, \sum\limits_s\; \lvert
  \mathscr{M} \rvert^{\,2}
\end{equation}
where in this case
\begin{equation}
  \frac{1}{2}\, \sum\limits_s\; \lvert \mathscr{M} \rvert^{\,2} =
  \frac{1}{2}\, \beta_H^2\, (A_L\,A_S)^2 \left( \lvert
  \mathscr{O}_{\widetilde{W}} \rvert^{\,2} + \lvert
  \mathscr{O}_{\tilde{h}} \rvert^{\,2} \right)\; \sum\limits_{s,s'}\;
  \lvert v^T_\ell (p_\ell,s)\, C^{-1}\, u_p (p_p,s') \rvert^{\,2}.
\end{equation}
\noindent The factors $A_L$ and $A_S$ arise due to the renormalization
of the $d=6$ dressed operators, from $M_p$ to $M_{\rm SUSY}$ and
$M_{\rm SUSY}$ to $M_{\rm U}$, respectively; their values have been
calculated in the literature as $A_L = 0.4$ and $A_S = 0.9 \mbox{-}
1.0$ \cite{hisano}. The spinor factor can be evaluated with the usual
trace methods; in the rest frame of the proton, where
$-\mathbf{p}_{\mathrm{M}} = \mathbf{p}_\ell \equiv \mathbf{p}$, and
utilizing $m_\ell^2 \ll \lvert \mathbf{p} \rvert^{\,2}$ (which is only
marginally valid for the muon but clearly so otherwise), the decay
width expression simplifies to
\begin{equation}
  \Gamma = \frac{1}{4\pi}\, \beta_H^2\, (A_L\,A_S)^2 \left( \lvert
  \mathscr{O}_{\widetilde{W}} \rvert^{\,2} + \lvert
  \mathscr{O}_{\tilde{h}} \rvert^{\,2} \right)\; \mathrm{p},
  \label{eq:gtotOs}
\end{equation}
where
\begin{equation}
  \mathrm{p} \equiv \lvert \mathbf{p} \rvert \simeq
  \frac{M_p}{2}\left(1 - \frac{m^2_{\mathrm{M}}}{M_p^2} \right).
\end{equation}
Note that p $\sim M_p/2$ for pion modes, but that value is
reduced by a factor of $\sim$\,25\% for kaon modes.

I now have all the pieces needed to write the working formulae for the
partial decay widths of the proton. Let me first define
$\mathrm{C}^{\mathcal{A}}$ as extended forms of the $C_{ijkl}$
by
\begin{align}
  \mathscr{C}_{\widetilde{W}}^{\mathcal{A}} =
  \mathrm{C}_{\widetilde{W}}^{\mathcal{A}} (qq)(q\ell) \nonumber \\
  \mathscr{C}_{\tilde{h}^\pm}^{\mathcal{A}} =
  \mathrm{C}_{\tilde{h}^\pm}^{\mathcal{A}} (qq)(q\ell) \nonumber \\
  \mathscr{C}_{\tilde{h}^0}^{\mathcal{A}} =
  \mathrm{C}_{\tilde{h}^0}^{\mathcal{A}} (qq)(q\ell),
\end{align}
so that these coefficients contain the $U^f$ or $y^f$ factors as well
as the $C^{L,R}$ of the $\mathscr{C}^{\mathcal{A}}$ operators in
(\ref{eq:CWops})-(\ref{eq:Ch0ops}). Now I can easily translate an
operator expression like
\begin{equation}
  \mathscr{O}_{\widetilde{W}} (K^+ \bar{\nu}) \simeq 
  \left(\frac{\alpha_2}{4\pi}\right) \frac{1}{M_\mathcal{T}}\,
  \left(\frac{M_{\widetilde{W}}}{m_{\tilde{q}}^2}\right)
  \{\mathscr{C}_{\widetilde{W}}^I + \mathscr{C}_{\widetilde{W}}^{I\!V}
  \}
  \label{eq:OKnuW}
\end{equation}
into a partial decay width statement,
\begin{equation}
  \Gamma_{\widetilde{W}} (p \rightarrow K^+ \bar{\nu}) \simeq
  \frac{1}{4\pi} \left(\frac{\alpha_2}{4\pi}\right)^2
  \frac{1}{M^2_{\mathcal{T}}} \left(\frac{M_{\widetilde{W}}}
  {m^2_{\tilde{q}}}\right)^2 \beta_H^2\,(A_L A_S)^2\, \mathrm{p}\:
  \lvert\, \mathrm{C}_{\widetilde{W}}^I +
  \mathrm{C}_{\widetilde{W}}^{I\!V}\, \rvert^2, 
  \label{eq:gKnuW}
\end{equation}
without losing either information or readability. Note though there is
still a ``black-box'' nature to the $\mathrm{C}^{\mathcal{A}}$ (it was
there in the $\mathscr{C}^{\mathcal{A}}$ operators as well), in that
without specifying the generation indices of the external $d_{j,i'}$
quarks, the sums in eqs.\,(\ref{eq:OKnuW}) and (\ref{eq:gKnuW}) could
just as easily apply to $\pi^+ \bar{\nu}$. Furthermore, there are at
least several channels present in each $\mathscr{C}^{\mathcal{A}}$
operator that contribute to any one mode, which are determined
uniquely by the generations of the internal sfermions in addition to
those of the external quarks.\footnote{Indeed I could have defined the
coefficients with six indices: $\mathrm{C}^{\mathcal{A}}_{ijklmn}$,
thereby creating a means of alleviating all degeneracy, but I do not
expect such information-dense objects to be so enlightening to
readers, especially since for most modes, at least the
Higgsino-dressed expression would devolve into an entire pageful of
terms corresponding to the individual channels.} If the reader wishes
to examine the decay widths at the full level of detail, he or she
should utilize these expressions along with the operators in
eqs.\,(\ref{eq:CWops})-(\ref{eq:Ch0ops}) and the diagrams in Appendix
\ref{fds}.

All remaining limitations aside, I can now present relatively compact
and intelligible expressions for the Wino- and Higgsino-dressed
partial decay widths of the proton for generic mode $p \rightarrow
\mathrm{M} \bar\ell$:
\begin{align}
  \label{eq:gammaW}
  \Gamma_{\widetilde{W}} (p \rightarrow \mathrm{M} \bar\ell) \simeq
  \frac{1}{4\pi} \left(\frac{\alpha_2}{4\pi}\right)^2
  \frac{1}{M^2_{\mathcal{T}}} \left(\frac{M_{\widetilde{W}}}
  {m^2_{\tilde{q}}}\right)^2 \beta_H^2\,(A_L A_S)^2\, \mathrm{p}\:
  \Big\lvert\! \sum\limits_{\mathcal{A} \in \mathrm{M} \bar\ell}\!
  \mathrm{C}_{\widetilde{W}}^{\mathcal{A}}\, \Big\rvert^{2} \\
  \Gamma_{\tilde{h}} (p \rightarrow \mathrm{M} \bar\ell) \simeq
  \frac{1}{4\pi} \left(\frac{1}{16\pi^2}\right)^2
  \frac{1}{M^2_{\mathcal{T}}} \left(\frac{\mu}
  {m^2_{\tilde{q}}}\right)^2 \beta_H^2\,(A_L A_S)^2\, \mathrm{p}\:
  \Big\lvert\! \sum\limits_{\mathcal{A} \in \mathrm{M} \bar\ell}\!
  \mathrm{C}_{\tilde{h}}^{\mathcal{A}} \Big\rvert^{2}. 
  \label{eq:gammah}
\end{align}
For the numerical analysis, I used the generic values $M_{\cal T} = 2
\!\times\! 10^{16}$\,GeV, $M_{\widetilde{W}} = \mu = 100$\,GeV, and
$m_{\tilde{q}} = 3$\,TeV. Also, let me repeat here that because of the
two SUSY Yukawa coupling factors in the
$\mathrm{C}_{\tilde{h}}^{\mathcal{A}}$, which always come in opposite
flavor,
\begin{equation}
  \Gamma_{\tilde{h}} \propto \left(\frac{1}{v^2
  \sin\beta\cos\beta} \right)^2 \sim
  \frac{\tan^2\beta}{v^4}. \nonumber
\end{equation}

Before moving on to the fermion sector fit results, let me remark that
because the Higgsinos vertices change the chiralities of the outgoing
fermions, there can be no interference between Wino- and
Higgsino-dressed diagrams, as implied by the notation in
eq.\,(\ref{eq:gtotOs}); however, since diagrams for the right-handed
$C^R$ operators have outgoing \emph{left-handed} fermions by the same
Higgsino mechanism, diagrams for $C^R$- and $C^L$-type operators with
the same external particles of matching chiralities \emph{do}
interfere with each other, and so all such contributions to a given
mode do in fact go into the same absolute-squared sum factor, as
suggested by eq.\,(\ref{eq:gammah}). \newpage


%
%

\chapter{Results of the Analysis}\label{6-results}

\section{Fitting the Fermion Mass Matrices} \label{fit} Diagonalizing
the mass matrices given in eq.\,(\ref{eq:mass}), with the Yukawa
textures shown in (\ref{eq:y}), gives the GUT-scale fermion masses and
mixing angles for a given set of values for the mass matrix parameters
$h_{ij}$, $f_{ij}$, $r_i$, etc. In order to find the best fit to the
experimental data, I used the {\tt Minuit} tool library for Python
\cite{minuit,python} to minimize the sum of chi-squares for
the mass-squared differences $\Delta m_{21}^2$ (aka $\Delta
m_\odot^2$) and $\Delta m_{32}^2$ (aka $\Delta m_{\rm atm}^2$) and the
PMNS mixing angles in the neutrino sector as well as the mass
eigenvalues and CKM mixing angles in the charged-fermion sector.
Type-I and type-II seesaw neutrino masses were each fit independently,
so I report the results for each separately.

Note that throughout the analysis, I have taken $v_u = 117.8$\,GeV,
which is calculated with $\tan\beta = 55$ and for $v$ run to
the GUT scale \cite{das}. The corresponding value for the down-type
vev is $v_d = 2.26$\,GeV.

Threshold corrections at the SUSY scale are $\propto\tan\beta$, and so
should be large in this analysis \cite{poko}. The most substantial
correction is to the bottom quark mass, which is dominated by gluino
and chargino loop contributions; this correction also induces changes
to the CKM matrix elements involving the third generation. The
explicit forms of these corrections can be seen in a previous work on
a related model \cite{ddms}. Additionally, smaller off-diagonal
threshold corrections to the third generation parts of $\mathcal{M}_d$
result in small corrections to the down and strange masses as well as
further adjustments to the CKM elements. All such corrections can be
parametrized in the model by
\begin{align}
  \mathcal{M}'_d = \mathcal{M}_d + 
    \frac{r_1}{\tan\beta}\left(\begin{array}{ccc}
    0 & 0 & \delta V_{ub}\\ 0 & 0 & \delta V_{cb}\\
    \delta V_{ub} & \delta V_{cb} & \delta m_b \end{array}\right),
\end{align}
\smallskip
where $\mathcal{M}_d$ is given by eq.\,(\ref{eq:mass}). If I simply
take this augmented form for $\mathcal{M}_d$ as part of the model
input, the $\delta$ parameters are fixed by the mass matrix fitting,
which results in implied constraints on certain SUSY parameters and
the mass values that depend on them, namely, the Higgs and the light
stop and sbottom masses. This entire prescription and its implications
were considered in detail in \cite{ddms}, and in comparing to that
work, one can see that for large $\tan\beta$ and relatively small
threshold corrections, the resulting constraints on the Higgs and
squark masses are less interesting, so I will not consider them in
more detail for this analysis.

\subsection{Fit Results for Type II Seesaw} If one breaks $SO(10)$ and
$B-L$ together at $v_R \gsim 10^{17}$\,GeV, and sets the vev $v_L \sim
1$\,eV through a tuning of the $SU(5)$ {\bf 15} mass term for
$\bar\Delta_L$, then the $v_L$ term in eq.\,(\ref{eq:neu}) dominates
over the type-I contribution by 2-4 orders of magnitude in the
neutrino mass matrix; therefore eq.\,(\ref{eq:neu}) reduces to
\begin{equation}
  {\cal M}_\nu \simeq v_L f
\end{equation}
Using this prescription, I find a fairly large parameter space for
which the sum of chi-squares is quite low, although some of the output
values, such as $\theta_{13}$ and the down and bottom masses, are
quite sensitive to the variation in the minima. This is problematic
for $\theta_{13}$ especially, since it is known to high experimental
precision \cite{dayabay}. Tables \ref{table:paramsII} and
\ref{table:fitII} display the properties of one of the more favorable
fits; Table \ref{table:paramsII} gives the values for the adjusted
model input parameters, and Table \ref{table:fitII} gives the
corresponding output values for the fermion parameters, with
experimentally measured values included for comparison. Note that the
down quark mass is seemingly a bit low, which seems to be a general
feature in this model, but I will discuss in the next section why this
is not a problem. The precise value of $v_L$ for this fit is
1.316\,eV, which sets the overall neutrino mass scale at
$m_3 \sim 0.05$\,eV.
\begin{table}[t]
\begin{center}
  \begin{tabular}{||c|c||c|c||}\hline\hline
    $M$ (GeV) &               106.6& $r_1/ \tan\beta$ &    0.014601\\ 
    $f_{11}$ (GeV) &      -0.045564& $r_2$ &              0.0090315\\
    $f_{12}$ (GeV) &       0.048871& $r_3$ &                 1.154 \\ 
    $f_{13}$ (GeV) &       -0.59148& $c_e$ &                -2.5342\\ 
    $f_{22}$ (GeV) &       -2.06035& $c_\nu$ &                  n/a\\     
    $f_{23}$ (GeV) &        -1.4013& $\delta m_b$ (GeV) &   -22.740\\
    $f_{33}$ (GeV) &       -1.40644& $\delta V_{cb}$ (GeV) & 1.2237\\
    $g_{12}$ (GeV) &       0.018797& $\delta V_{ub}$ (GeV) & 4.2783\\
    $g_{13}$ (GeV) &       -0.92510& & \\
    $g_{23}$ (GeV) &        -3.8353& & \\
    \hline\hline
  \end{tabular}
  \caption[Best fit values for the model parameters at the GUT scale
  with type-II seesaw.]{Best fit values for the model parameters at
  the GUT scale with type-II seesaw. Note that $c_\nu$, which appears
  in the Dirac neutrino mass contribution to the type-I term, is not
  relevant for type-II.}
\label{table:paramsII}
\end{center}
\end{table}
\begin{table}[t]
\begin{center}
  \begin{tabular}{||c|c|c||c|c|c||}\hline\hline
    & best fit & exp value & & best fit & exp value\\ \hline
    $m_u$ (MeV) &       0.7172 &    $0.72^{+0.12}_{-0.15}$ &
    $V_{us}$ &          0.2245 &    $0.2243\pm 0.0016$\\
    $m_c$ (MeV) &       213.8 &    $210.5^{+15.1}_{-21.2}$ &
    $V_{ub}$ &          0.00326&    $0.0032\pm 0.0005$\\
    $m_t$ (GeV) &       106.8  &    $95^{+69}_{-21}$ &
    $V_{cb}$ &          0.0349 &    $0.0351\pm 0.0013$\\
    $m_d$ (MeV) &       0.8827 &    $1.5^{+0.4}_{-0.2}$ &
    $J \times 10^{-5}$ & 2.38 &    $2.2\pm 0.6$\\
    $m_s$ (MeV) &       34.04 &    $29.8^{+4.18}_{-4.5}$ &
    $\Delta m_{21}^2 / \Delta m_{32}^2$ & 0.03065& $0.0309\pm 0.0015$\\
    $m_b$ (GeV) &       1.209 &    $1.42^{+0.48}_{-0.19}$ &
    $\theta_{13}~(^\circ)$ & 9.057 &$8.88\pm 0.385$\\
    $m_e$ (MeV) &       0.3565 &    $0.3565^{+0.0002}_{-0.001}$ &
    $\theta_{12}~(^\circ)$ & 33.01 &$33.5\pm 0.8$\\
    $m_\mu$ (MeV) &     75.297 &    $75.29^{+0.05}_{-0.19}$ &
    $\theta_{23}~(^\circ)$ & 47.70 &$44.1\pm 3.06$\\
    $m_\tau$ (GeV) &    1.635 &    $1.63^{+0.04}_{-0.03}$ &
    $\delta_{\rm CP}~(^\circ)$ &   -7.506 & \\
    \hline
    & & & $\sum \chi^2$ & 6.0 & \\
    \hline\hline
  \end{tabular}
  \caption[Best fit values for the charged fermion masses,
  solar-to-atmospheric mass squared ratio, and CKM and PMNS mixing
  parameters for the fit with Type-II seesaw.]{Best fit values for the
  charged fermion masses, solar-to-atmospheric mass squared ratio, and
  CKM and PMNS mixing parameters for the fit with Type-II seesaw. The
  $1\sigma$ experimental values are also shown for comparison
  \cite{das}, \cite{pdg}, where masses and mixings are extrapolated to
  the GUT scale using the MSSM RGEs. Note that the fit values for the
  bottom quark mass and the CKM mixing parameters involving the third
  generation shown here include the SUSY-threshold corrections} 
\label{table:fitII}
\end{center}
\end{table}

In order to calculate the $C_{ijkl}$ proton decay coefficients, as
well as for use in the neutrino mass matrix (\ref{eq:neu}), I needed
to determine the ``raw'' Yukawa couplings, $h,f,g$, from the
dimensionful couplings, $\tilde{h},\tilde{f},\tilde{g}$, of the mass
matrices given in eq.\,(\ref{eq:mass}), which are obtained directly
from the fit; to do so I need to extract the absorbed vev $v_u$ and
doublet mixing parameters $f({\cal U}_{I\!J}, {\cal V}_{I\!J})$
discussed in section \ref{model}. There is some freedom in the values
of those mixing elements from the viewpoint of this predominantly
phenomenological analysis, but they are constrained by both unitarity
and the ratios $r_i$ and $c_\ell$, which have been fixed by the
fermion fit. Again, see \cite{olddmm} for details, or see \cite{ddms}
for an example of such a calculation. The resulting dimensionless
couplings corresponding to this type-II fit are
\smallskip
\begin{gather}
  h = \left( \begin{array}{ccc}
     0 & &\\ & 0 & \\ & & 1.207
  \end{array}\right); \qquad
  f = \left( \begin{array}{ccc}
    -0.00053748 &  0.00057649 & -0.0069772 \\
     0.00057649 & -0.024304 &  -0.016530 \\
    -0.0069772 &  -0.016530 & -0.0165906 \\
  \end{array} \right) \nonumber \\[4mm]
  g = i \left(\begin{array}{ccc}
     0          & 0.00033485 & -0.016480 \\
    -0.00033485 & 0          & -0.0683214 \\
     0.016480   & 0.0683214  &  0        \end{array} \right) 
\label{hfgII}
\end{gather}
Note that in addition to $f_{11} \sim f_{12} \sim 0$, this
fit satisfies $g_{12},f_{13} \ll 1$ as is desired for proton decay.

\subsection{Fit Results for Type I Seesaw} If one instead takes $v_R
\lsim 10^{16}$\,GeV and $v_L \sim 1$\,meV, then the type-I contribution
is dominant over the type-II contribution, and eq.\,(\ref{eq:neu})
becomes
\begin{equation}
  {\cal M}_\nu \simeq - {\cal M}_{\nu_D}\left(v_R f\right)^{-1}\left(
  {\cal M}_{\nu_D}\right)^T,
\end{equation}
In this case, initial searches again showed that certain output
parameters were quite sensitive to the input and were often in
contention with each other or with the {\it de facto} upper bounds on
the $f_{ij}$ needed for proton decay. In the first cluster of minima
found by the fitting, the output values for one or more of charm mass,
bottom mass, or $\theta_{23}$ was much too small; furthermore, those
results came with odd, large tunings of certain input parameters, such
as $c_{e,\nu} \sim {\cal O}(100)$ or $\delta m_b > 40$\,GeV. The
addition of a small type-II correction to the neutrino matrix led me
to a new swath of parameter space, and ultimately I found a new
cluster of minima that did not require the correction. Table
\ref{table:paramsI} gives the values for the adjusted model input
parameters for one such pure type-I fit, and Table \ref{table:fitI}
gives the corresponding output values for the fermion parameters. Fits
in this swath of parameter space still have $c_\nu \sim 50$ and
$\delta m_b \sim 25$\,GeV, but this value for $c_\nu$, while slightly
strange, can be accommodated by the freedom in the doublet mixing
parameters, and such a value for the largest SUSY threshold correction
is actually quite moderate for large $\tan\beta$. The precise value
for the $\bar\Delta_{R}$ vev in this fit is $v_R = 1.21\!
\times\! 10^{15}$\,GeV.

Note also that the top and strange masses are quite a bit lower than
in the type-II fit; however, note I have also quoted different
experimental values with which agreement is maintained. The
differences here come from an update to the work in \cite{das} in
determining two-loop MSSM RGEs for fermion masses. The update
\cite{bora} reports notably lower masses for all the quarks at
$\tan\beta = 55$ and $\mu = 2.0 \!\times\! 10^{16}$\,GeV, especially
for the up, down, strange, and top masses, due to updates in initial
values and methodology. Hence, one should not give the specific values
too much weight in such a fit, and I do not consider the reported
differences to be significant. This same thinking applies for the
type-II down mass value in Table \ref{table:fitII}.
\begin{table}[t]
\begin{center}
  \begin{tabular}{||c|c||c|c||}\hline\hline
    $M$ (GeV) &           76.10& $r_1/ \tan\beta$ &      0.024701\\
    $f_{11}$ (GeV) &   0.010130& $r_2$ &                  0.24414\\
    $f_{12}$ (GeV) &  -0.089576& $r_3$ &                  0.00600\\
    $f_{13}$ (GeV) &    0.93973& $c_e$ &                  -3.3279\\
    $f_{22}$ (GeV) &     0.8659& $c_\nu$ &                 45.218\\
    $f_{23}$ (GeV) &     1.4884& $\delta m_b$ (GeV) &     -28.000\\
    $f_{33}$ (GeV) &     3.5495& $\delta V_{cb}$ (GeV) & -0.84394\\
    $g_{12}$ (GeV) &    0.20048& $\delta V_{ub}$ (GeV) &  0.51486\\
    $g_{13}$ (GeV) &    0.05352& & \\
    $g_{23}$ (GeV) &    0.35153& & \\
    \hline\hline
  \end{tabular}
  \caption[Best fit values for the model parameters at the GUT scale
  with type-I seesaw.]{Best fit values for the model parameters at the
  GUT scale with type-I seesaw.}
\label{table:paramsI}
\end{center}
\end{table}
\begin{table}[t]
\begin{center}
  \begin{tabular}{||c|c|c||c|c|c||}\hline\hline
    & best fit & exp value & & best fit & exp value\\ \hline
    $m_u$ (MeV) &       0.72155&    $0.72^{+0.12}_{-0.15}$ &
    $V_{us}$ &          0.2240 &    $0.2243\pm 0.0016$\\
    $m_c$ (MeV) &        212.2 &    $210.5^{+15.1}_{-21.2}$ &
    $V_{ub}$ &          0.00310&    $0.0032\pm 0.0005$\\
    $m_t$ (GeV) &        76.97 &    $80.45^{+2.9\,*}_{-2.6}$ &
    $V_{cb}$ &          0.0352 &    $0.0351\pm 0.0013$\\
    $m_d$ (MeV) &        1.189 &    $0.930\pm 0.38^*$ &
    $J \times 10^{-5}$ & 2.230 &    $2.2\pm 0.6$\\
    $m_s$ (MeV) &        20.81 &    $17.6^{+4.9\,*}_{-4.7}$ &
    $\Delta m_{21}^2 / \Delta m_{32}^2$ & 0.0309 & $0.0309\pm 0.0015$\\
    $m_b$ (GeV) &        1.278 &    $1.24\pm 0.06^*$ &
    $\theta_{13}~(^\circ)$ & 8.828 &$8.88\pm 0.385$\\
    $m_e$ (MeV) &       0.3565 &    $0.3565^{+0.0002}_{-0.001}$ &
    $\theta_{12}~(^\circ)$ & 33.58 &$33.5\pm 0.8$\\
    $m_\mu$ (MeV) &      75.29 &    $75.29^{+0.05}_{-0.19}$ &
    $\theta_{23}~(^\circ)$ & 41.76 &$44.1\pm 3.06$\\
    $m_\tau$ (GeV) &     1.627 &    $1.63^{+0.04}_{-0.03}$ &
    $\delta_{\rm CP}~(^\circ)$ &   -46.3 & \\
    \hline
    & & & $\sum \chi^2$ & 1.75& \\
    \hline\hline
  \end{tabular}
  \caption[Best fit values for the charged fermion masses,
  solar-to-atmospheric mass squared ratio, and CKM and PMNS mixing
  parameters for the fit with Type-I seesaw.]{Best fit values for the
  charged fermion masses, solar-to-atmospheric mass squared ratio, and
  CKM and PMNS mixing parameters for the fit with Type-I seesaw. The
  $1\sigma$ experimental values are shown \cite{das} ($^*$\,-\,from
  \cite{bora} instead), \cite{pdg}; masses and mixings are
  extrapolated to the GUT scale using the MSSM RGEs.  Note that again
  that pertinent fit values include threshold corrections.} 
\label{table:fitI}
\end{center}
\end{table}

Again I need to determine the raw Yukawa couplings for proton decay
analysis. The resulting couplings corresponding to this type-I fit
are \smallskip
\begin{gather}
  h = \left( \begin{array}{ccc}
     0 & &\\ & 0 & \\ & & 1.6152
  \end{array}\right) \qquad
  f = \left( \begin{array}{ccc}
     0.0001623 &  -0.00143525 &  0.01505699 \\
    -0.00143525 &  0.01387415 &  0.02384774 \\
     0.01505699 &  0.02384774 &  0.05687217
  \end{array} \right) \nonumber \\[4mm]
  g = i \left(\begin{array}{ccc}
     0          & 0.0068081  & 0.0018175 \\
    -0.0068081  & 0          & 0.0119376 \\
    -0.0018175  &-0.0119376  & 0        \end{array} \right) 
\label{eq:hfgI}
\end{gather} 

\noindent Here, one still finds $f_{11} \sim 0$, but each of $f_{12}$,
$f_{13}$, and $g_{12}$ is larger by an order of magnitude than in the
type-II case, which is thought to be unfavorable for proton decay. At
the same time, $g_{13}$ and $g_{23}$ are smaller by an order of
magnitude, so it is not clear that the net benefit lost is
substantial. In the end, a different distinction will give way to
success for this type-I fit; I will discuss those details in the next
section.

\section{Results of Calculating Proton Partial Lifetimes} \label{pfit}
In order to give an actual number for any decay width, in addition to
choosing representative values for the triplet, sfermion, and Wino or
Higgsino masses, I also need values for the $x_i$ and $y_i$ triplet
mixing parameters in order to calculate the $C_{ijkl}$ values. Recall
that the {\bf 10} mass parameter $x_0$ must be fixed at ${\cal O}(1)$
to allow the SUSY Higgs fields to be light; the remaining mixing
parameters are functions of many undetermined GUT-scale masses and
couplings found in the full superpotential for the heavy Higgs fields,
the details of which can be seen in \cite{aulgarg}. There are nearly
as many of those GUT parameters as there are independent $x$s and
$y$s, so it is not unreasonable to simply treat the latter as free
parameters.

Ideally, one would find that the width for any particular mode would
be essentially independent of those parameter values, {\it i.e.}, that
for arbitrary choices $0 < |x_i|,|y_i| < 1$, devoid of unlucky
relationships leading to severe enhancements, all mode lifetimes would
be comfortably clear of the experimentally determined lower limits,
given in Table \ref{table:explims}. The reality is quite bleak in
comparison. For a typical GUT model, if the proton decay lifetimes can
be satisfied at all, one is required to choose $x$ and $y$ values very
carefully such that either individual $C_{ijkl}$ or $\Big\lvert\!
\sum \mathrm{C}^{\mathcal{A}} \Big\rvert$ are small through
cancellations among terms. These tunings may need to be several orders
of magnitude in size ({\it e.g.}, $\mathrm{C}^{\cal A} =
-\mathrm{C}^{\cal B} + {\cal O}(10^{-3})$), and many such
relationships may be needed. 
\begin{table}[t]
\begin{center}
  \begin{tabular}{||c|c||}\hline\hline
    decay mode & $\tau$ exp lower limit (yrs) \\ \hline 
    $p \rightarrow K^+ \bar\nu$   &  $6.0 \!\times\! 10^{33}$ \\
    $p \rightarrow K^0 e^+$       &  $1.0 \!\times\! 10^{33}$ \\
    $p \rightarrow K^0 \mu^+$     &  $1.3 \!\times\! 10^{33}$ \\
    $p \rightarrow \pi^+ \bar\nu$ &  $2.7 \!\times\! 10^{32}$ \\
    $p \rightarrow \pi^0 e^+$     &  $1.3 \!\times\! 10^{34}$ \\
    $p \rightarrow \pi^0 \mu^+$   &  $1.0 \!\times\! 10^{34}$ \\
    \hline\hline
  \end{tabular}
  \caption[Experimental lower limits on the partial lifetimes of
  dominant proton decay modes.]{Experimentally determined lower limits
  \cite{babuexp} on the partial lifetimes of dominant proton decay
  modes considered in this work.}
\label{table:explims}
\end{center}
\end{table}

The Yukawa textures shown in eq.\,(\ref{eq:y}) are intended to
naturally suppress the values of some crucial $C_{ijkl}$ values so
that the need for such extreme tuning is alleviated. In order to test
the ansatz, I ``simply'' needed to find a set of values for the mixing
parameters yielding partial decay widths that satisfy the experimental
constraints; the difficulty in determining those values inversely
corresponds to success of the ansatz. If the ansatz does indeed work
optimally, I should be able to choose arbitrary $x_i$ and $y_i$ values
as suggested above.  Realistically though, the authors of
\cite{dmm1302} and I expected some searching for a valid region of
parameter space to be required.

To perform that search, I designed a second Python program to find
maximum partial lifetimes based on user-defined mixing values as well
as the raw Yukawa couplings fixed by the fermion sector fitting.
Parameter values are defined on a per-trial basis for any number of
trials. I started with the most optimistic case by generating random
initial values for $x_i$ and $y_i$ (but $x_0 \sim 1$ fixed), with the
decay width for $K^+ \bar\nu$ minimized by adjusting those values in
each trial. The minimization was again performed using the {\tt
Minuit} tool library.

The search based on fully random initial values was unsuccessful, in
that the $K^+ \bar\nu$ mode lifetime consistently fell in the $10^{31
\mbox{-} 32}$\,year-range for the type-II solution and was 
\noindent typically $\sim\!1 \!\times\! 10^{33}$\,years for the type-I
case;\footnote{The {\tt Minuit} tool used, Migrad, works using a local
gradient-based algorithm, so that in large parameter spaces, initial
values are crucial in locating global minima.} at the same time
however all five other modes in question were usually near or above
their respective limits for those same arbitrary mixing values. Hence
it was clear even with the $K^+ \bar\nu$ mode failure that the ansatz
was having the desired effect to some extent. Also, note that this
type-I solution for $K^+ \bar\nu$ was short of the limit by only about
a factor of five.  This is surprising since the type-I-based Yukawas
reported in eq.\,(\ref{eq:hfgI}) fell short of meeting the ansatz
criteria. Given the differing behaviors of the two solutions, I will
report the remaining details in separate subsections once again.

\subsection{Proton Partial Lifetimes for Type II Seesaw } To further
explore the properties of the ``default behavior'' of the lifetime
values in the model, I considered the case in which $x_0 \sim 1$ and
all other $x_i$ and $y_i$ are set to zero; one can see this case as
defining a baseline for the partial lifetimes, in that any $x_0$
terms in the $C_{ijkl}$ not suppressed by the Yukawa textures are
necessarily large, and whereas problematic contributions from some
other $x_k$ with $k\neq 0$ may be suppressed simply by setting $x_k
\ll 1$, the $x_0$ contributions can be mitigated {\it only} through
cancellation. 
\begin{table}[t]
\begin{center}
  \begin{tabular}{||c|c|c||}\hline\hline
    decay mode & baseline for $\tau$ (yrs) & baseline in ref.
    \cite{altarelli} (yrs) \\ \hline 
    $p \rightarrow K^+ \bar\nu$   &  $8.29 \!\times\! 10^{31}$ &  
        $6.38 \!\times\! 10^{28}$ \\
    $p \rightarrow K^0 e^+$       &  $9.73 \!\times\! 10^{34}$ &
        $2.52 \!\times\! 10^{30}$ \\
    $p \rightarrow K^0 \mu^+$     &  $5.68 \!\times\! 10^{33}$ &
        $6.15 \!\times\! 10^{29}$ \\
    $p \rightarrow \pi^+ \bar\nu$ &  $4.25 \!\times\! 10^{33}$ &
        $4.45 \!\times\! 10^{29}$ \\
    $p \rightarrow \pi^0 e^+$     &  $1.08 \!\times\! 10^{36}$ &
        $3.90 \!\times\! 10^{30}$ \\
    $p \rightarrow \pi^0 \mu^+$   &  $6.45 \!\times\! 10^{34}$ &
        $6.00 \!\times\! 10^{29}$ \\
    \hline\hline
  \end{tabular}
  \caption[Hypothetical baseline partial lifetimes determined using
  type-II solution Yukawas.]{Hypothetical baseline partial lifetimes
  determined using type-II solution Yukawas and $x_0 = 0.95$ with all
  other $x_i,y_i = 0$. For comparison, I give the analogous results
  for calculation using type-II Yukawas from the 2010 paper by
  Altarelli and Blankenburg \cite{altarelli}, which use general Yukawa
  texture.  Note in comparing with Table \ref{table:explims} that for
  our model, only the $K^+ \bar\nu$ mode fails to satisfy the lower
  limit, while all modes are well below the limits for the model in
  \cite{altarelli}.}
\label{table:baselineII}
\end{center}
\end{table}

The corresponding baseline lifetimes for the dominant modes in the
type-II case are given in Table \ref{table:baselineII}. One can see
that the $K^+ \bar\nu$ mode decay width must be lowered by two orders
of magnitude through cancellation of $x_0$ terms by the others.  Since
it is $\lvert \mathrm{C} \rvert^{\,2}$ that appears in the decay width
expressions, the needed cancellation amounts to an ${\cal O}(10^{-1})$
tuning among the ${\rm C}^{\cal A}$ factors.  Furthermore, as it would
be equally unnatural to see $x_k \ll 1$ for all $k\neq 0$, one should
expect ${\cal O}(1)$ cancellations to be present anyway; therefore,
the needed ``tuning'' is little more than a very ordinary restriction
of parameter space.

In order to elucidate the significance of the improvement created by
the Yukawa ansatz, consider the outcome of this baseline calculation
for a case with more general Yukawa texture. The model from a 2010
paper by G. Altarelli and G. Blankenburg \cite{altarelli} has the same
{\bf 10}-{\bf 126}-{\bf 120} Yukawa structure but with general $h$ and
$g$ as in eq.\,(\ref{eq:y0}) and a tri-bimaximal $f$ having no
hierarchical texture.\footnote{This specific model has already been
ruled out due to $\theta_{13} \sim 6\mbox{-}7^\circ$ typical of
tri-bimaximal models.} Using the parameters reported to give a
successful fermion fit in the work (\emph{see footnote}), I obtain the
baseline results shown in the final column of Table
\ref{table:baselineII}. One can see here that lifetimes for all modes
are far below the experimental limits, by factors of ${\cal
O}(10^{3\mbox{-}5})$; hence cancellation among the ${\rm C}^{\cal A}$
factors must be ${\cal O}(10^{-2\mbox{-}4})$. Such sensitive
relationships among these factors are seemingly less natural than the
result from our model in the absence of some new symmetry.
\begin{figure}[t]
        \includegraphics[width=7.8cm]{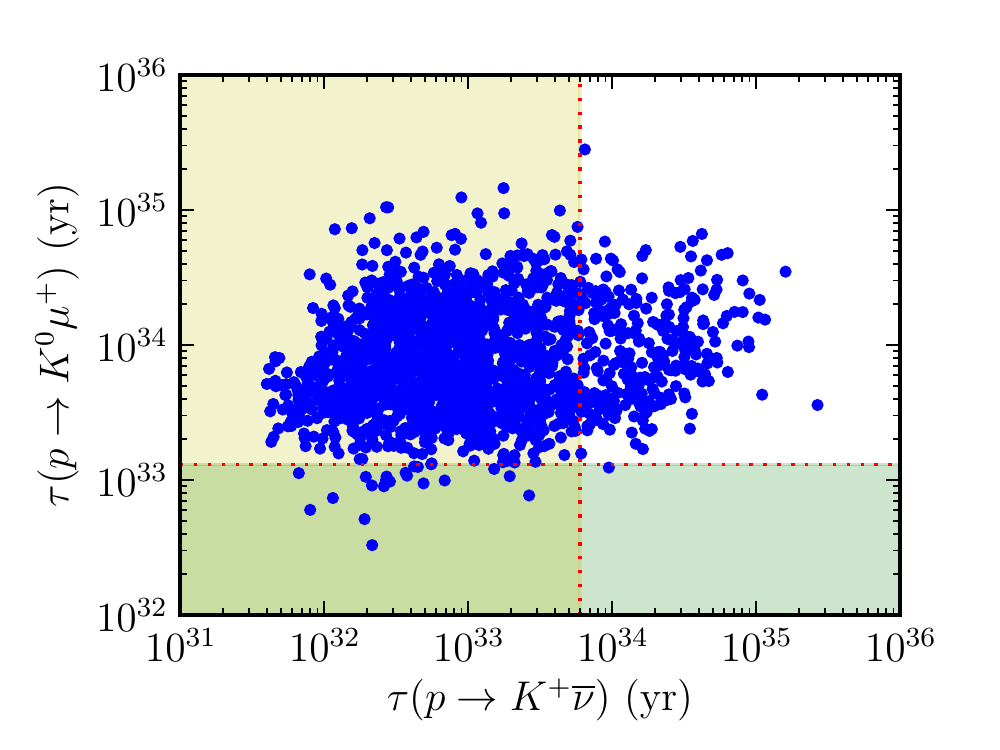}
        \includegraphics[width=7.8cm]{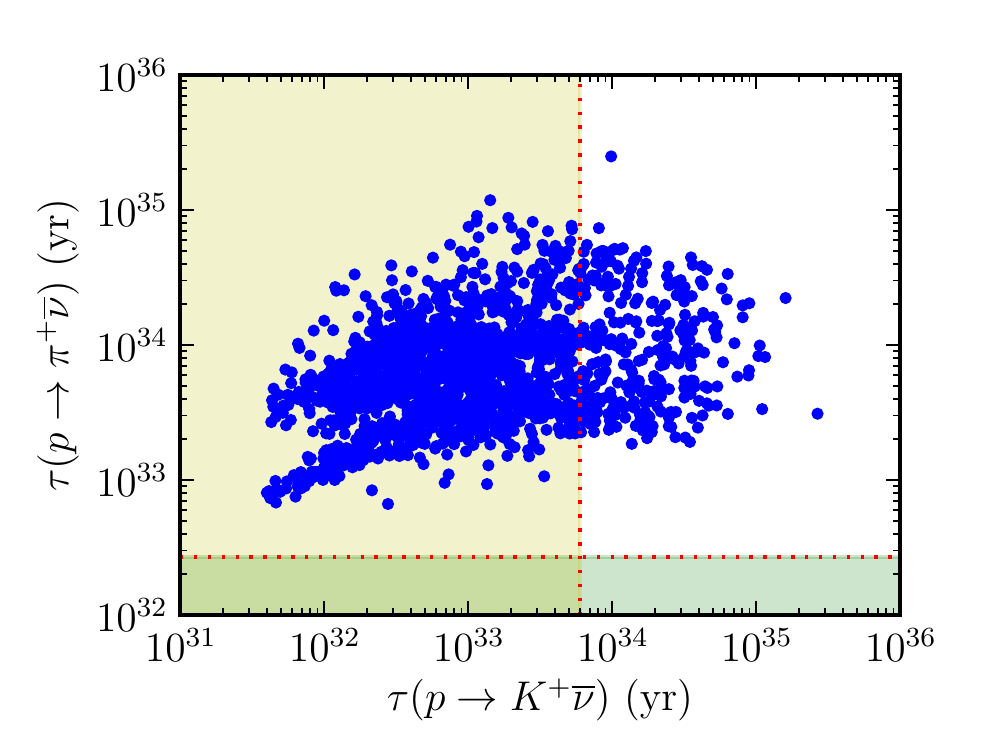}\\
        \includegraphics[width=7.8cm]{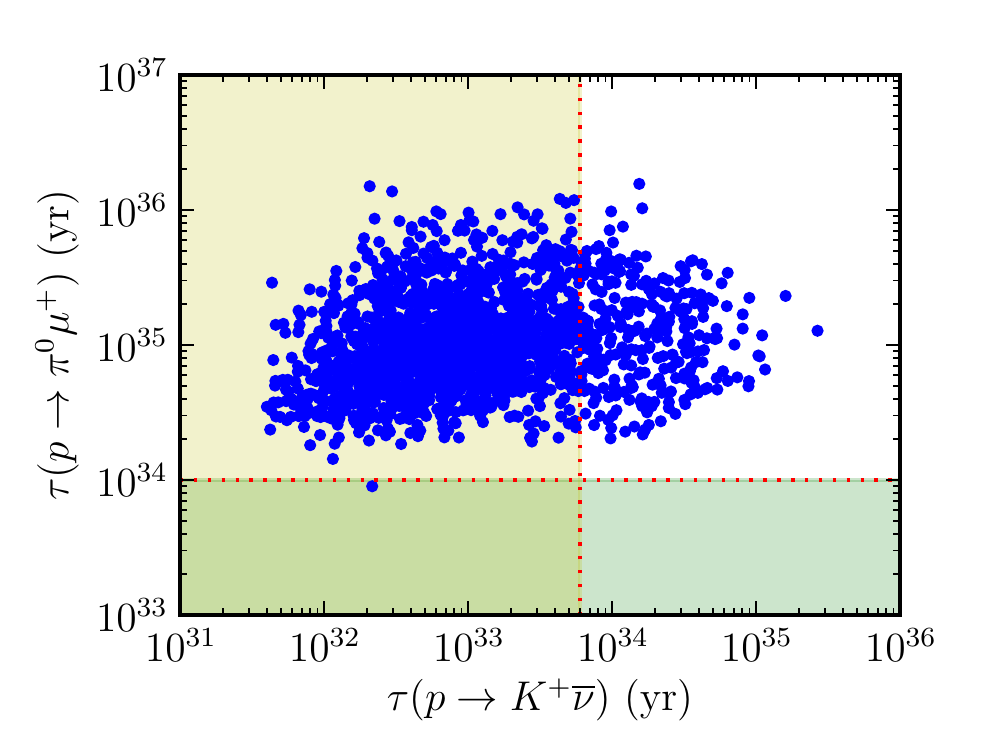}
        \includegraphics[width=7.8cm]{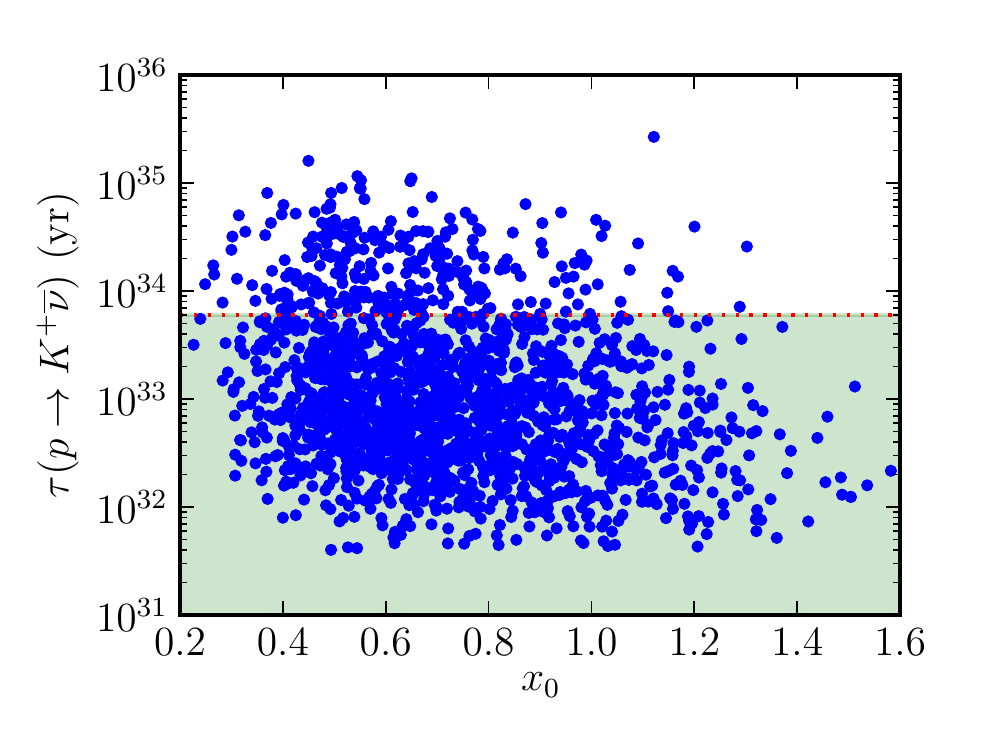}
        \caption[Comparisons of $K^+ \bar\nu$ partial lifetime to
        those of other dominant modes in the model, and that lifetime
        as a function of $x_0$, for the type-II case.]{Comparisons of
        $K^+ \bar\nu$ partial lifetime to those of other dominant
        modes in the model, and that lifetime as a function of the
        {\bf 10} mass parameter $x_0$, for the type-II case. Note the
        unsurprising preference for smaller $x_0$.}
	\label{fig:KnuplotsII}
\end{figure}

In order to locate an area of mixing parameter space which yields a
sufficient $K^+ \bar\nu$ lifetime, I wrote a supplementary Mathematica
code to search for minima among strongly abridged versions of
$\lvert\, \mathrm{C}_{\widetilde{W}}^I +
\mathrm{C}_{\widetilde{W}}^{I\!V}\, \rvert$ and $\lvert\,
\mathrm{C}_{\tilde{h}^\pm}^{I\!V}\, \rvert$ that contribute to the
decay width.\footnote{$\mathrm{C}_{\tilde{h}^\pm}^{I\!I\!I}$ and
$\mathrm{C}_{\tilde{h}^0}^{I\!I\!I}$ cancel identically for all
contributing channels of both the $K^+ \bar\nu$ and $\pi^+ \bar\nu$
modes.} Specifically I started with $x_0$ terms only, corresponding to
the baseline case, and then iteratively added back the largest
contributions one by one while readjusting the initial values each
time. Once all of the most important terms were present, I took the
resulting mixing parameters as my initial values in the Python code.
The resulting minimization gave a large percentage of trials with all
six modes exceeding the lifetime bounds.
\begin{figure}[t]
        \includegraphics[width=7.8cm]{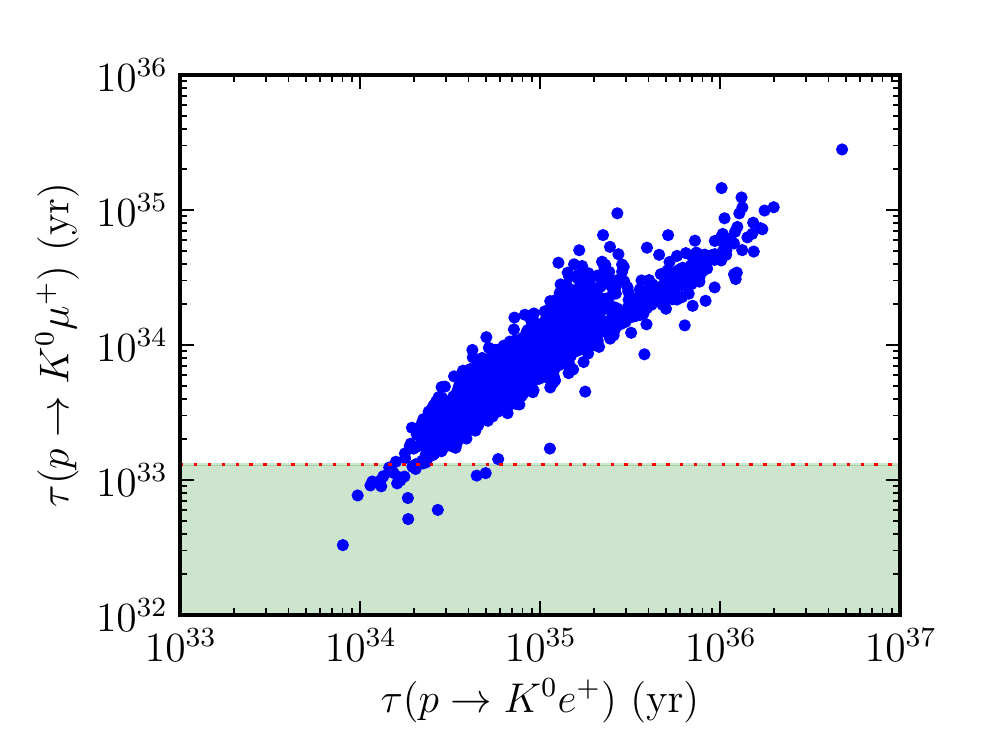}
        \includegraphics[width=7.8cm]{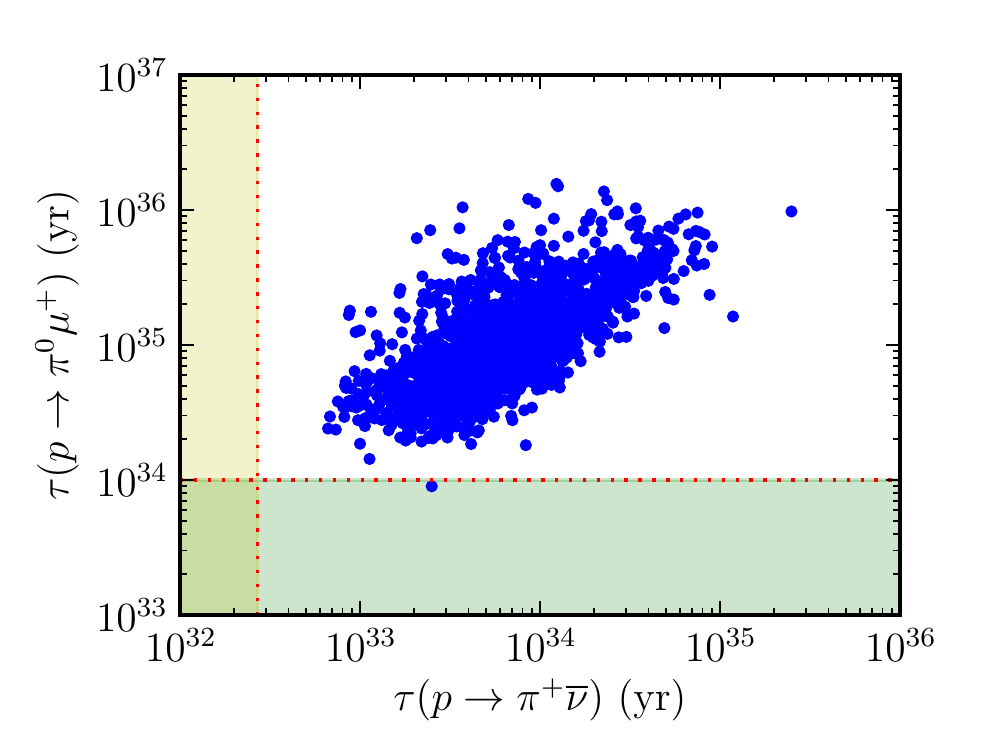}
        \caption[Comparisons of partial lifetimes among
        highly-correlated sub-dominant modes in the model for the
        type-II case.]{Comparisons of partial lifetimes among
        highly-correlated sub-dominant modes in the model for the
        type-II case.}
	\label{fig:otherplotsII}
\end{figure}

With an allowed region of parameter space found, I expanded my
searches to include a slightly wider range of values for the heavily
restricted $x_0$. Using six different ``seeds'' for parameter values,
all of which give every mode sufficient with $\tau(K^+ \bar\nu)$
roughly twice the experimental bound, I created a large number of
trials for which the initial values were distributed normally around
the seed values and with large standard deviations. The resulting data
for such a search is shown in scatter plots below. Figure
\ref{fig:KnuplotsII} gives the relationships between the $K^+ \bar\nu$
mode and other representative modes and also the distribution of $K^+
\bar\nu$ lifetime for varying $x_0$. Figure \ref{fig:otherplotsII}
shows the relationships between other more closely correlated modes
for completeness.

Note the strong correlation between $\pi^+ \bar\nu$ and $\pi^0 \mu^+$,
which are related by isospin, and the extreme correlation between $K^0
e^+$ and $K^0 \mu^+$. The latter is due to a manifestation of the
hierarchical nature of the Yukawas in the $C_{ijkl}$, as well as minor
features such $f_{11} \sim f_{12}$; similar structure is present in
the $y^f$ and $U^f$, which tend to also have $11 \sim 12$ or $11 \ll
12$; these properties result in a straightforward scaling under the
replacement $l: 1 \rightarrow 2$. Furthermore, the same relationship
is present between $\pi^0 e^+$ and $\pi^0 \mu^+$. These relationships
imply that the remaining plots I omitted differ only trivially from
the representatives present.

I also performed simple scans in search of a maximum value for
$\tau(K^+ \bar\nu)$, as well as taking note of any especially large
values in the previous searches. While there does not seem to be any
analytically-enforced maximum present in the model, I did consistently
find that $\tau > 10^{35}$\,years was extremely rare, and I never saw
a value higher than $\sim 6\!\times\! 10^{35}$\,yr. Given those
findings, combined with the apparent smallness of the swath of
parameter space yielding the above results and the low likelihood of a
more global minimum based on my search methods, I believe that
$\tau(K^+ \bar\nu) \gsim 10^{36}$\,yr is statistically infeasible in
this model for type-II seesaw. If such a value does exist, it is
likely contained in a vanishingly small area of allowed parameter
space and accomplished through truly extreme tuning. Therefore I will
take $10^{36}$\,years as a {\it de facto} upper limit on $\tau(K^+
\bar\nu)$ for the type-II case, which will not be accessible by
Hyper-K and similar experiments \cite{hyperk,dune} in the near future,
but should nonetheless allow the model to be tested eventually.

The other modes of course have similar limits, but it would seem that
all the others are substantially higher and thus either far beyond the
reach of the forthcoming experiments or beyond the contributions from
gauge boson exchange, if not both, with the possible exception of
$\tau(\pi^+ \bar\nu)$, which is rather highly correlated with $K^+
\bar\nu$ in this model. Determining that value is tricky though
because if I simply maximize the $\pi^+ \bar\nu$ mode, then the $K^+
\bar\nu$ mode will be below its bound; thus, there is some question as
to how one defines the maximization.

\subsection{Proton Partial Lifetimes for Type I Seesaw} I begin
again by examining the same baseline case for the partial lifetimes,
with $x_0 \sim 1$ and all other $x_i,y_i = 0$. The resulting values
for the dominant modes in the type-I case are given in Table
\ref{table:baselineI}. Here I find a much more favorable situation, in
that even the $K^+ \bar\nu$ mode decay width is sufficient, and in
fact the other modes exceed the bounds by 2-4 orders of magnitude.
Hence I expect that virtually all solutions will be adequate for
modes other than $K^+ \bar\nu$, and as long as there is no {\it
enhancement} due to (de)tuning among the ${\rm C}^{\cal A}$ factors,
that mode will be adequate as well.

This is of course a remarkable improvement over traditional models,
yet it seems to contradict our expectations given the properties of
the fermion fit. Why then is the model successful? There are two
primary reasons, both of which are quite subtle. The first reason is
that the smaller values for $g_{13}$ and $g_{23}$ seen in
eq.\,(\ref{eq:hfgI}) do in fact improve the situation, as I
suggested, while the larger $f_{12}$ and $g_{12}$ seem to have less
impact. Since $M\;(h_{33})$ is such an extremely dominant factor in
the Yukawas, it is generally the case that contributions involving
third generation are larger and more important than the others.
\begin{table}[t]
\begin{center}
  \begin{tabular}{||c|c||}\hline\hline
    decay mode & baseline for $\tau$ (yrs) \\ \hline 
    $p \rightarrow K^+ \bar\nu$   &  $7.87 \!\times\! 10^{33}$ \\
    $p \rightarrow K^0 e^+$       &  $5.93 \!\times\! 10^{35}$ \\
    $p \rightarrow K^0 \mu^+$     &  $2.45 \!\times\! 10^{35}$ \\
    $p \rightarrow \pi^+ \bar\nu$ &  $2.37 \!\times\! 10^{36}$ \\
    $p \rightarrow \pi^0 e^+$     &  $6.11 \!\times\! 10^{38}$ \\
    $p \rightarrow \pi^0 \mu^+$   &  $2.27 \!\times\! 10^{38}$ \\
    \hline\hline
  \end{tabular}
  \caption[Hypothetical baseline partial lifetimes determined using
  type-I solution Yukawas.]{Hypothetical baseline partial lifetimes
  determined using type-I solution Yukawas and $x_0 = 0.95$ with all
  other $x_i,y_i = 0$. Note in comparing with Table
  \ref{table:explims} that all modes satisfy the lower limits, and
  most do so by several orders of magnitude.}
\label{table:baselineI}
\end{center}
\end{table}

The second reason is even more unexpected, to the point that it was
not even examined in the preceding works on this ansatz. The unitary
matrices $U^f$ for the charged fermions are generally $\sim 1$, just
as one would expect, given the texture of CKM and the absence of any
known mixing among charge leptons. This model is no exception, with
off-diagonal terms generally ${\cal O}(10^{-1 \mbox{-} 3})$; however,
with such sparse or hierarchical (flavor basis) Yukawas due to the
ansatz, these ``small'' off-diagonal elements lead to ``small''
rotations of $h,f,g$ resulting in relatively substantial changes to
the textures of $\hat{h},\hat{f},\hat{g}$. Especially noteworthy are
the changes in $h \rightarrow \hat{h}$, where some previously-zero
off-diagonal elements are replaced by the same ${\cal O}(10^{-1
\mbox{-} 3})$ values seen in the $U^f$.

In light of the surprising non-triviality of the basis rotations, if
one compares $U^{u,d}$ for the type-I case: 
\begin{align}
  U^u = &\left( \begin{array}{ccc}
    0.994 & -0.1085 + 0.0057i &  0.00298 + 10^{-5}i \\
    0.1084 + 0.0057i &  0.994 &  0.0047 + 10^{-5}i  \\
   -0.0035 - 10^{-5}i & -0.0044 + 10^{-5}i &  0.99998
  \end{array} \right) \nonumber \\[2mm]
  U^d = &\left(\begin{array}{ccc}
    0.967 & -0.1087 + 0.2309i & 0.00175 + 0.001175i \\
    0.1086 + 0.2308i & 0.966  & 0.03935 + 0.00690i \\
   -0.0076 - 0.0072i & -0.0381 + 0.00613i & 0.9992
  \end{array} \right), 
\label{eq:VudI}
\end{align}
to those for the type-II case:
\begin{align}
  U^u = &\left( \begin{array}{ccc}
    0.972 &  0.2098 - 0.1044i & -10^{-5} - 0.010i \\
   -0.210 - 0.1043i &  0.971 & -0.00012 - 0.0414i  \\
   -0.0043 - 0.001i & -0.001 - 0.0423i &  0.999
  \end{array} \right) \nonumber \\[2mm]
  U^d = &\left(\begin{array}{ccc}
    0.9998 & 0.00633 - 0.0095i &  0.00765 - 0.01117i \\
   -0.00708 - 0.0095i &  0.9983 & 0.03386 - 0.04514i\\
   -0.00785 - 0.01054i & -0.03401 - 0.04514i &  0.9983        
  \end{array} \right), 
\label{eq:VudII}
\end{align}
one sees that the off-diagonal entries are the same size or smaller for
the type-I case in every entry except $U^d_{12},U^d_{21}$; furthermore,
several of the elements involving the third generation are smaller by
an order of magnitude. These differences may seem rather benign, but
in fact each of these slightly suppressed values individually
translates into a factor of 10 suppression in most of the dominant
$C_{ijkl}$, which all tend to involve third generation elements. In some
cases, two or even three such suppressions may affect a single ${\rm
C}^{\cal A}$ factor. The squaring of factors in the decay width then
gives suppressions of generally 2-4 orders of magnitude in the
lifetimes, which is precisely what one can see when comparing Tables
\ref{table:baselineII} and \ref{table:baselineI}.
\begin{figure}[t]
        \includegraphics[width=7.8cm]{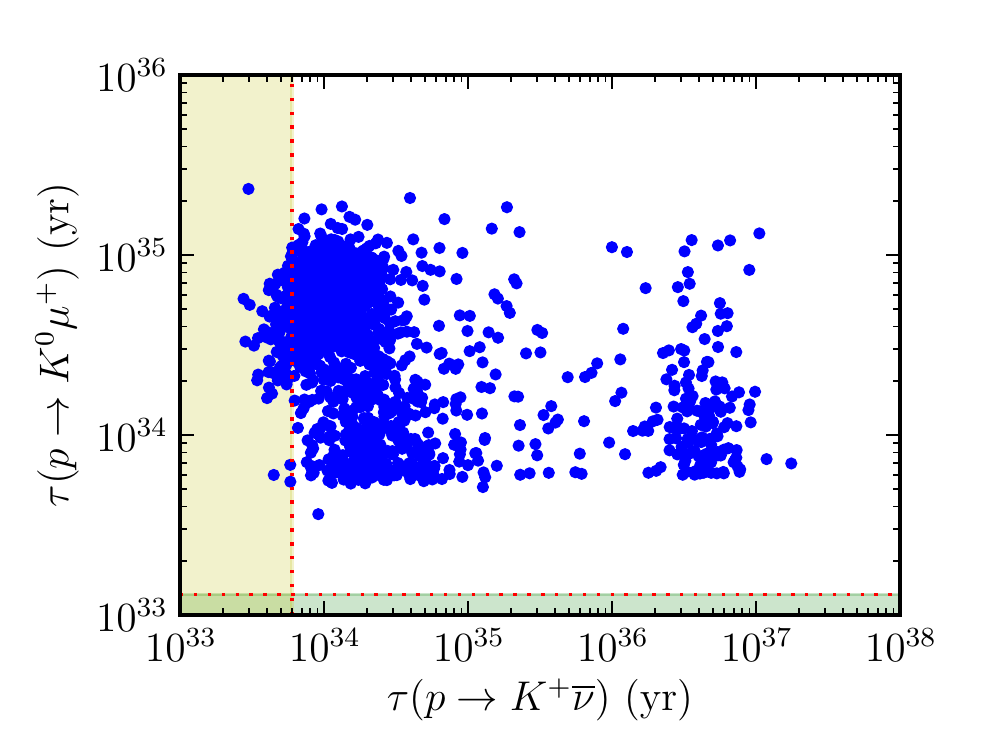}
        \includegraphics[width=7.8cm]{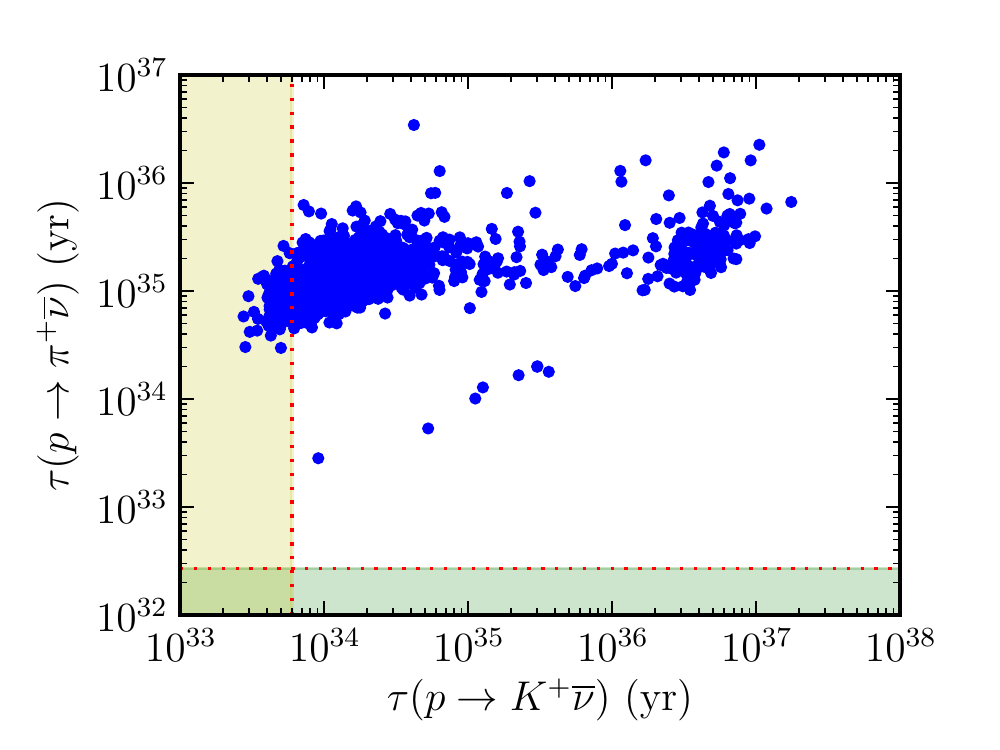}\\
        \includegraphics[width=7.8cm]{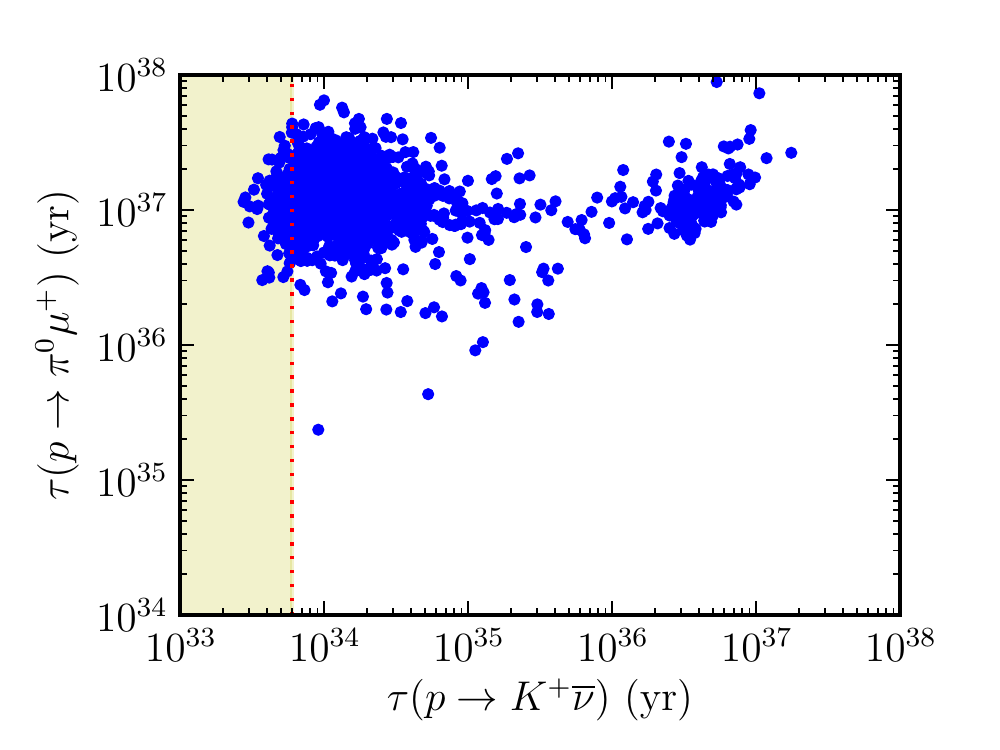}
        \includegraphics[width=7.8cm]{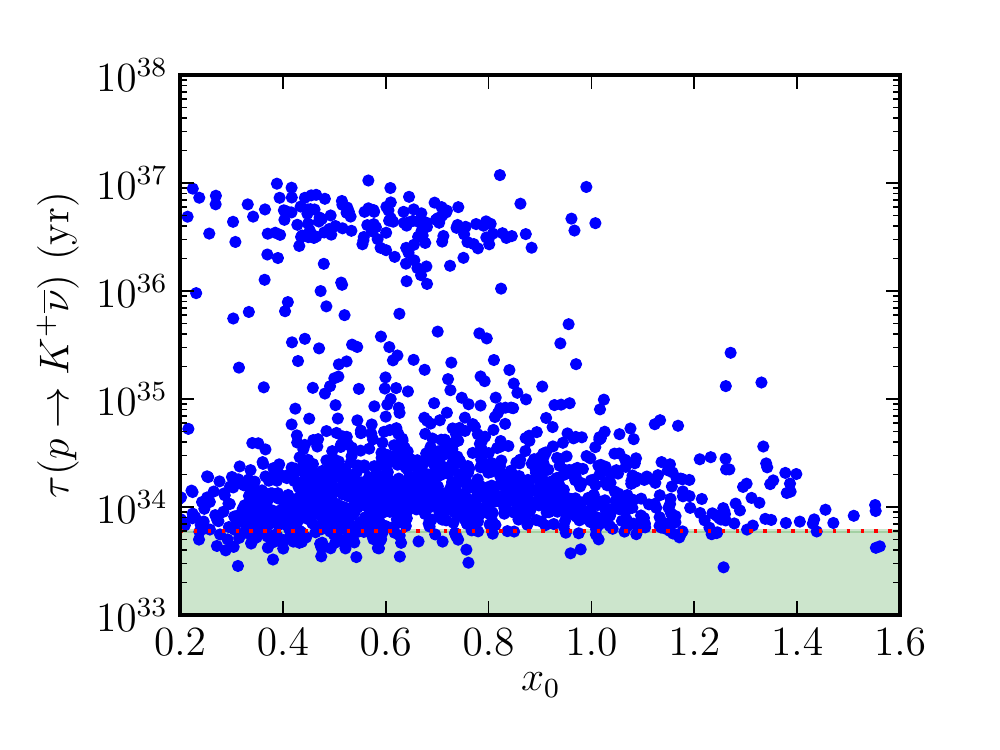}
        \caption[Comparisons of $K^+ \bar\nu$ partial lifetime to
        those of other dominant modes in the model, and that lifetime
        as a function of $x_0$, for the type-I case.]{Comparisons of
        $K^+ \bar\nu$ partial lifetime to those of other dominant
        modes in the model, and that lifetime as a function of the
        {\bf 10} mass parameter $x_0$, for the type-I case. Note the
        unsurprising preference for smaller $x_0$.}
	\label{fig:KnuplotsI}
\end{figure}
\begin{figure}[t]
        \includegraphics[width=7.8cm]{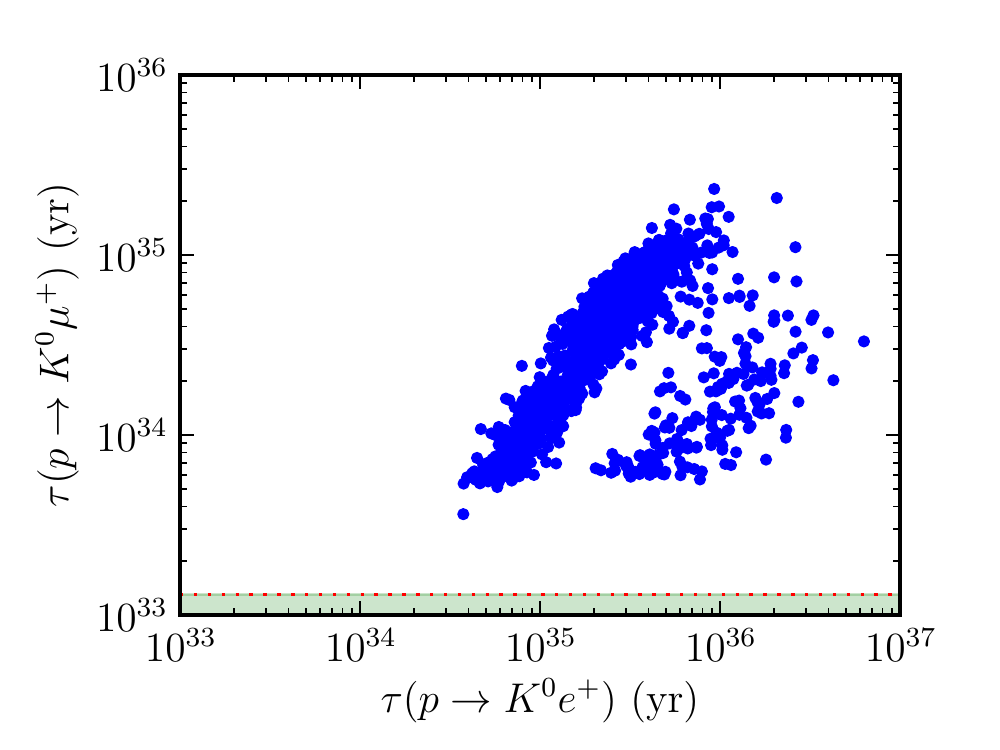}
        \includegraphics[width=7.8cm]{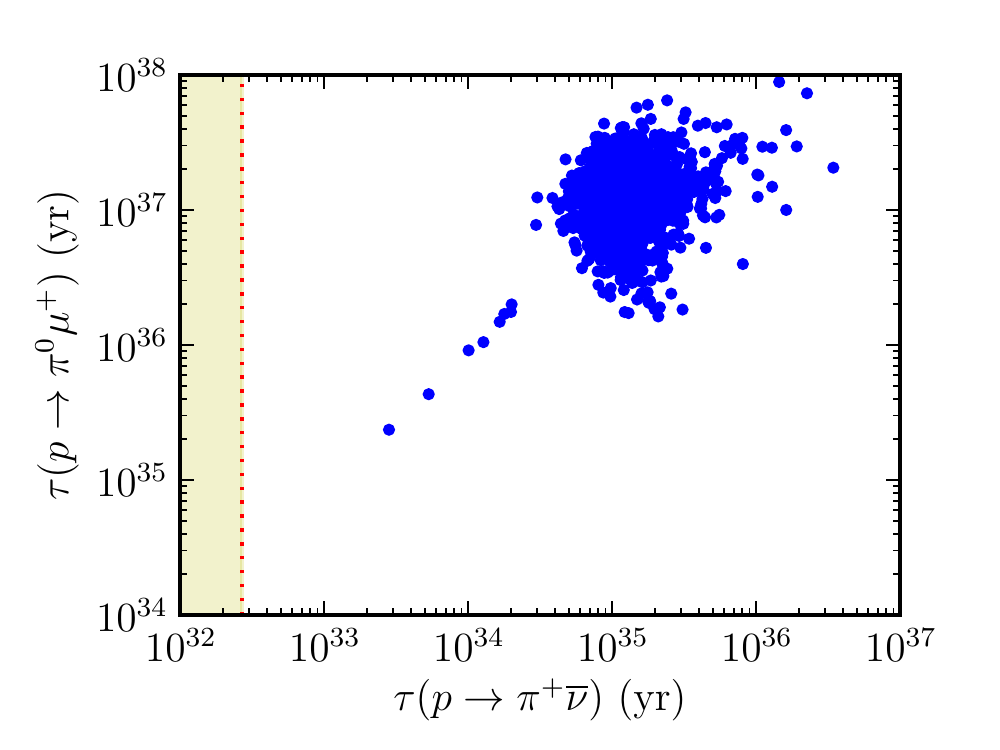}
        \caption[Comparisons of partial lifetimes among
        highly-correlated sub-dominant modes for the type-I
        case.]{Comparisons of partial lifetimes among
        highly-correlated sub-dominant modes in the model for the
        type-I case.}
	\label{fig:otherplotsI}
\end{figure}

Due to the more favorable circumstances, I was able to locate an
allowed region of parameter space for type-I simply by running a
large number of trials with the type-II parameter seeds. I repeated
the process of expanding the range of $x_0$ by again choosing five
seeds that gave every mode as sufficient and $\tau(K^+ \bar\nu)$ roughly
twice the experimental bound, and I again used those seeds to create
scatter plots for a large number of trials. Figure \ref{fig:KnuplotsI}
gives the relationships between the $K^+ \bar\nu$ mode and other
representative modes and the distribution of $\tau(K^+ \bar\nu)$ as a
function of $x_0$, and Figure \ref{fig:otherplotsI} shows the
relationships between other more closely related modes. Note the
bifurcation of the solution set in each plot; I have not yet been able
to discover the cause of this behavior.

Again I performed scans to determine a statistical upper bound for the
value of $\tau(K^+ \bar\nu)$ in the model. I consistently found that
$\tau > 10^{37}$\,years was rare and did not see a value higher than
$\sim 3\!\times\! 10^{37}$\,yr. Given those findings, I suspect that
the {\it de facto} upper limit on $\tau(K^+ \bar\nu)$ for the type-II
case is slightly lower than $10^{38}$\,years for the type-I seesaw
case.  Such a value is certainly out of reach of Hyper-K and other
imminent experiments. Note that as values for the neutral Kaon and
pion lifetimes often exceeded $10^{38}$\,years in my findings
involving $K^+ \bar\nu$ minimization, the upper limits for those modes
are surely sub-dominant to gauge exchange as well as out of reach of
experiments and so not of interest.


%
%

\chapter{Conclusion}\label{7-conclusion}

In this work I have presented a full analysis of the nature of proton
decay in an $SO(10)$ model that has {\bf 10}, $\overline{\bf{126}}$,
and {\bf 120} Yukawa couplings with restricted textures intended to
naturally give favorable results for proton lifetime as well as a
realistic fermion sector. The model is capable of supporting either
type-I or type-II dominance in the neutrino mass matrix, and I have
analyzed both types throughout. 

Using, numerical minimization of chi-squares, I was able to obtain
successful fits for all fermion sector parameters, including the
$\theta_{13}$ reactor mixing angle, and for both seesaw types. Using
the Yukawa couplings fixed by those fermion sector fits as input, I
then searched the parameter space of the heavy triplet Higgs sector
mixing for areas yielding adequate partial lifetimes, again using
numerical minimization to optimize results. For the case with type-II
seesaw, I found that lifetime limits for five of the six decay modes
of interest are satisfied for nearly arbitrary values of the triplet
mixing parameters, with an especially mild ${\cal O}(10^{-1})$
cancellation required in order to satisfy the limit for the $K^+
\bar\nu$ mode. Additionally, I deduced that partial lifetime values of
$\tau(K^+ \bar\nu) \gsim 10^{36}$\,years are vanishingly unlikely in
the model, implying the value can be taken as a {\it de facto}
lifetime for the mode, which makes the model ultimately testable. For
the case with type-I seesaw, I found that limits for {\it all six}
decay modes of interest are satisfied for values of the triplet mixing
parameters that do not result in substantial enhancement, with limits
for modes other than $K^+ \bar\nu$ satisfied for nearly arbitrary
parameter values; furthermore, I deduced a statistical maximum
lifetime for $K^+ \bar\nu$ of just under $10^{38}$\,years.

Given these results, I conclude that the well-motivated Yukawa texture
ansatz proposed by Dutta, Mimura, and Mohapatra is a remarkable
phenomenological success, capable of suppressing proton decay without
the usual need for cancellation, and without compromising any aspect of
the corresponding fermion mass spectrum. This result stands out among
similar analyses and perhaps represents a generally more favorable
approach for understanding the suppression of proton decay in grand
unified theory models. 


\singlespacing

\begin{appendices}
%
%

\chapter[Feynman Diagrams for Operators Contributing to Proton
Decay]{\Large Feynman Diagrams for Dimension-6 Operators Contributing
to Proton Decay} \label{fds}

$\tilde\phi_{\cal T}$ is the Higgsino component of a heavy
color-triplet Higgs superfield; $\phi = {\rm H},\bar\Delta,\Sigma$.

\paragraph*{Channels for $\boldsymbol{p \rightarrow \pi^+
\bar\nu}$} 

\mbox{ } \\ $i,l = 1,2,3$.  \vspace{6mm}
\begin{figure}[h]
\begin{center}
  \includegraphics{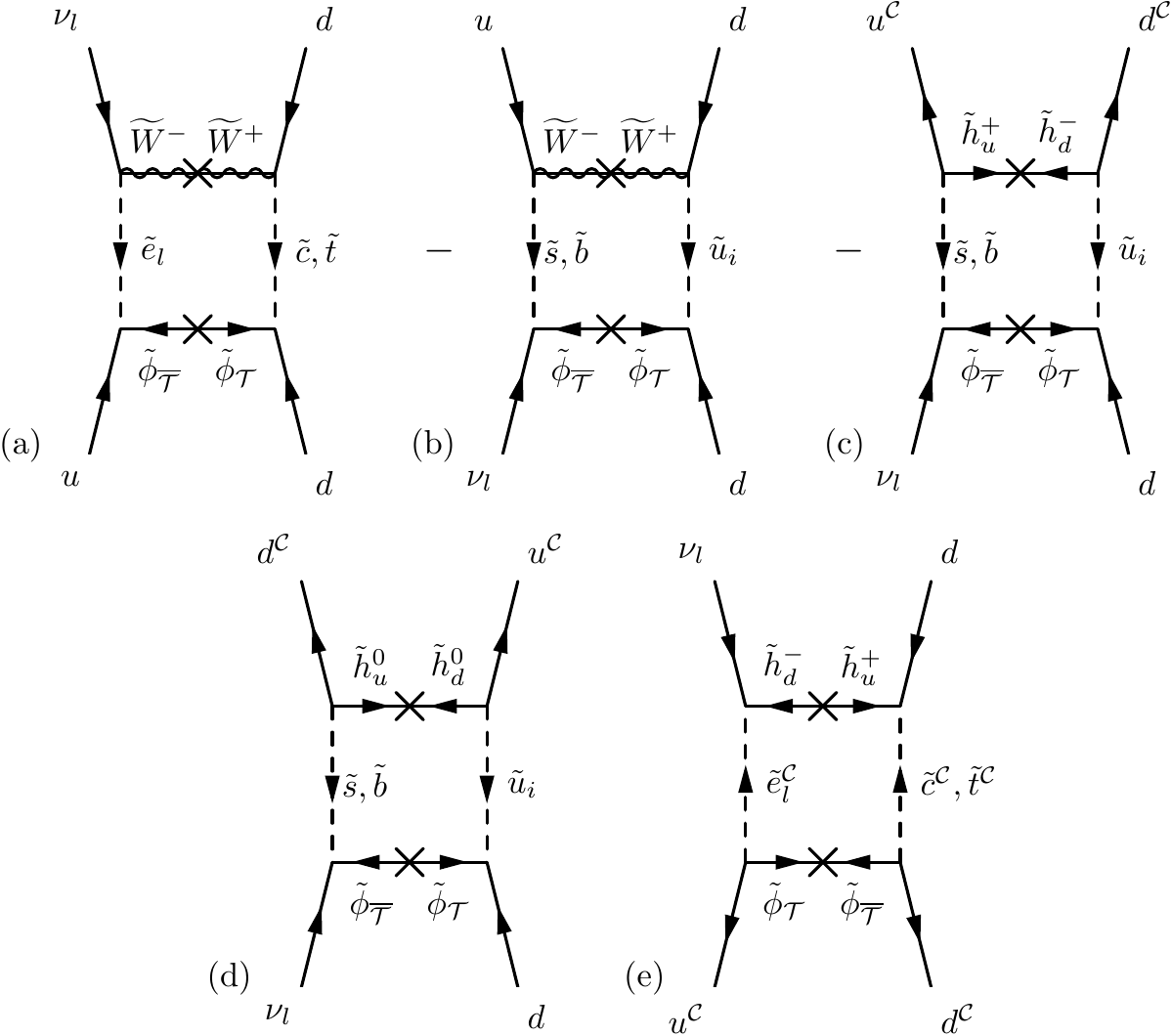}
  \vspace{-8mm}
\end{center}
\end{figure}

\newpage
\paragraph*{Channels for $\boldsymbol{p \rightarrow
\pi^0 \ell^+}$}

\mbox{ } \\ $j = 1,2,3$;\, $l = 1,2$ ($\leftrightarrow \ell = e,\mu$),
or for diagrams including $l'$, instead $l = 1,2,3$ and $l' = 1,2$.

\vspace{6mm}
\begin{figure}[h]
\begin{center}
  \includegraphics{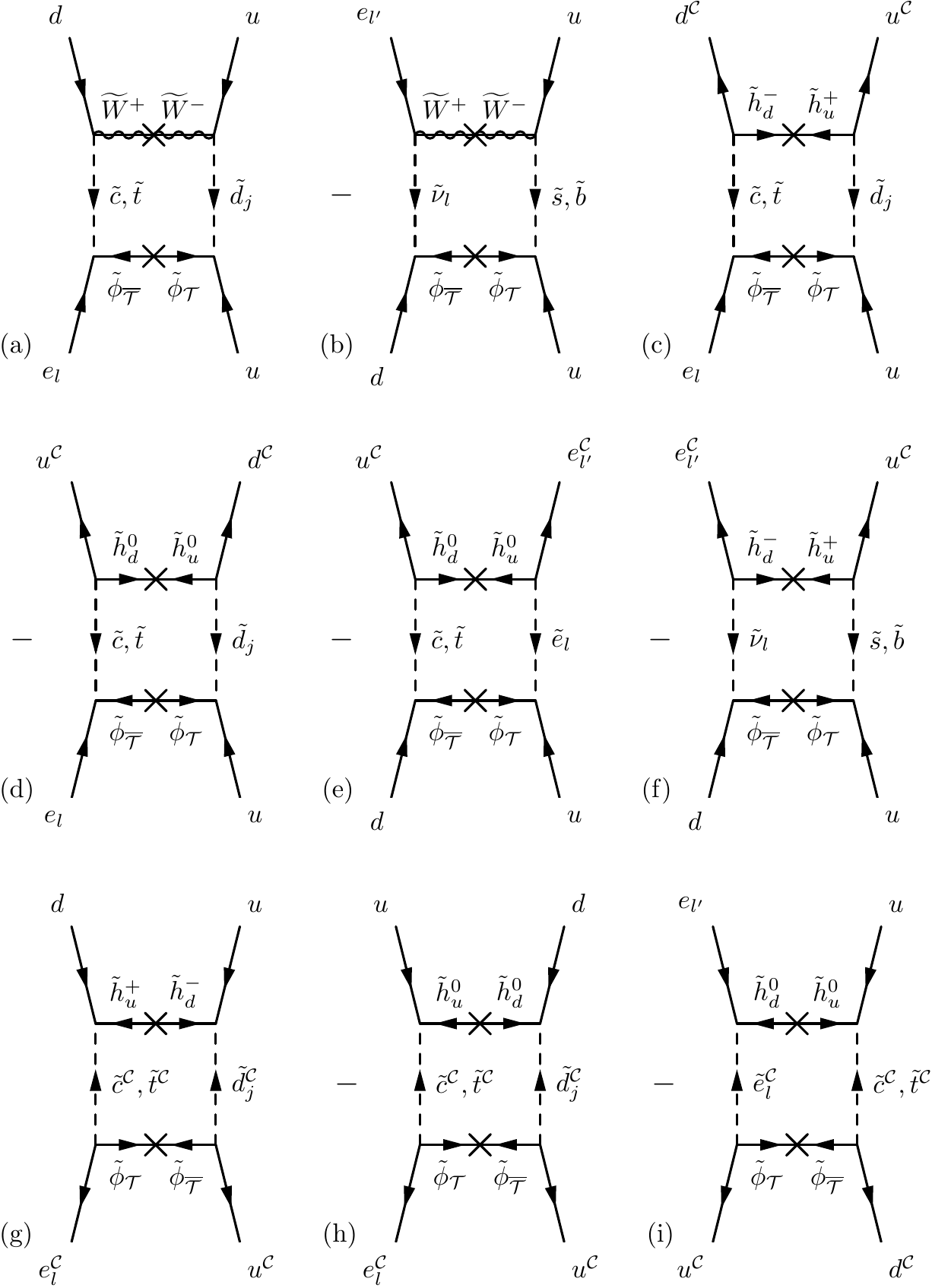}
  \vspace{-8mm}
\end{center}
\end{figure}

\newpage
\paragraph*{Channels for $\boldsymbol{p \rightarrow K^+
\bar\nu}$} 

\mbox{ } \\ $i,l = 1,2,3$; parentheses indicate coupled choices; 
absence of diagrams for $\tilde{u} d u \tilde{e}$ dressed by
$\tilde{h}^\pm$ and $u \tilde{d} d \tilde{\nu}$ dressed by
$\tilde{h}^0$ is due to resulting external $\nu^{\cal C}$.

\vspace{6mm}
\begin{figure}[h]
\begin{center}
  \includegraphics{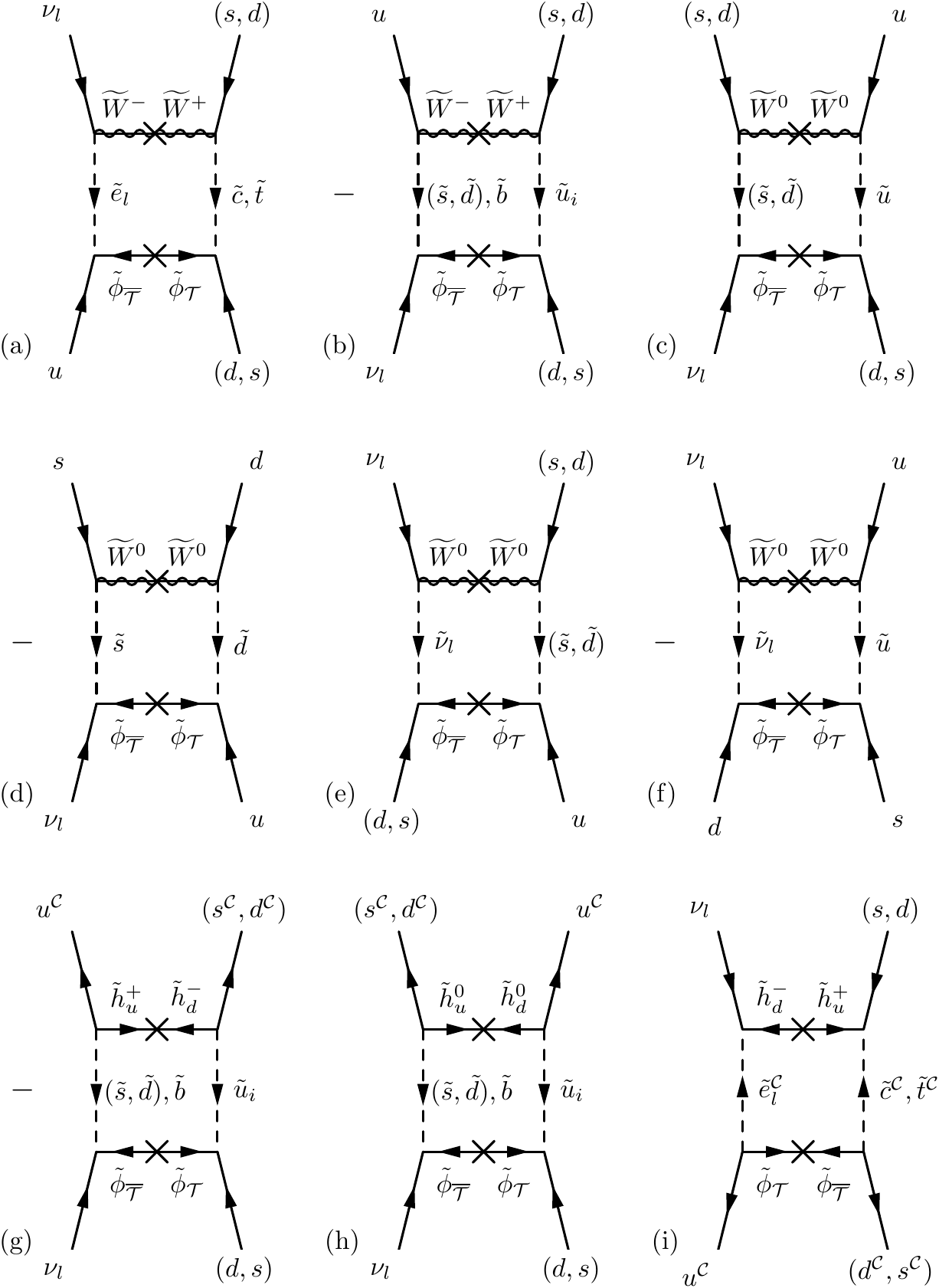}
  \vspace{-8mm}
\end{center}
\end{figure}

\newpage
\paragraph*{Channels for $\boldsymbol{p \rightarrow K^0
\ell^+}$}

\mbox{ } \\ $j = 1,2,3$;\, $l = 1,2$ ($\leftrightarrow \ell = e,\mu$),
or for diagrams including $l'$, instead $l = 1,2,3$ and $l' = 1,2$.

\vspace{6mm}
\begin{figure}[h]
\begin{center}
  \includegraphics{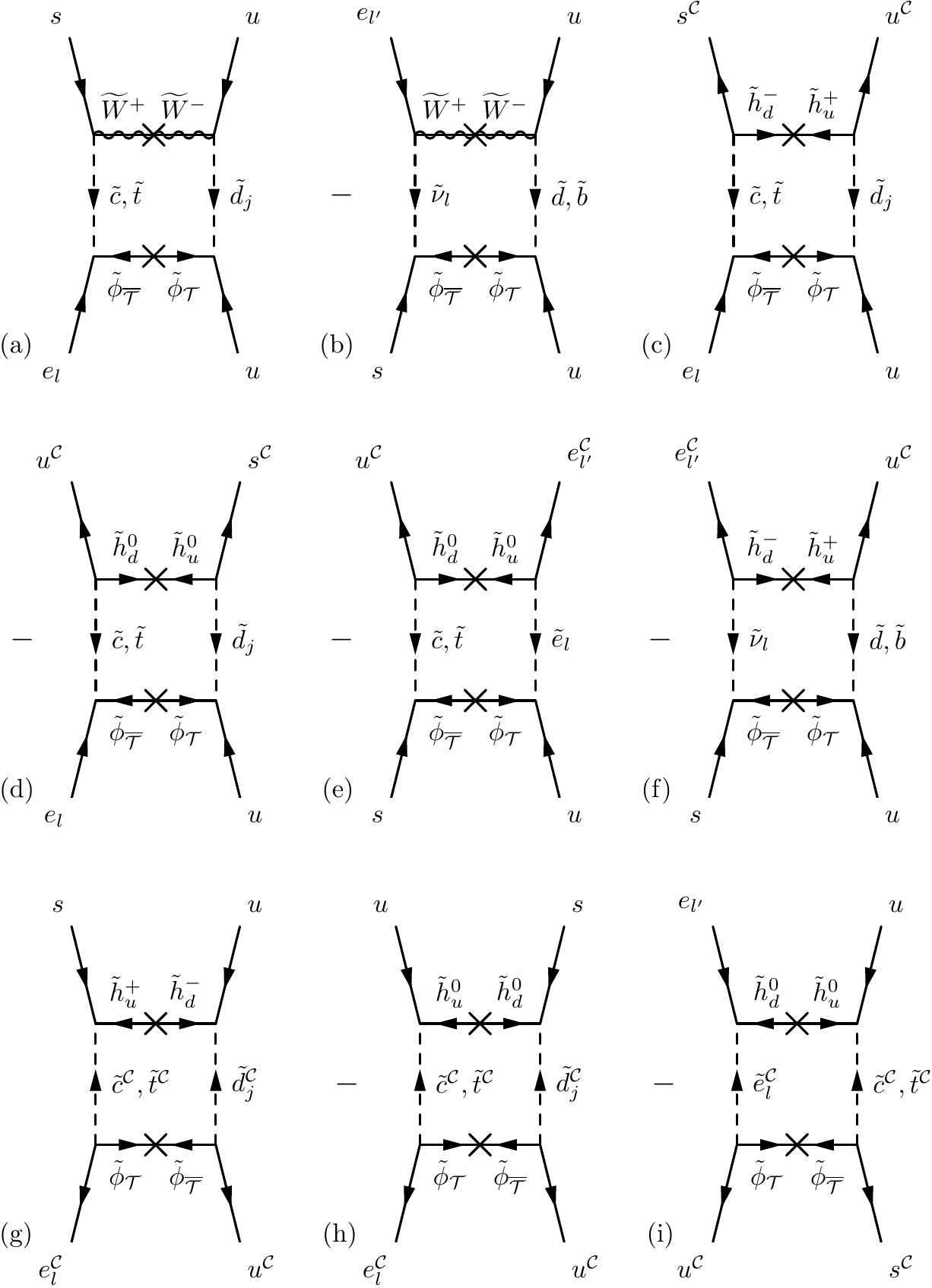}
  \vspace{-8mm}
\end{center}
\end{figure}


\end{appendices}

\bibliographystyle{unsrtnat}
\bibliography{thesis}

\end{document}